\documentclass[12pt,preprint]{aastex}

\newcommand{\Msun}{{\rm M}_{\sun}}
\newcommand{\Lsun}{{\rm L}_{\sun}}
\newcommand{\Ha}{\mathrm{H\alpha}}
\newcommand{\HI}{H~{\sc{i}} }
\newcommand{\HII}{H~{\sc{ii}} }
\newcommand{\SNII}{SN~{\sc{ii}}}

\shorttitle{Massive stars and the energy balance of the ISM}
\shortauthors{Freyer, Hensler, and Yorke}

\begin{document}


\title{Massive stars and the energy balance of the ISM. \\
       I. The impact of an isolated 60\,$\Msun$ star}


\author{Tim Freyer and Gerhard Hensler}
\affil{Institut f\"ur Theoretische Physik und Astrophysik
       der Universit\"at, D-24098 Kiel, Germany}
\email{freyer@astrophysik.uni-kiel.de,hensler@astrophysik.uni-kiel.de}

\and

\author{Harold W. Yorke}
\affil{Jet Propulsion Laboratory, MS 169-506, 4800 Oak Grove Drive,
       Pasadena, CA 91109, USA}
\email{Harold.Yorke@jpl.nasa.gov}




\begin{abstract}
  We present results of numerical simulations carried out with a 2D radiation
  hydrodynamics code in order to study the impact of massive stars on their
  surrounding interstellar medium. This first paper deals with the evolution
  of the circumstellar gas around an isolated 60\,$\Msun$ star.
  The interaction of the photoionized \HII region with the stellar wind bubble
  forms a variety of interesting structures like shells, clouds, fingers, and
  spokes. These results demonstrate that complex
  structures found in \HII regions are not necessarily relics from the time
  before the gas became ionized but may result from dynamical processes during
  the course of the \HII region evolution. We have also analyzed the transfer
  and deposit of the stellar wind and radiation energy into the circumstellar
  medium until the star explodes as a supernova.
  Although the total mechanical wind
  energy supplied by the star is negligible compared to the accumulated energy
  of the Lyman continuum photons, the
  kinetic energy imparted to the circumstellar gas over the star's lifetime
  is 4 times higher than for a comparable windless
  simulation. Furthermore, the thermal energy of warm photoionized 
  gas is lower by some 55\,\%. Our results document the necessity to consider
  both ionizing radiation and stellar winds for an
  appropriate description of the interaction of OB stars with their
  circumstellar environment.
\end{abstract}

\keywords{
  galaxies: evolution---\HII regions---hydrodynamics---instabilities---ISM:
  bubbles---ISM: structure
}

\section{Introduction}
\label{sec_intro}

  Massive stars play an important role in the evolutionary history of
  galaxies. They are the primary source of metals and dominate
  the turbulent energy input into the interstellar medium (ISM)
  by stellar winds, radiation, and supernova explosions. The radiation field
  of a massive star
  first dissociates the ambient molecular gas and forms a so-called
  photodissociation region of neutral hydrogen. Subsequently, the Lyman
  continuum photons of the star ionize the \HI gas and produce an \HII region
  that expands into the neutral ambient medium. A fast stellar wind creates
  shocks that form a so-called stellar wind bubble (SWB) filled with hot
  plasma, which expands into the \HII region. Finally, the star explodes as a
  supernova of type\,{\sc ii} (\SNII), creating a supernova remnant (SNR) that
  sweeps up the ambient medium. The SWBs and SNRs of neighboring stars can
  overlap and form a superbubble with a diameter of order 1\,kpc.

  This paper pursues two major goals. First, we want to examine the combined
  influence of wind and ionizing radiation on the dynamical evolution of
  circumstellar matter around massive stars, i.e.~we are interested in the
  interaction processes between the photoionized \HII region and the SWB that
  evolves into the ionized gas. The second goal is to improve
  our knowledge of the energy transfer efficiency between massive stars and
  the ISM: How and to what fraction is the energy of stellar radiation and
  the stellar wind converted into kinetic, thermal, and ionization energy of
  the ISM? How does the formation of the SWB influence the 
  transfer of stellar radiation? To what extent does all this depend on the
  evolutionary state of the star?

  To investigate these effects we perform numerical 2D radiation hydrodynamic
  simulations of the interaction between an isolated massive star and its
  surrounding ISM via stellar hydrogen-ionizing
  photons and a stellar wind. We calculate the hydrodynamical evolution of the
  circumstellar gas coupled with radiation transfer, time-dependent ionization
  of hydrogen, and a realistic description of cooling. The stellar mass-loss
  rate, the terminal velocity of the wind, the effective temperature, and the
  luminosity of the star are specified as time-dependent boundary conditions. 
  We examine the evolution of the circumstellar material starting from the
  main sequence (MS) phase of the star until it explodes as a supernova.
  We do not consider the SNR formation. In this paper we
  present results of these calculations for a star with an initial mass of
  60\,$\Msun$, extending the work of \citet*[hereafter GML1]{garcia96a}
  to a more precise analysis of the energetic aspect during the whole
  evolution of the star.

  The remainder of this paper is structured as follows:
  In section \ref{sec_theo_and_obs} we briefly review theoretical
  and observational studies of \HII regions and SWBs around massive stars
  for later comparison.  Our numerical procedure as well as the 
  initial conditions and time-dependent boundary conditions 
  are described in section \ref{sec_numerics}.
  In section \ref{sec_results} we present the numerical results of our
  calculations and discuss them in the context of previous analytical
  approaches, other numerical investigations, and observations.
  Finally, in section \ref{sec_conclusions} we summarize our main results
  and draw some conclusions from the work presented in this paper.

\section{Theoretical concepts and observational constraints}
\label{sec_theo_and_obs}

\subsection{\HII regions}
\label{sec_HII_regions}

  \HII regions are observed in various sizes and shapes. Depending on
  their geometrical extent, they are labeled as ultracompact, compact,
  extended, or giant. The ultracompact (linear size $\lesssim$ 0.1\,pc) and
  partly also the compact (0.1--0.3\,pc) \HII regions are still deeply embedded
  in their birthplaces, mature molecular clouds, and due to dust obscuration 
  generally are only observable in the IR and radio wavelengths. On the other
  hand, extended \HII regions with sizes up to several parsecs and giant
  \HII regions which are composed of individual \HII regions and have sizes of
  a few 100\,pc are often easily discernible in the optical by their bright
  Balmer and forbidden metal lines.
  Large holes and shells make it obvious that the giant \HII regions are not
  only powered by stellar radiation but also by stellar winds and supernova
  explosions in the underlying OB star cluster \citep[see e.g.][]{yang96}.

  Within the framework of the simplest theoretical approach an O star
  suddenly ``turns on'' in a constant density, motionless medium and begins to
  ionize its surroundings. The reaction of the medium to the stellar photons
  is well-known and has been described in detail in standard textbooks
  \citep[see e.g.][]{spitzer78}. For comparison purposes with our
  numerical results we are particularly interested in the time evolution of
  the location $ r_{\mathrm{i}}$ of the ionization front:
  \begin{eqnarray}
    r_{\mathrm{i}} &=&
      \left(\frac{7}{4}\,c_{\mathrm{s,\scriptscriptstyle{II}}}\,
      r_0^{3/4}\tau + r_0^{7/4} \right)^{4/7}  \\
    r_0 &=& \left({3 L_\mathrm{LyC} \over 4 \pi n_{\mathrm{0}}^2
         \beta^{(2)}(T)} \right)^{1/3} \ ,
    \label{eq_r_i}
  \end{eqnarray}
  where $c_{\mathrm{s,\scriptscriptstyle{II}}}$ is the isothermal
  sound speed of the ionized gas, $r_0$ the initial Str\"omgren radius,
  $L_\mathrm{LyC}$ the photon luminosity in the Lyman continuum,
  $n_{\mathrm{0}}$ the hydrogen number density in the neutral
  ambient medium, $\beta^{(2)}(T)$ the coefficient for recombinations
  of hydrogen into levels 2 and higher,
  and $\tau$ the age of the star. According to \citet{lasker67} the
  kinetic energy of the expanding swept-up shell is
  \begin{equation}
    E_{\mathrm{k}}
       = \frac{4}{3}\,\pi\,n_{\mathrm{0}}\,m_{\mathrm{H}}\,
          c_{\mathrm{s,\scriptscriptstyle{II}}}^2\,r_0^{3/2}
          \left[\left(\frac{7}{4}\,
          c_{\mathrm{s,\scriptscriptstyle{II}}}\,r_0^{3/4}\,
          \tau + r_0^{7/4}\right)^{6/7}
          - r_0^{3/2}\right]\ ,
    \label{eq_E_k}
  \end{equation}
  where $m_{\mathrm{H}}$ is the mass of a hydrogen atom.
  Because this swept-up shell can be expected
  to remain cool and neutral due to strong radiative cooling,
  we can also estimate the ionization energy stored in the \HII region:
  \begin{equation}
    E_{\mathrm{i}}
        = \frac{4}{3}\,\pi\,n_{\mathrm{0}}\,
          \left(\frac{7}{4}\,c_{\mathrm{s,\scriptscriptstyle{II}}}\,
          r_0^{5/2}\tau + r_0^{7/2} \right)^{6/7}
          \chi_0\ .
      \label{eq_E_i}
  \end{equation}
  Here we have used equation (\ref{eq_r_i}) for the time-dependent
  radius of the ionization front and $\chi_0$ is the ionization
  potential of hydrogen in the ground state.

  The same ansatz can then be used to estimate the thermal energy of warm gas
  in the \HII region:
  \begin{equation}
    E_{\mathrm{t}}
      = 4\,\pi\,n_{\mathrm{0}}\,
          \left(\frac{7}{4}\,c_{\mathrm{s,\scriptscriptstyle{II}}}\,
          r_0^{5/2}\tau + r_0^{7/2} \right)^{6/7}
          k_{\mathrm{B}}\,T_{\mathrm{\scriptscriptstyle{II}}}\ ,
      \label{eq_E_t}
  \end{equation}
  where we will insert $T_{\mathrm{\scriptscriptstyle{II}}} = 8000\,\mathrm{K}$
  as an appropriate approximation to the temperature in the \HII region.
  $k_{\mathrm{B}}$ is the Boltzmann constant.

\subsection{Theory of stellar wind bubbles}
\label{sec_bubbles}

\begin{figure}
  \plotone{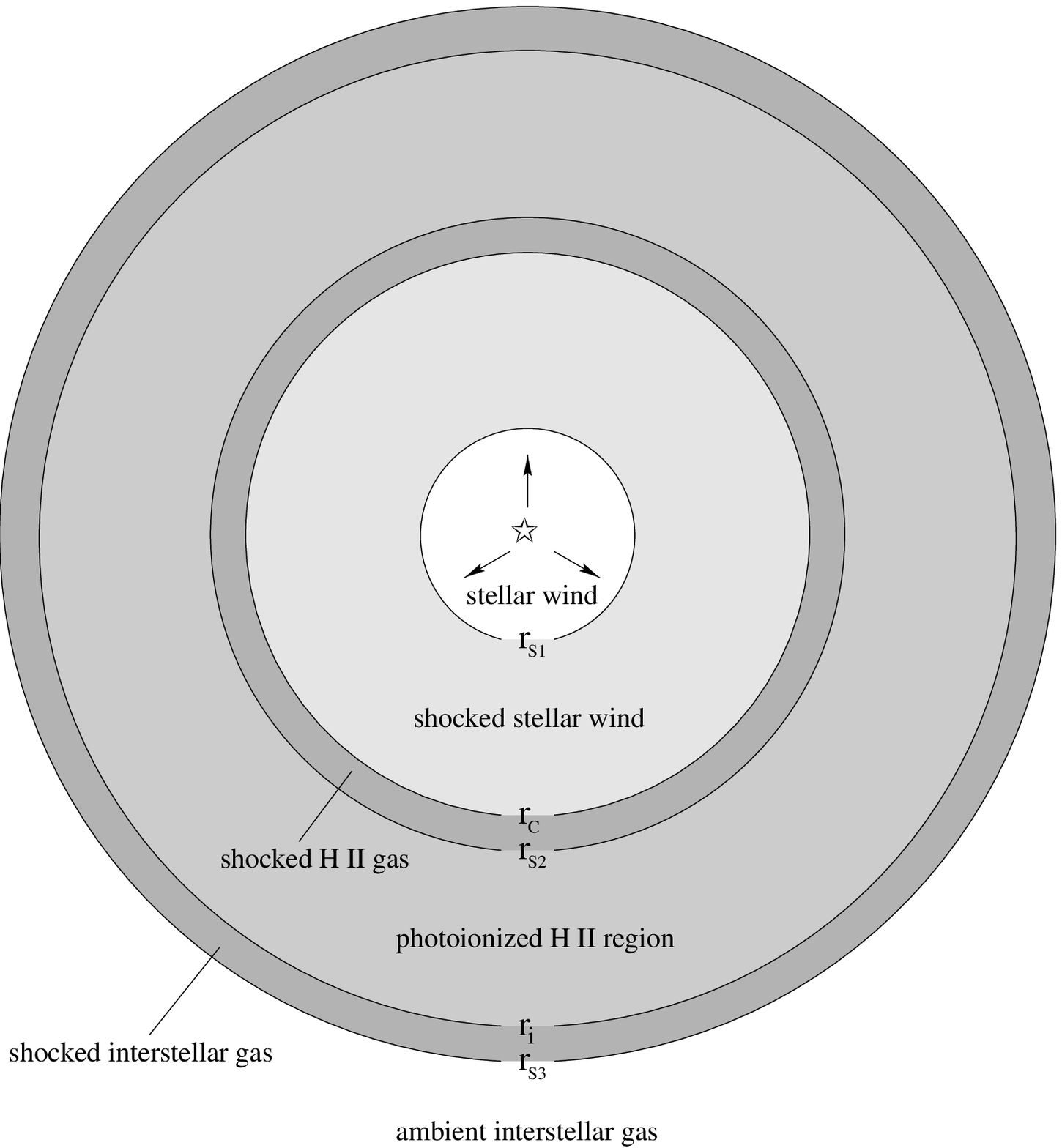}
  \caption{Schematic structure of an SWB: $r_{\mathrm{s1}}$ marks the
           position of the reverse shock, $r_{\mathrm{c}}$ the contact
           discontinuity, $r_{\mathrm{s2}}$ the forward shock of the stellar
           wind bubble, $r_{\mathrm{i}}$ the ionization front, and
           $r_{\mathrm{s3}}$ the forward shock of the \HII region expansion.
           \label{bubble_structure_new_up.eps}
          }
\end{figure}

  First considerations about the inner structure of SWBs go back to
  \citet{pikel'ner68}, \citet{avedisova72}, \citet{dyson72}, and
  \citet{dyson73}, but the work of
  \citet*{castor75} and \citet{weaver77} set a milestone in the
  approach of understanding SWBs. They presented a fairly complete
  picture of the structure and evolution of SWBs together
  with a set of equations that describes the evolution under the simplifying
  assumptions of a point source of a constant and spherically symmetric
  strong wind that interacts with a homogeneous ambient ISM. The global
  structure that arises from such a wind-ISM interaction is depicted in
  Figure \ref{bubble_structure_new_up.eps} together with the expanding
  \HII region into which the SWB evolves.
  It consists of a free-flowing supersonic wind that is heated to about
  $\mathrm{10^6-10^8\,K}$ when it passes the inner reverse shock at
  $r_{\mathrm{s1}}$.
  The pressure of this hot rarefied gas that normally fills most of the
  volume of the SWB is typically much higher than the pressure in the
  photoionized ambient medium. As a consequence, the hot gas bubble expands
  into the \HII region producing a forward shock at $r_{\mathrm{s2}}$ that
  sweeps up the gas from the \HII region in a shell which is separated from
  the hot bubble interior by a contact discontinuity at $r_{\mathrm{c}}$. 

  Three phases in the evolution of such a bubble can be distinguished:
  The first is the adiabatic phase that lasts until the shock speed
  $v_{\mathrm{s2}}$ drops
  below $\mathrm{200\,km\,s^{-1}}$ \citep{falle75}, typically a few 10 to a
  few 1000\,yr, depending on the mechanical luminosity of the wind and on the
  ambient density. The bubble is expanding so fast that radiative cooling
  does not play a significant role for the dynamical behavior.
  The second stage of evolution is characterized by strong cooling in the
  shell of swept-up material (between $r_{\mathrm{c}}$ and $r_{\mathrm{s2}}$),
  allowing it to be compressed into a geometrically thin
  ($r_{\mathrm{c}}(t) = r_{\mathrm{s2}}(t)$, where $t$ denotes time), dense
  shell, whereas for the hot bubble (between $r_{\mathrm{s1}}$ and
  $r_{\mathrm{c}}$) cooling is still negligible. This phase lasts much longer
  than the first, so-called adiabatic phase. 
  Thermal conduction from the hot bubble interior to the collapsed shell may
  become important and may modify the structure of the bubble in this stage.
  \citet{weaver77} assume that an equilibrium between the
  conductive energy flux and the mechanical energy flux due to evaporation of
  shell mass in the reverse direction is established. Under the assumptions
  that in the hot bubble between $r_{\mathrm{s1}}$ and
  $r_{\mathrm{c}} = r_{\mathrm{s2}}$
  the pressure is everywhere the same and that the thermal energy contained
  within this region is much higher than the kinetic energy, 
  \citet{weaver77} derived the following equations for the
  temporal dependence of the radius $r_{\mathrm{s2}}(t)$
  and the pressure in the hot bubble $P_{\mathrm{b}}(t)$ in this stage: 
  \begin{eqnarray}
    r_{\mathrm{s2}}(t) & = & r_{\mathrm{c}}(t)~=~\left(\frac{125}{154\pi}
               \right)^{1/5}\,L_{\mathrm{w}}^{1/5}\,\rho_{0}^{-1/5}\,t^{3/5}\ ,
    \label{eq_bubble_ph2_R2} \\
    P_{\mathrm{b}}(t) & = & \frac{7}{\left(3850\pi\right)^{2/5}}\,
                    L_{\mathrm{w}}^{2/5}\,\rho_{0}^{3/5}\,t^{-4/5}\ .
    \label{eq_bubble_ph2_P}
  \end{eqnarray}
  $\rho_{0} = n_{0}\,m_{\mathrm{H}}$ is the density of the ambient medium and 
  $L_{\mathrm{w}} = \frac{1}{2}\dot{M}_{\mathrm{w}} v_{\mathrm{w}}^2$
  the stellar wind luminosity, with $\dot{M}_{\mathrm{w}}$ and
  $v_{\mathrm{w}}$ being the mass-loss rate and the terminal velocity of
  the star. 

  It is also possible to obtain an analytical result for the energy that is
  deposited in the ISM by SWBs. We neglect the first (fully adiabatic) stage
  of bubble evolution because it is very short. Cooling in the hot bubble
  strongly depends on the efficiency of heat conduction, or, how efficiently
  heat conduction is suppressed by magnetic fields. We will assume for the
  analytical theory that there is no heat conduction and thus cooling in the
  hot bubble is unimportant during the lifetime of the star. 
  The hot bubble acts like a piston on the ambient medium and analogous
  to the case of an \HII region it can be shown that the $P\mathrm{d}V$ work
  is equally distributed on kinetic energy of shell motion and thermal energy
  in the shell (which gets immediately lost due to cooling). If we furthermore
  assume that the kinetic energy of the stellar wind is completely
  transformed into thermal energy at the reverse shock (the strong-jump
  conditions give 15/16, but clumpiness of the stellar wind, which we
  completely neglect, might reduce this value), we get for the kinetic
  energy:
  \begin{equation}
    E_{\mathrm{k}}
      = \frac{1}{2}\,\int_0^{V_{\mathrm{s2}}(\tau)}
        P_{\mathrm{b}}\,\mathrm{d}V_{\mathrm{s2}}\ . \nonumber
  \end{equation}
  $V_{\mathrm{s2}}$ is the volume inside $r_{\mathrm{s2}}$.
  Using equation (\ref{eq_bubble_ph2_R2}) for $r_{\mathrm{s2}}$ and
  equation (\ref{eq_bubble_ph2_P}) for the pressure in the hot bubble yields
  \begin{equation}
    E_{\mathrm{k}} = \frac{3}{11}\,L_{\mathrm{w}}\,\tau 
    \label{eq_bubble_E_k}\ .
  \end{equation}
  The thermal energy of the hot gas is thus
  \begin{equation}
    E_{\mathrm{t}}
      = L_{\mathrm{w}}\,\tau - \int_0^{V_{\mathrm{s2}}(\tau)}
        P_{\mathrm{b}}\,\mathrm{d}V_{\mathrm{s2}}
      = \frac{5}{11}\,L_{\mathrm{w}}\,\tau
      \label{eq_bubble_E_t}\ .
  \end{equation}
  Besides heat conduction and cooling in the hot bubble 
  we have also neglected the fact that part of the thermal energy might be
  used for collisional ionization of gas. Thus, the results for
  $E_{\mathrm{k}}$ and $E_{\mathrm{t}}$ are upper limits.

  Though the analytical and semi-analytical solutions for the evolution of
  SWBs have been improved in recent years
  \citep*{koo92a, koo92b, garcia95a, pittard01a, pittard01b}
  a variety of physical effects remains to be included in order to
  achieve a better agreement of models and observations. For example, the
  discrepancy between models and observations with regard to the evolution of
  the hot phase in bubbles has recently been reviewed by \citet{maclow00}.
  See also \citet{chu00}. It has become evident that
  the stellar parameters such as the mass-loss rate, the terminal velocity, the
  effective temperature and the luminosity of the star which drive the
  evolution of the circumstellar matter vary strongly over time.
  Because most previous studies dealt with either the
  evolution of \HII regions or of SWBs, little is known about the
  interaction of these two structures.

\subsection{Observations of stellar wind bubbles}
\label{sec_observations}

  Hot gas ($\mathrm{10^6-10^8\,K}$) is
  expected to be a crucial indicator for the existence of SWBs. However,
  there have been only two successful X-ray observations of SWBs so far:
  NGC\,6888 \citep*{bochkarev88, wrigge94} and
  S308 \citep{wrigge99}. Both are actually Wolf-Rayet (WR) bubbles
  (i.e. the wind-driving star has already entered its WR stage).
  The fact that no MS bubble has yet been observed in X-rays
  may be due to their physical properties; they are expected to be
  large, diffuse, and consequently dim (compared with WR
  bubbles). There may also be a problem with the theoretical models.
  More sensitive X-ray telescopes of the next generation
  should be able to illuminate this problem.

  The shells of MS bubbles are also difficult to observe in the optical
  because they are large and dim \citep*{mckee84}. Nevertheless,
  shells around MS stars as well as shells around evolved massive stars, which
  --- based on their radius, expansion velocities, and shell mass ---
  are considered to
  originate from the MS phase of these stars, have been observed in the optical
  \citep{lozinskaya82, oey94, marston97}, in the IR \citep{marston96},
  and in the \HI 21\,cm line \citep{cappa96, cappa98, benaglia99}.

  The MS shells are typical results of the interaction of massive stars with
  their surrounding ISM over a time period of a few Myr. There are also
  young objects which have influenced their interstellar environment so
  strongly that they can already be identified as star-gas interactions.
  One example that should certainly be mentioned here is the so-called
  Becklin-Neugebauer-Kleinman-Low (BN-KL) nebula in the Orion molecular cloud
  1 (OMC1) behind the Orion nebula. It shows a spectacular, finger-like
  outflow of molecular hydrogen \citep{taylor84, allen93, mccaughrean97,
  salas99}. Due to extremely high visual extinction
  ($A_{\mathrm{V}} \approx 20-50\,\mathrm{mag}$) it can only be observed in
  the infrared or at longer wavelengths.
  Most models for the gas outflow proposed so far assume that it is driven by
  mass ejection from a young star, namely IRc2 and/or the
  Becklin-Neugebauer (BN) object which are both very close to the projected
  center of the mass outflow. IRc2 has a mass-loss rate
  $\dot{M}_{\mathrm{w}} \approx 10^{-(3\ldots 4)}\,\Msun\,\mathrm{yr^{-1}}$
  \citep{downes81} and a luminosity
  $L_{\mathrm{phot}} \approx 2 \ldots 10 \times 10^{4}\,\Lsun$
  \citep{genzel89}. The mass-loss rate of BN is
  $\dot{M}_{\mathrm{w}} \approx 4 \times 10^{-7}\,\Msun\,\mathrm{yr^{-1}}$ and
  the luminosity $L_{\mathrm{phot}} \approx 1 \ldots 2 \times 10^{4}\,\Lsun$
  \citep{scoville83}.

  Another fairly well studied region where formation of stars and their
  interaction with the ambient medium takes place in an early phase
  is the Eagle Nebula (M16). Imaging with the WFPC2 onboard the Hubble Space
  Telescope \citep{hester96} as well as observations in the
  infrared with ISOCAM \citep{pilbratt98} and NIRSPEC
  \citep{levenson00} reveal a number of interesting insights
  into the structure and possible formation history of this active region.
  The picture of M16 in the optical is dominated by the appearance of the
  so-called ``elephant trunks'', impressive pillars of dense molecular gas
  protruding into the \HII region. Their surface is
  partly illuminated by the ionizing light of nearby (distance to the pillars
  $\approx 2\,\mathrm{pc}$) massive stars so that photoevaporative flows from
  the surface can be identified on the well resolved pictures
  \citep{hester96} as striations extending normally from the
  surface of the cloud. The flow
  is driven by the pressure gradient that arises when the ultraviolet photons
  from the massive star heat the dense gas at the surface of the trunk to a
  few thousand Kelvin, resulting in a pressure which is higher than in the rest
  of the (lower density) \HII region.

  An interesting question arises in connection with M16: Is the elephant
  trunk structure a remnant of the original molecular cloud from which the
  young stars in the region were formed or has this structure actively been
  molded by the massive stars as a shell swept-up by stellar winds and folded
  by hydrodynamical instabilities? More detailed observations and better
  theoretical models are needed to answer this question.

\subsection{Previous numerical models of stellar wind bubbles: an overview}
\label{sec_numerics_so_far}

  Since effects associated with inhomogeneous ambient media or
  time-dependent stellar parameters as well as the formation of gas dynamical
  instabilities are difficult to handle within an analytical framework,
  numerical simulations have become a powerful tool to study the evolution of
  \HII regions and SWBs. First 2D numerical calculations of SWBs have been
  performed by \citet{rozyczka85a} and \citet{rozyczka85b, rozyczka85c} in a
  series of papers. For the expansion of the bubble into a homogeneous medium
  the authors described the formation of clumps in the thin shell but due to
  the lack of resolution they were unable to completely resolve the dynamics
  of the shell and to compare it with theoretical predictions. The authors
  also studied how a change of the wind luminosity (an increase between two
  fixed levels in a certain time) influences the shell stability and how
  SWBs break out from a plane-parallel stratified disk.
  The shell fragmentation as result of an increase of the wind luminosity has
  also been studied by \citet*{stone95} and has been compared
  with the observation of ``bullets'' in the vicinity of young stars.

  \citeauthor{garcia96a} and \citet*{garcia96b}
  focused on the history of the stellar mass loss and the implications for
  the formation of WR bubbles. They calculated the evolution of the SWB in
  one dimension until the star enters the LBV phase (for the model with an
  initial stellar mass of 60\,$\Msun$) or until the star leaves the
  RSG stage (for the 35\,$\Msun$ model). The resulting 1D profiles of
  circumstellar density, pressure and velocity were used to set up the ambient
  conditions for 2D calculations of the following stage when the WR wind
  interacts with the surrounding structure. The authors found that in both
  cases the slowly expanding shells originating from prior mass loss phases
  are heavily eroded when they are overtaken by the faster
  WR wind. This is also the phase in which the resulting nebula shell is
  considered to be most visible due to the high emission measure 
  in the high density filaments resulting from the collision.
  The nebula NGC\,6888 with
  its filamentary structure would fit in this picture as being driven by a WR
  star that has undergone MS and RSG evolution and the shell swept up by
  the WR wind is currently colliding with the RSG wind shell.

  The approach of \citet{garcia96c} is slightly
  different because they were mainly concerned with the very early evolution of
  \HII regions. They presented 2D gas dynamical calculations for the evolution
  of \HII regions in constant and power-law density profiles and showed the
  effect of radiative cooling on the thickness of the swept-up shell and
  therefore also on the development of instabilities. They found that the
  ionization front can reinforce the growth of the thin-shell instability
  and in a power-law density fall-off with exponent equal to two it can even
  lead to violent shell disruption.
  
  \citet{brighenti97} investigated the formation of WR
  shells resulting from fast WR winds evolving into slow anisotropic RSG
  winds. They found that the shell swept up by the WR wind becomes
  Rayleigh-Taylor unstable and fragments even before it hits the shell that
  had previously been piled
  up by the RSG wind. X-ray maps computed from the numerical models show that
  the anisotropy of the RSG wind leads to a ``two-lobe'' X-ray morphology which
  is qualitatively in agreement with the observation of NGC\,6888
  \citep{wrigge94}, but cannot reproduce the small filling factor
  of the X-ray emitting gas.

  \citet*{frank98} tried a different approach: They
  modeled the interaction of an anisotropic fast wind with an isotropic slow
  wind to explain the morphology of LBV bubbles. They found that anisotropic
  fast winds can indeed produce strongly bipolar outflows without assuming
  that the fast wind collides with a slowly expanding disk or torus as
  previously postulated.

  \citet{strickland98} calculated simple MS bubbles
  blown by a wind with constant mass-loss rate and terminal velocity into a
  uniform ISM. Their main goal was to calculate synthetic X-ray spectra from
  these numerical models as they would be observed with the ROSAT satellite,
  in order to analyze these spectra according to standard procedures and to
  compare the inferred properties with the original numerical models.
  Surprisingly (or not), they found that the inferred properties of the bubble
  can considerably deviate from the ``real'' properties of the model bubble.
  This implies that detailed X-ray emission models are necessary  --- instead
  of simple one- or two-temperature spectral fits --- in order to derive
  reliable properties of bubbles and superbubbles.

  In Table \ref{table_num_models} we compare the
  numerical hydrodynamic models of SWBs / \HII regions quoted above with
  respect to the included physics, the covered range of parameters, and some
  technical properties.

\section{Numerical method}
\label{sec_numerics}

\subsection{The radiation-hydrodynamics scheme}
\label{sec_num_rad_hydro}
  The numerical calculations presented in this paper have been performed using
  a 2D radiation-hydrodynamics code with a fairly sophisticated treatment of
  time-dependent ionization, radiation transfer, and heating and cooling
  processes. Most features of the code and results of test calculations have
  already been described in \citet{yorke95} and \citet{yorke96}.
  Therefore, here we only give a short summary of the code's main properties.

  The hydrodynamical equations for the conservation of mass, momentum, and
  energy in an ideal, inviscid, single fluid have been formulated in
  cylindrical coordinates ($r$,$z$), assuming axial symmetry around the $z$
  axis. This set of equations is solved numerically on an Eulerian grid in the
  quadrant with $r \ge 0$ and $z \ge 0$, i.e. we additionally assume mirror
  symmetry with respect to the equatorial plane. The differencing scheme used
  to discretize the equations is second order accurate in space. Due to
  operator splitting the accuracy in time is greater than first order.
  The advection scheme of \citet{vanLeer77} is employed. The star that
  influences the gas is located in the center of the coordinate system at
  $r = z = 0$. We use square grids with constant mesh size and insert multiply
  nested grids in the corner at $r = z = 0$ to enhance the spatial resolution
  close to the star. All the nested grids are self-similar and the linear
  spatial resolution is improved by a factor of two for each level of nesting. 
  The basic code is explicit, and therefore the Courant-Friedrichs-Lewy (CFL)
  condition determines the maximum time step except when changes
  in the stellar parameters take place on even smaller time scales.
  Von Neumann--Richtmyer artificial viscosity is used for the treatment
  of shocks. We use the equation of state for an ideal gas (in our case a
  mixture of molecular, neutral, and ionized hydrogen with electrons).
  The formation and dissociation of molecular hydrogen is not yet considered
  in the energy balance, because it is assumed that continuum radiation below
  the Lyman threshold dissociates the molecules before the gas is heated up
  to temperatures where thermal dissociation occurs.

  Our integration order starts with the finest grid. The sizes of the time 
  step satisfies the CFL condition for the finest grid as well as for all
  coarser grids (normalized by the appropriate power of 2).  After two time
  steps have been done on the finest grid and the solution has been advanced to
  $t + \delta t_{\mathrm{1}} + \delta t_{\mathrm{2}}$, one time step
  $\delta t_{\mathrm{1}} + \delta t_{\mathrm{2}}$ is done on the next coarser
  grid to advance the solution there to the same time. Because of our 
  prior careful choice
  of $\delta t_{\mathrm{1}}$ and $\delta t_{\mathrm{2}}$, their sum obeys
  the CFL condition on the next coarser grid. After the solutions on the two
  grid levels have been advanced to the same point in time, all coarse grid
  values which have underlying fine grid values (not from the boundary) are
  replaced by weighted averages from the fine grid.
  The outer boundary conditions of the fine grid are interpolated from the
  corresponding grid values of the coarse grid. After this cycle is repeated
  (taking into account the new CFL time-step constraints) the first
  integration on the next coarser (3rd level) grid
  can be performed; 3rd level values are partly replaced by 2nd level values
  and 2nd level outer boundaries are imposed from 3rd level values, and
  so on. We use this recursive scheme for 7 grid levels, which means that
  $\sum_{i=0}^6 2^i = (2^7 - 1) = 127$ single integrations have to be done
  before one time step on the coarsest grid is completed.

  The ionization structure of hydrogen is calculated for each hydrodynamical
  time step by including the effects of photoionization,
  collisional ionization, and
  spontaneous recombination. The absorption of the stellar point-source Lyman
  continuum by neutral hydrogen is calculated along radial lines-of-sight.
  We use $N_{\mathrm{r}} + N_{\mathrm{z}} - 1$ rays to ensure that every
  grid cell is traversed by at least one ray, where $N_{\mathrm{r}}$ and
  $N_{\mathrm{z}}$ are the numbers of grid cells in $r$ and $z$ direction,
  respectively, without ghost cells.
  The photoionizing radiation field is assumed to have a black body spectrum
  at the effective temperature of the star. We use the grey approximation for
  the transfer of the stellar photons, i.e. we calculate the absorption cross
  section and excess energy only for the mean energy of photons above the Lyman
  threshold and, therefore, do not account for
  selective absorption processes. To account for the diffuse Lyman continuum
  that originates from electron recombinations directly into the ground state
  of hydrogen we simply use the on-the-spot approximation. 

  Radiative heating and cooling are considered as source and sink
  terms in the energy equation and are calculated in an additional substep.
  Photoionization of hydrogen is responsible for gas heating.
  The treatment of cooling depends on the gas temperature.
  Below 15\,000\,K the contributions by the following processes
  are explicitly calculated: Bremsstrahlung,
  collisional ionization of hydrogen, thermal energy lost in hydrogen
  recombinations, collisional excitation of neutral hydrogen followed by
  $\mathrm{Ly\alpha}$ emission,
  collisional excitation of the low-lying $^1D$ terms of $\mathrm{N^+}$ and
  $\mathrm{O^{++}}$ and the $^2D$ term of $\mathrm{O^+}$.
  Simple assumptions have been made about the degree of ionization of nitrogen
  and oxygen: Due to the similar ionization potentials and comparable
  recombination coefficients of $\mathrm{H^0}$, $\mathrm{N^0}$, and
  $\mathrm{O^0}$, we set the ratios of ionized-to-neutral species for
  nitrogen and oxygen equal to the respective number calculated for hydrogen,
  i.e. $\rho_{\mathrm{N^+}} / \rho_{\mathrm{N}} = X$ and   
  $(\rho_{\mathrm{O^+}} + \rho_{\mathrm{O^{++}}}) / \rho_{\mathrm{O}} = X$,
  where $X$ is the hydrogen ionization fraction. It is furthermore assumed
  that the oxygen ions are equally distributed on the ionization stages
  $\mathrm{O^+}$ and $\mathrm{O^{++}}$, i.e.
  $\rho_{\mathrm{O^+}}    / \rho_{\mathrm{O}} =
   \rho_{\mathrm{O^{++}}} / \rho_{\mathrm{O}} = X / 2$.

  The chemical composition is solar everywhere in our computational
  domain. Helium is not yet considered because it is much less abundant than
  hydrogen and has no low-lying energy levels. For temperatures above
  $10^5\,\mathrm{K}$ we use the interstellar cooling function from
  \citet{sarazin87} with the correction given by \citet{soker90}.
  This cooling function is based on the assumption of
  collisional ionization equilibrium. In the intermediate temperature range
  between $15\,000\,\mathrm{K}$ and $10^5\,\mathrm{K}$ we calculate both
  values and use a weighted average. We have not included dust in the
  calculations presented here.

  \subsection{Initial conditions}
  \label{subsec_ini_cond}

  We start all our calculations with the turn-on of the ZAMS stellar radiation
  field and stellar wind in a homogeneous and quiescent ambient medium. This 
  is definitely a gross oversimplification of what can be expected for the
  structure of the circumstellar gas shortly after the star has been formed at
  the end of the pre-MS phase. We have two reasons for our
  approach: The first is that, in order to understand what happens in the
  circumstellar medium, it is important to start with a well defined and
  simple initial configuration. As we will show below, even with this
  initially homogeneous ambient density and temperature distribution, the
  interaction of the SWB with the ionizing radiation field
  produces a variety of interesting morphological structures whose formation
  processes have to be understood in detail before studies with a more
  realistic initial setup can be performed. The second reason is that we want
  to keep our simulations comparable with those of other authors who use
  different algorithms or incorporate different physical effects in their
  codes.
  We choose $n_{\mathrm{0}} = 20\,\mathrm{cm^{-3}}$ and
  $T_{\mathrm{0}} = 200\,\mathrm{K}$ for our quiescent ambient medium.
  The density is perhaps high for a star that has already left its parental
  molecular cloud, but we want to keep our results comparable to those of
  \citeauthor{garcia96a} who also used $n_{\mathrm{0}} = 20\,\mathrm{cm^{-3}}$
  in order to reduce computational expenses. The temperature
  $T_{\mathrm{0}} = 200\,\mathrm{K}$ in the ambient medium was chosen to yield 
  a thermal pressure ``typical'' for the ISM at the solar galacto-centric
  distance.

  We use a spherical 1D solution after 1000\,yr as initial model for the 2D
  simulations in order to prevent boundary effects from influencing the
  formation of the very small wind bubble while the development of the
  self-similar structure is not completed. We switch from 1D to 2D before
  the swept-up gas collapses into a thin shell that is subsequently subject to
  instabilities.

  \subsection{Boundary conditions}
  \label{subsec_bound_cond}

  In our models the stellar parameters for mass-loss rate
  ($\dot{M}_{\mathrm{w}}$), terminal velocity of the wind ($v_{\mathrm{w}}$),
  effective temperature ($T_{\mathrm{eff}}$), and photon luminosity in the
  Lyman continuum ($L_{\mathrm{LyC}}$) are time-dependent boundary
  conditions which drive and govern the evolution of the circumstellar gas.
  For the 60\,$\Msun$ star we adopt the stellar parameters given by
  \citeauthor{garcia96a}. They were obtained from stellar evolutionary models
  and from observations and are shown in Figure \ref{60Msun_input4_up.eps}.
  The 60\,$\Msun$ star is supposed to undergo the following evolution:
  MS O star
  $\rightarrow$ H-rich WN star
  $\rightarrow$ P Cygni-type LBV
  $\rightarrow$ H-poor WN star
  $\rightarrow$ H-free WN star
  $\rightarrow$ WC star
  $\rightarrow$ SN.
  
  According to the mass-loss rate and terminal velocity at time $t$,
  appropriate values for density and velocity of the gas are set within the
  ``wind generator region'' on the finest grid (a small sphere around the
  center of the coordinate system where the star is located). The radius of
  the ``wind generator region'' is $3.5 \times 10^{17}\,\mathrm{cm}$ for the
  calculations shown in this paper.

  Reflecting boundary conditions are used at the $r$- and $z$-axis. The outer
  boundaries of the nested grids were taken from the corresponding values
  on the next coarser grid. The outer boundaries of the outermost (coarsest)
  grid are semi-permeable, i.e. mass, energy, and momentum can flow out of
  the computational domain, but not in. Of course, outflow would conflict
  with our intention to carefully take stock of the energetic processes in
  our calculations. Thus, we prevent outflow by choosing the coarsest grid
  large enough that neither the ionization front nor moving material can reach
  the outer boundary before the star dies and the calculation stops.
  
  \subsection{Geometry and resolution}
  \label{subsec_geometry}

  The size of the computational domain is
  $r_{\mathrm{max}} = z_{\mathrm{max}} = 64\,\mathrm{pc}$. We use 6
  nested grids within the coarsest grid, resulting in 7 grid levels.
  The simulation has been performed with $125 \times 125$ cells on each grid
  level (excluding ghost cells).  The linear resolution
  ranges from $8\times 10^{-3}\,\mathrm{pc}$ close to the
  star to approximately 0.5\,pc in the outermost parts of the coarsest grid
  which were affected only during the late stages of the bubble evolution.
  For a resolution study we repeat the calculations
  with $61 \times 61$ cells and for the first Myr
  of evolution with $253 \times 253$ cells.

\section{Results and discussion}
\label{sec_results}

  \subsection{A resolution study}
  \label{subsec_validity}

  Before we present our results we briefly discuss their validity. It is
  currently impossible to spatially resolve all details of the entire dynamical
  evolution when modeling an SWB in two or more
  dimensions over such a long time with all the physics we have included.
  Especially the modes of the thin-shell instability
  \citep{vishniac83, ryu88, vishniac89} are
  extremely difficult to resolve, because a sufficient number of grid cells
  across the thin shell is required to follow the tangential flow of material.
  \citet{maclow93} performed purely hydrodynamical
  simulations of the thin-shell instability and confirmed the linear stability
  analysis of \citet{vishniac83}. Since they completely focused on
  the hydrodynamical evolution of the thin shell, they made no efforts to
  treat much more complex systems with more sophisticated physics.   

  To check the impact of this resolution effect on our results for the
  energetic evolution of the SWB, we performed a resolution study.
  We calculated the same 60\,$\Msun$ model with three different
  resolutions, medium (125 cells in each dimension), low (61 cells in each
  dimension), and high (253 cells in each dimension). 
  We compare only the first Myr of the
  evolution (because the high-resolution model is too CPU intensive).
  The results are displayed in Figure \ref{comp_resol_184_226_227_up.eps}.
  The change of kinetic energy due to different resolutions is systematic
  (at times later than $0.2\,\mathrm{Myr}$) in the sense that a higher
  resolution increases the kinetic energy that remains in the system. For most
  of the time the difference in kinetic energy between medium and high
  resolution ($\lesssim$ 0.05 dex) is smaller than
  between low and medium resolution,
  which is expected as the value converges toward the limit. The variation of
  the thermal energy with resolution is only weak, the maximum difference
  between low and high resolution is $\lesssim$ 0.05 dex at any time during the
  first Myr. The scatter in ionization energy for the three different
  resolutions is also $\lesssim$ 0.05 dex for
  $\mathrm{0.5\,Myr \le {\it{t}} \le 1.0\,Myr}$
  and $\lesssim$ 0.1 dex for $\mathrm{{\it{t}} < 0.5\,Myr}$. In the latter
  case, the deviation of ionization energy between the medium and high
  resolution runs varies, but is considerably smaller than between low and
  medium resolution, indicating that the value of the ionization energy is
  already close to the actual value.

  These results show that the errors in our energetic analysis due to
  resolution effects are within an acceptable range, although some details
  of the evolution are not yet completely resolved in our calculations.

\subsection{The evolution of the 60\,$\Msun$ model}
\label{subsec_60msun_model}

  Figures \ref{ion_uchii184.001.001.3.med.mono.eps} to 
  \ref{tem_uchii227.805.001.1.med.mono.eps} show the evolution of the
  circumstellar gas around our 60\,$\Msun$ model star. The plots show
  grey-scale coded physical quantities in a quadrant of the $r$-$z$-plane
  through the star. The size of the
  displayed area varies with time as the expansion proceeds outward.
  The star is located in the center of the coordinate system. We display the
  mass density together with the velocity field, the gas temperature, and the
  degree of hydrogen ionization, respectively, for certain stages of
  evolution. The degree of ionization is not shown for stages after 1\,Myr,
  when the ionized region spatially almost coincides with the hot gas.
  To prevent confusion, velocity arrows are skipped in the free-flowing wind
  and (except for Figure \ref{ion_uchii184.001.001.3.med.mono.eps}) in the
  hot bubble. 

  Figure \ref{ion_uchii184.001.001.3.med.mono.eps} shows the initial model for
  the 2D calculation that has been set up from the 1D solution after 1000\,yr.
  The fundamental structure of the combined SWB / \HII region is well
  described by the analytical solutions. The ionization front at a distance of
  about $\approx 8\,\mathrm{pc}$ from the star separates the ionized gas at
  about $8000\,\mathrm{K}$ from the undisturbed medium.
  The front has not yet reached the initial Str\"omgren radius at about
  12.7\,pc and is still R-type. Thus, there is little dynamical response of the
  heated gas at this point.
  The SWB evolves into the pre-ionized and pre-heated but
  otherwise undisturbed medium. The free-flowing wind is heated by the reverse
  shock at $r \approx 0.16\,\mathrm{pc}$ to very high temperatures of about
  $10^8\,\mathrm{K}$. Though the density in the hot gas is fairly low
  ($\approx 4 \times 10^{-25}\,\mathrm{g\,cm^{-3}}$), the thermal pressure is
  still much higher than in the \HII region and the hot bubble expands
  supersonically (with respect to the isothermal sound speed in the \HII
  region of
  $\approx 12\,\mathrm{km\,s^{-1}}$) at roughly $200\,\mathrm{km\,s^{-1}}$.
  The photoionized \HII region gas becomes shocked by the expanding hot bubble
  and compressed into a shell with a density of more than
  $10^{-22}\,\mathrm{g\,cm^{-3}}$ and a temperature of nearly
  $10^6\,\mathrm{K}$. At this early time, the SWB and the
  photoionized \HII region still evolve independently.

  Figure \ref{ion_uchii184.008.001.3.med.mono.eps} depicts the evolution after
  $2 \times 10^4\,\mathrm{yr}$. The photoionized region has reached its
  initial radiative equilibrium size and gas at the outer edge of the \HII
  region starts to accelerate outward due to the pressure gradient between
  the warm ionized gas and the cold neutral ambient medium. The hot bubble has
  grown to about 2\,pc in radius. The shell has been decelerated to
  approximately $60\,\mathrm{km\,s^{-1}}$. Due to the high density and high
  temperature, strong cooling has led to compression of the swept-up gas
  into a thin shell. The thin-shell instability
  initially produces some dense knots in the shell
  that imply variations in the optical depth along radial lines-of-sight
  indicated by the rippling of the photoionization front. The initial
  evolution of the SWB (as long as it is independent of the \HII region
  evolution) is very similar to what is described by \citet{strickland98},
  except for the deviations due to the different
  ambient density and stellar parameters.

  On the basis of Figure \ref{ion_uchii184.030.001.3.med.mono.eps}
  ($t = 4 \times 10^4\,\mathrm{yr}$) and
  Figure \ref{ion_uchii184.103.001.3.med.mono.eps} ($t = 0.1\,\mathrm{Myr}$)
  the structuring impact of the interaction processes between the SWB and the
  \HII region can be well identified: The photoionized \HII region has started
  to expand into the ambient medium at 6 to $7\,\mathrm{km\,s^{-1}}$.
  The isothermal sound speed of the cold neutral gas is about
  $1\,\mathrm{km\,s^{-1}}$. Thus, an outer radiative forward shock occurs
  that sweeps up the ambient gas into a shell. The hot
  gas bubble has grown to about 5\,pc in radius and the shell expansion
  velocity is still about $25\,\mathrm{km\,s^{-1}}$ ($t = 0.1\,\mathrm{Myr}$).
  The ionization plots nicely illustrate the
  influence of the density knots in the wind bubble shell on the evolution of
  the \HII region: The gas clumps cast shadows into the \HII region.
  The optical depth for hydrogen ionizing photons along lines-of-sight through
  the clumps is higher than for lines-of-sight which pass the shell between
  clumps. Due to the weakening of the radiation field, the gas behind the
  clumps starts to recombine, forming neutral spokes within the \HII region.
  The spokes are less prominent in temperature than in degree of ionization,
  because the cooling efficiency breaks down with the
  disappearance of the free electrons and the shadowed regions remain at
  several thousand K for some time.

  One could imagine that the shadowed regions are artificial due to the use of
  the on-the-spot approximation for the diffuse radiation field that
  originates from recombinations directly into the ground state of hydrogen.
  But a test calculation that used flux limited diffusion instead of the
  on-the-spot approximation yielded qualitatively the same results with
  respect to the shadowed regions. Only slight morphological differences were
  obtained due to the transfer of the diffuse radiation field.

  Considerations about the influence of shadow patterns on ionized regions are
  not completely new: \citet{capriotti73} already discussed the
  impact of shadows by shell fragments on the formation and appearance of
  planetary nebulae (PNe). \citet{williams99} examined the corrugation of
  R-type ionization fronts by density inhomogeneities with subsequent
  formation of dense clumps. \citet{soker98} studied the formation of
  compressed tails in the shadow zones behind dense clumps which are present
  in the vicinity of the central star of a PN before the
  ionization starts. And \citet{canto98} presented an
  analytical model for the partial ionization of the shadow regions behind
  neutral clumps by the diffuse radiation produced in the nebula along with
  some gas dynamic simulations. The model is not restricted to PNe
  and can be generalized to photoionized regions.

  \citet{richling00} considered the effects of UV 
  dust scattering and calculated the evolution of disks 
  externally illuminated by an O star.  Here, also, shadows
  are cast which appear cometary in shape with tails 
  pointing away from the ionizing source.  Had we considered
  UV scattering by dust in this investigation it is very
  likely that our clumps' shadows would also have had a
  cometary shape.

  For the combined \HII region / SWB described in our work we show that the
  density inhomogeneities needed to cast the shadows can be produced by the
  action of the stellar wind even after the \HII region has already been
  formed in a completely uniform medium. In our case the growth of the
  ionized fingers of the \HII region is therefore not only due to the
  ionization front instability described by \citet{garcia96c}. It is
  strongly triggered by the redistribution of mass (and thus opacity) by the
  action of the stellar wind shell.

  The real nature of the spokes, clumps, and instabilities is 3D rather than
  2D, a fact which certainly should influence their size distribution, shapes,
  and longevity.  3D simulations, though extremely expensive, would be useful
  to study these structures in more detail.  These 2D simulations should
  thus be regarded as qualitative indicator that such instabilities exist,
  a feature missing from 1D calculations.

  Another consequence for the structure of the \HII region and the shell of
  the SWB can be seen on Figure \ref{ion_uchii184.149.001.3.med.mono.eps} at
  $0.22\,\mathrm{Myr}$. The disturbed structure of the \HII region,
  especially the formation of the neutral spokes and the ionized fingers,
  has induced several flow patterns with velocities up to
  $\approx 20\,\mathrm{km\,s^{-1}}$. Due to the pressure gradient between the
  neutral spokes and the ionized gas, the spokes are compressed by the ionized
  gas. At the same time the expanding SWB compresses the \HII region
  (and its spokes) as a whole.  The clumps in the stellar wind shell
  originally responsible for the shadows grow in mass and partially merge
  during the further course of evolution.
  Photoevaporation of the stellar wind shell and its embedded clumps does not
  strongly influence the dynamics, because the gas in these structures is
  still warm and its photoionization does not produce large pressure gradients.
  However, where stellar photons directly impinges upon
  the cold outer shell through the optically thinner ionized fingers 
  photoevaporation occurs and leads to significant dynamical effects, as
  evident in Figures 8 through 16.

  In the top panel of Figure \ref{ion_uchii184.149.001.3.med.mono.eps}
  and also in some of the other density plots are swirls of material visible
  in the hot bubble. These vortices are sometimes generated at the $r$- and
  $z$-axes, possibly as an artifact of the reflecting boundary conditions.
  Since the total mass involved in these flow patterns is very small, they
  do not influence the morphological or the energetic evolution significantly.

  Proceeding to $0.5\,\mathrm{Myr}$ 
  (Fig.~\ref{ion_uchii184.244.001.2.med.mono.eps}, note the different
  scale) we see that the entire \HII region has been swept
  up by the SWB, which has grown to approximately 12\,pc in
  radius. The photoionized \HII region (it can best be distinguished from the
  hot bubble using the temperature plot) consists of
  3-7\,pc long pillars, the remnants of the ionized fingers. 
  The stellar wind shell loses its structure, becomes more or less
  dissolved, and merges into the ``remnant'' \HII region, because
  it sweeps up a highly nonuniform medium with an expansion speed that becomes
  comparable to the sound speed of the \HII gas.

  Figures \ref{ion_uchii184.473.001.2.med.mono.eps} and
  \ref{ion_uchii227.283.001.2.med.mono.eps} compare the results of our high
  and medium resolution runs with respect to the morphological structure after
  1\,Myr. One can easily see that there are significant deviations in the
  structure of the photoionized region between the two runs. This is not very
  surprising, because, as mentioned above, the formation of density
  fluctuations due to the thin-shell instability is very sensitive to the
  spatial resolution, especially as we have started with a completely
  undisturbed ambient medium and have not artificially excited any particular
  unstable mode. And we have also seen that exactly these clumps in the
  stellar wind shell strongly influence the evolution of the \HII region due
  to their opacity for Lyman continuum radiation.

  Nevertheless, the general structure is very similar for both resolutions:
  The radius of the hot bubble is roughly 17\,pc (strong deviations from
  spherical shape appear in both cases). The photoionized gas forms a highly
  irregular shell around the hot bubble with a mean thickness of
  $\approx 5\,\mathrm{pc}$, bound on the outside by a swept-up shell of
  shocked ambient material. 
  It appears as if the phenomenon of ``break-out'' of the hot gas into the
  photoionized gas has occurred between the spokes in both cases.  However,
  we see no evidence for further break-out into the undisturbed constant density
  medium.  Had we chosen an ambient environment of limited size --- i.e.
  included cloud edges of decreasing density within the computational grid ---
  we could expect a more dramatic demonstration of break-out.

  We now jump to $t = 3.30\,\mathrm{Myr}$ and show in
  Figure \ref{tem_uchii227.758.002.1.med.mono.eps} the evolutionary state of
  the circumstellar material shortly before the star enters the LBV phase
  with its strongly enhanced mass-loss rate. (Please note that we once again
  changed the plot scale to an appropriate size.) The radius of the hot bubble
  has grown to approximately 40\,pc. Due to the growth of the bubble
  (and the slight decrease of H-ionizing photon luminosity after 2\,Myr) the
  photoionized \HII region has more or less collapsed into a skin with
  $\approx 2\,\mathrm{pc}$ thickness at the inside of the outer shell.
  The outer shell itself is still corrugated, and so is the photoionized gas.
  Some small clumps have become detached from the shell.

  Figure \ref{tem_uchii227.760.003.1.med.mono.eps} depicts the structure
  during the LBV
  eruption. Compared with the earlier plots one can detect a much higher
  density in the free-flowing wind zone. That is due to the higher mass-loss
  rate of the star and the smaller terminal velocity. The rest of the structure
  is largely unchanged, because the time-step difference is small.

  The LBV phase lasts only for about 10\,000\,yr and is followed by the final
  WR phases of the star, which are once again associated with wind terminal
  velocities of a few $1000\,\mathrm{km\,s^{-1}}$. \citeauthor{garcia96a}
  describes what happens next: The WR wind becomes shocked when it hits
  the much slower LBV wind and the hot, rarefied gas accelerates the dense
  LBV wind (Fig.~\ref{tem_uchii227.761.002.1.med.mono.eps}). This is now a
  highly Rayleigh-Taylor unstable situation and the shocked WR wind
  breaks through the LBV material, fragments it, and blows it into the
  MS bubble (Fig.~\ref{tem_uchii227.761.008.1.med.mono.eps}) where it mixes
  with the hot, rarefied gas, enhancing the density and lowering the
  temperature there for some time.

  The last set of pictures (Fig.~\ref{tem_uchii227.805.001.1.med.mono.eps})
  shows the structure of the
  circumstellar gas at the end of our simulation. This is
  the pre-SN configuration immediately before the star explodes as \SNII.
  The radius of the hot bubble has reached nearly 50\,pc and the
  photoionized \HII region is limited to the illuminated inner skin of the
  outer shell of swept-up ambient gas.

  A ``re-ionization'' phase \citep*[cf.][]{beltrametti82, tenorio82} cannot be
  found in our model, because the dynamical evolution of the circumstellar gas
  at late stages is dominated by the influence of the stellar wind rather than
  the stellar radiation field.

  \subsection{The energy balance in the circumstellar gas}
  \label{subsec_e_balance}

  \subsubsection{Numerical results}

  Figure \ref{e_distri227_up.eps} shows the total energy contributions in the
  circumstellar gas of the 60\,$\Msun$ model as a function of time.
  We follow the kinetic energy of bulk motion throughout the computational
  domain, the ionization energy (13.6\,eV per ionized hydrogen atom),
  the thermal energy of cold ($T \le 10^3\,\mathrm{K}$),
  warm ($10^3\,\mathrm{K} < T < 10^5\,\mathrm{K}$),
  and hot ($T \ge 10^5\,\mathrm{K}$) gas. As we have seen before, the
  ionization and heating of the \HII region is the fastest process at the
  beginning of the evolution. Thus, the ionization energy reaches
  $1.1 \times 10^{50}\,\mathrm{ergs}$ and the thermal energy of warm gas
  around $1.5 \times 10^{49}\,\mathrm{ergs}$ very soon after the stellar
  turn-on. 

  The evolution of the thermal energy of warm gas follows that of the
  ionization energy over the lifetime of the star with a shift of 0.7-0.9 dex.
  This is due to the fact that photoionization by the
  stellar radiation field is responsible for the bulk production of ionized
  gas at typically $8000\,\mathrm{K}$. The contribution by shock heating
  is small, because the reverse shock does not heat much mass
  and the thermal energy of this gas is accounted under ``hot gas'' rather than
  ``warm gas''. Therefore, we will discuss only the ionization energy bearing
  in mind that the thermal energy of warm gas behaves quite similarly.

  An interesting feature is the drop of ionization energy between its first
  maximum and $t \approx 0.2\,\mathrm{Myr}$. This is due to the
  formation of the dense structures (shell, spokes) in the \HII region which
  reduce the recombination time and increase the cooling of the gas
  (see also Fig.~\ref{E_with_without_wind60_up.eps}).
  Between $t \approx 0.2\,\mathrm{Myr}$ and 
  $t \approx 1.9\,\mathrm{Myr}$ the denser structures dissolve and
  the pressure decay within the hot bubble together with the \HII 
  region expansion result in a decrease of density in the \HII region.
  Thus, the recombination time increases and the ionization
  energy stored in the system increases accordingly. The ionization energy
  reaches $2.7 \times 10^{50}\,\mathrm{ergs}$ and the thermal energy of
  warm gas $3.7 \times 10^{49}\,\mathrm{ergs}$.
 
\begin{figure}
  \plotone{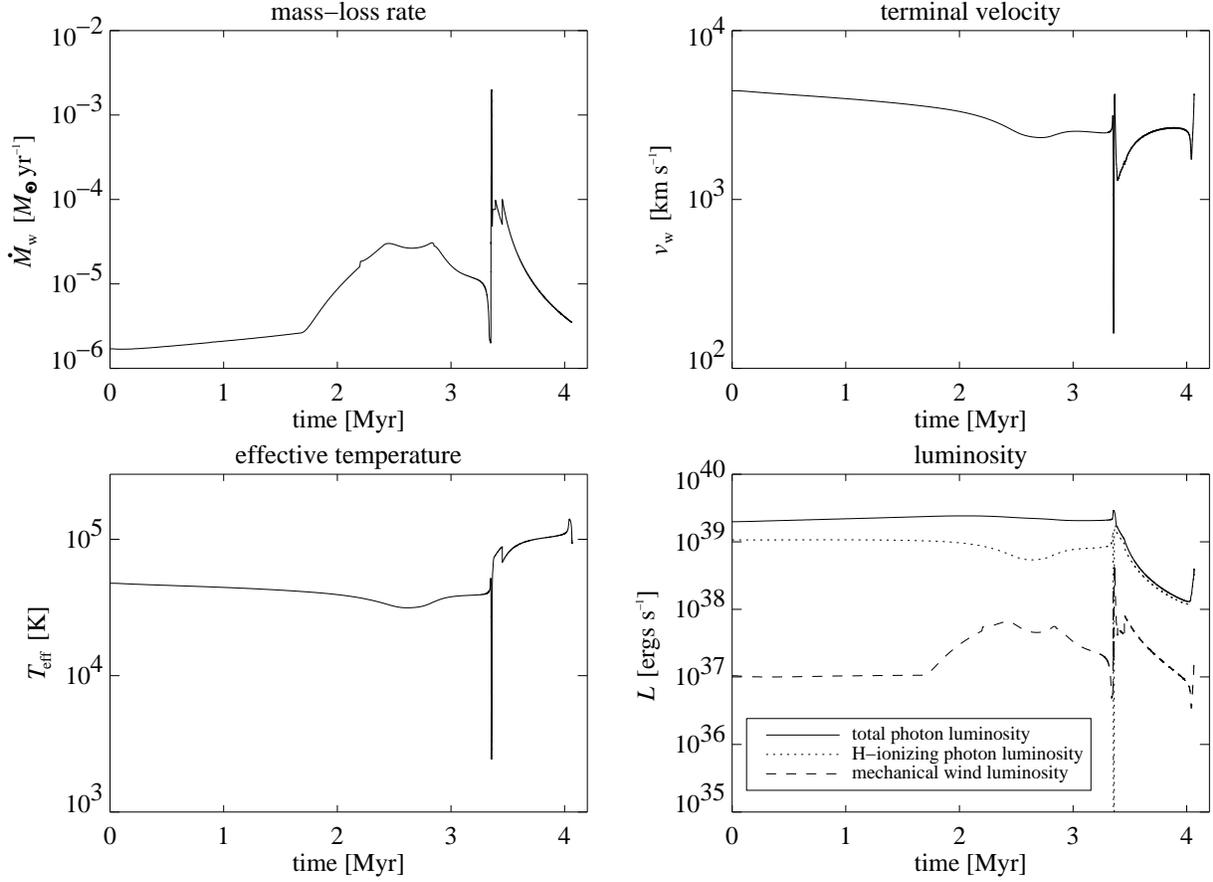}
  \caption{Time-dependent stellar parameters used as boundary conditions
           for the calculation of the 60\,$\Msun$ model. Mass-loss
           rate (upper left diagram), terminal velocity of the wind
           (upper right), effective temperature (lower left), and
           luminosity (lower right). All parameters are adopted
           from GML1.
           \label{60Msun_input4_up.eps}
          }
\end{figure}

\begin{figure}
  \plotone{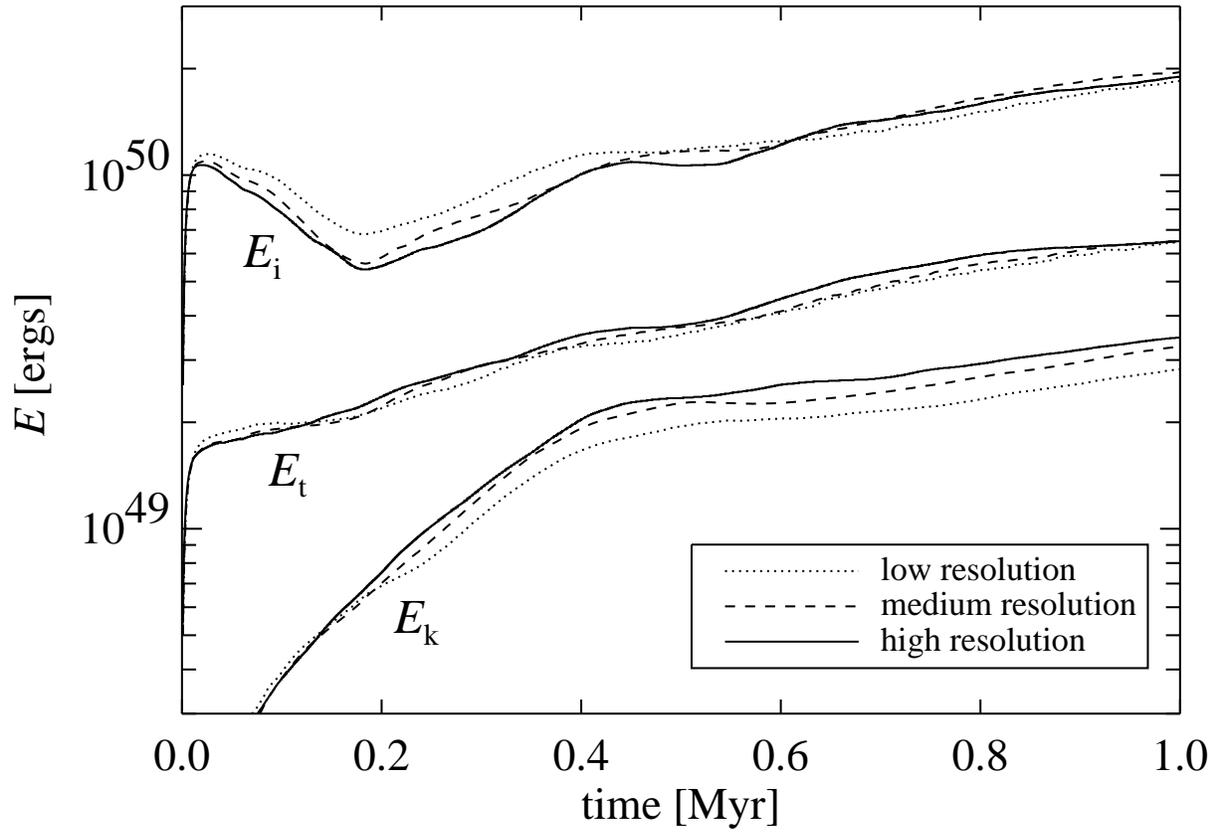}
  \caption{Resolution study for the 60\,$\Msun$ model.
           $E_{\mathrm{k}}$ is the total kinetic energy of bulk motion in the
           system, $E_{\mathrm{t}}$ the thermal energy, and
           $E_{\mathrm{i}}$ the ionization energy
           (13.6\,eV per ionized hydrogen atom). 
           \label{comp_resol_184_226_227_up.eps}
          }
\end{figure}

\begin{figure}
  \epsscale{0.50}
  \plotone{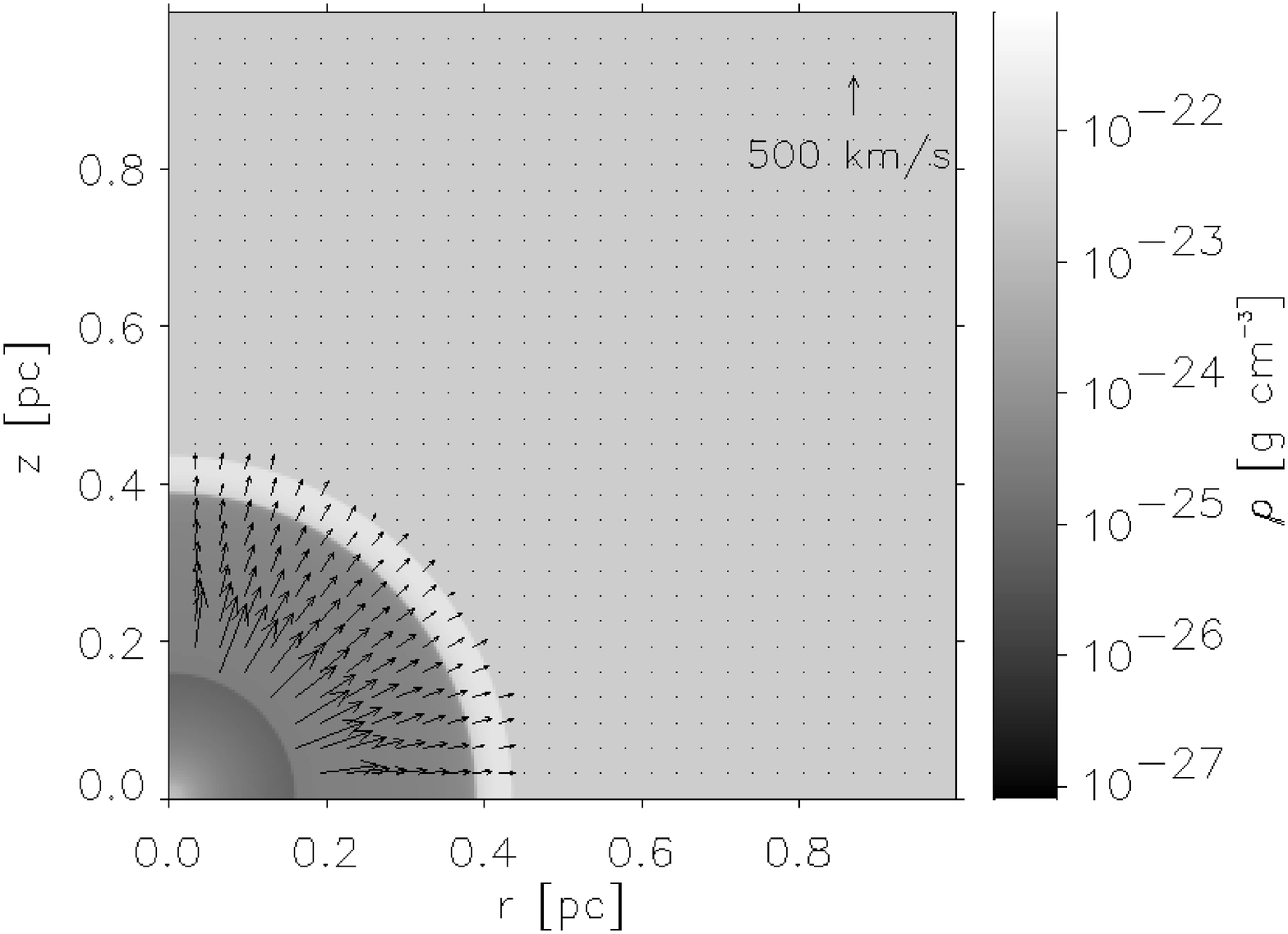}
  \plotone{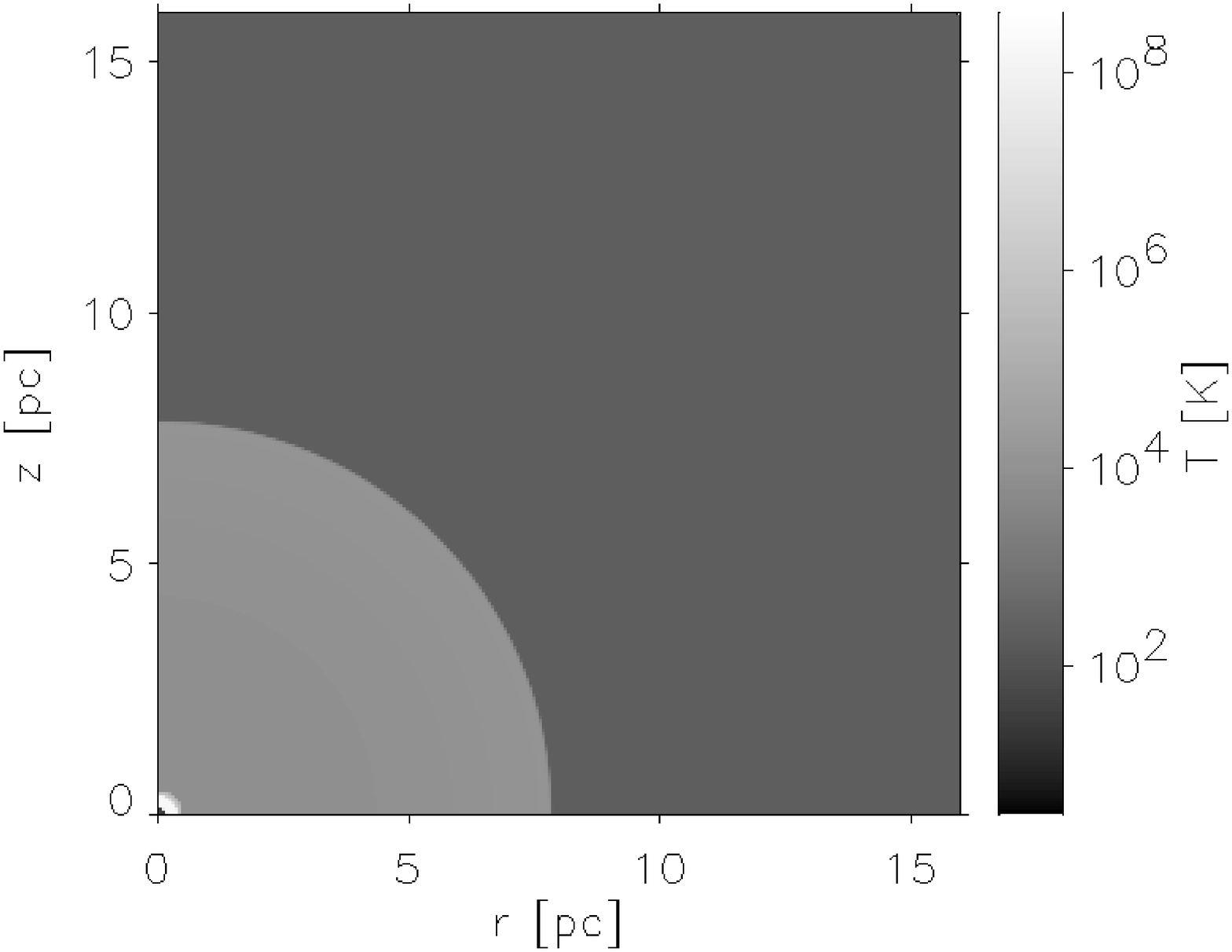}
  \plotone{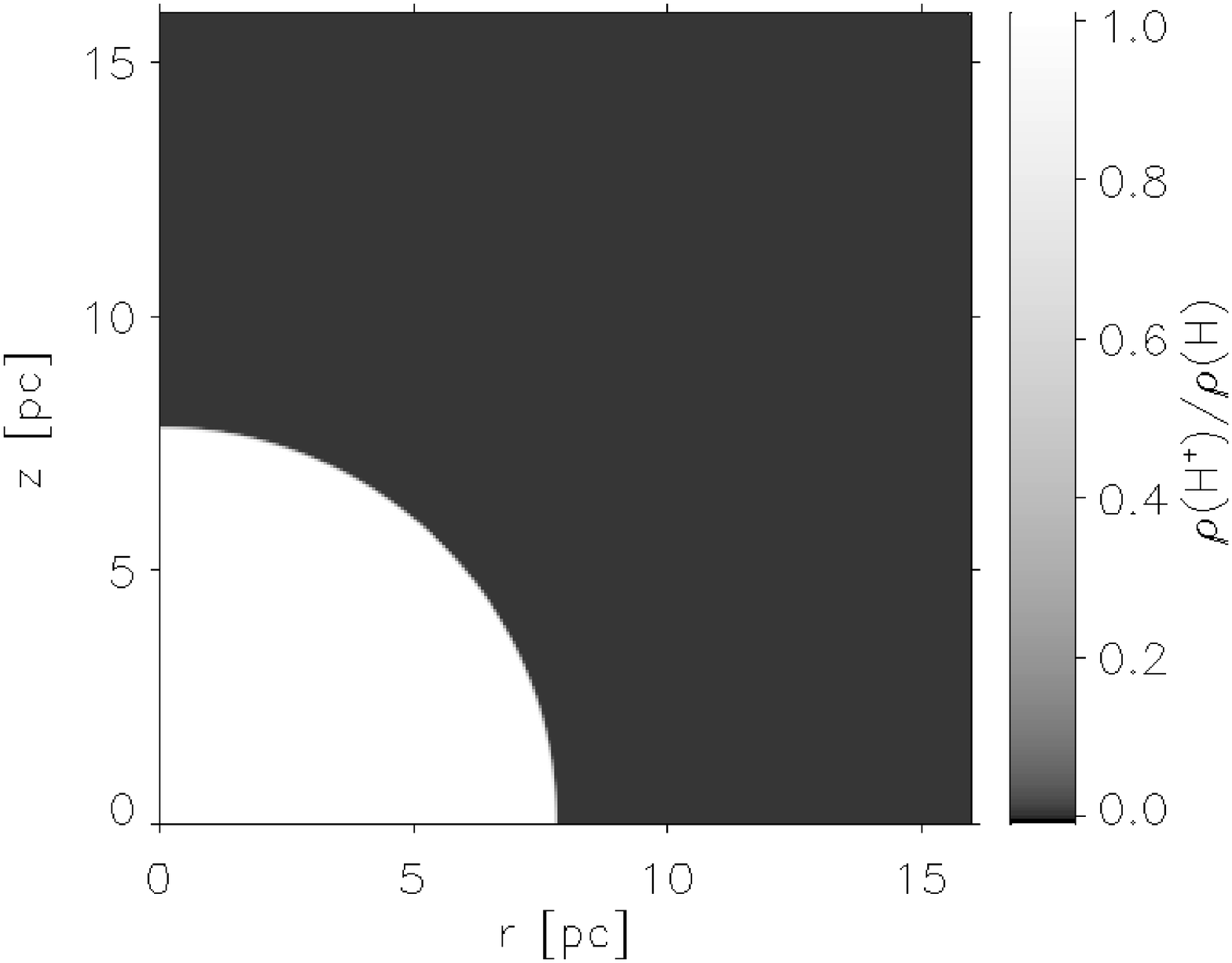}
  \caption{Circumstellar mass density and velocity field (top panel),
           temperature (center panel), and degree of hydrogen ionization
           (lower panel) for the 60\,$\Msun$ model at age
           $10^3\,\mathrm{yr}$ (high resolution run). The velocity arrows
           in the free-flowing wind zone have been omitted to prevent
           confusion. The star is located in the center of the coordinate
           system. Please note the different length scales.
           \label{ion_uchii184.001.001.3.med.mono.eps}
          }
\end{figure}

\begin{figure}
  \epsscale{0.50}
  \plotone{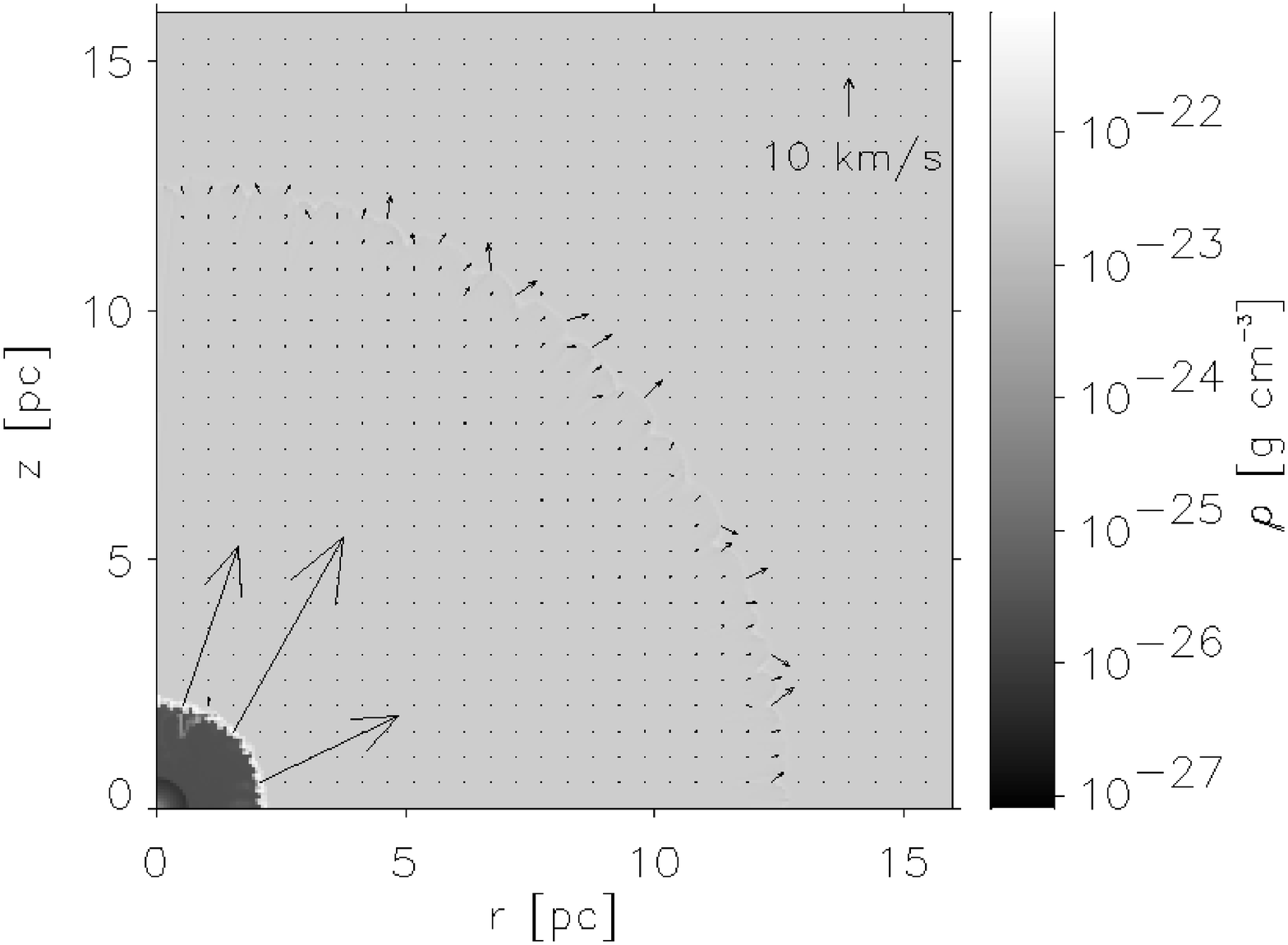}
  \plotone{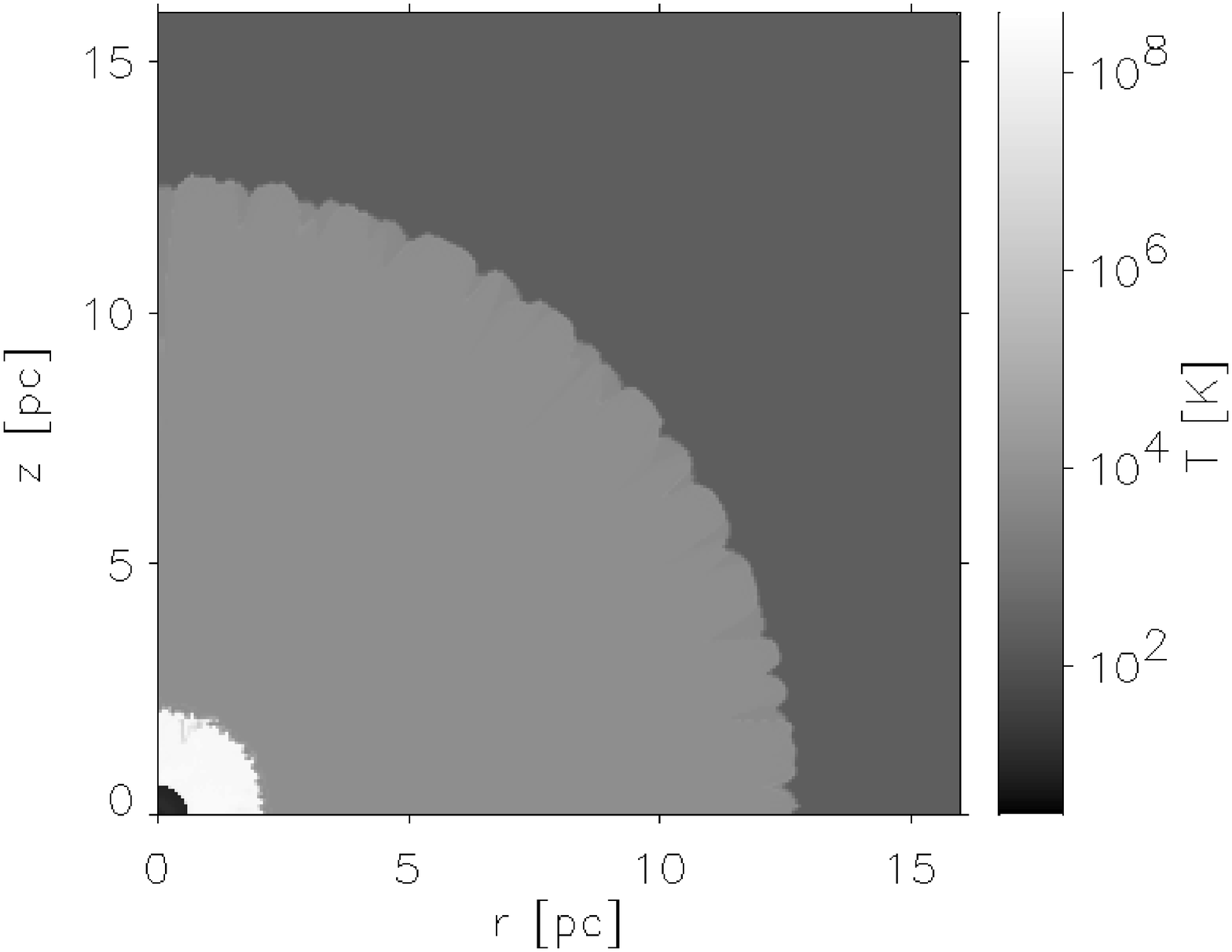}
  \plotone{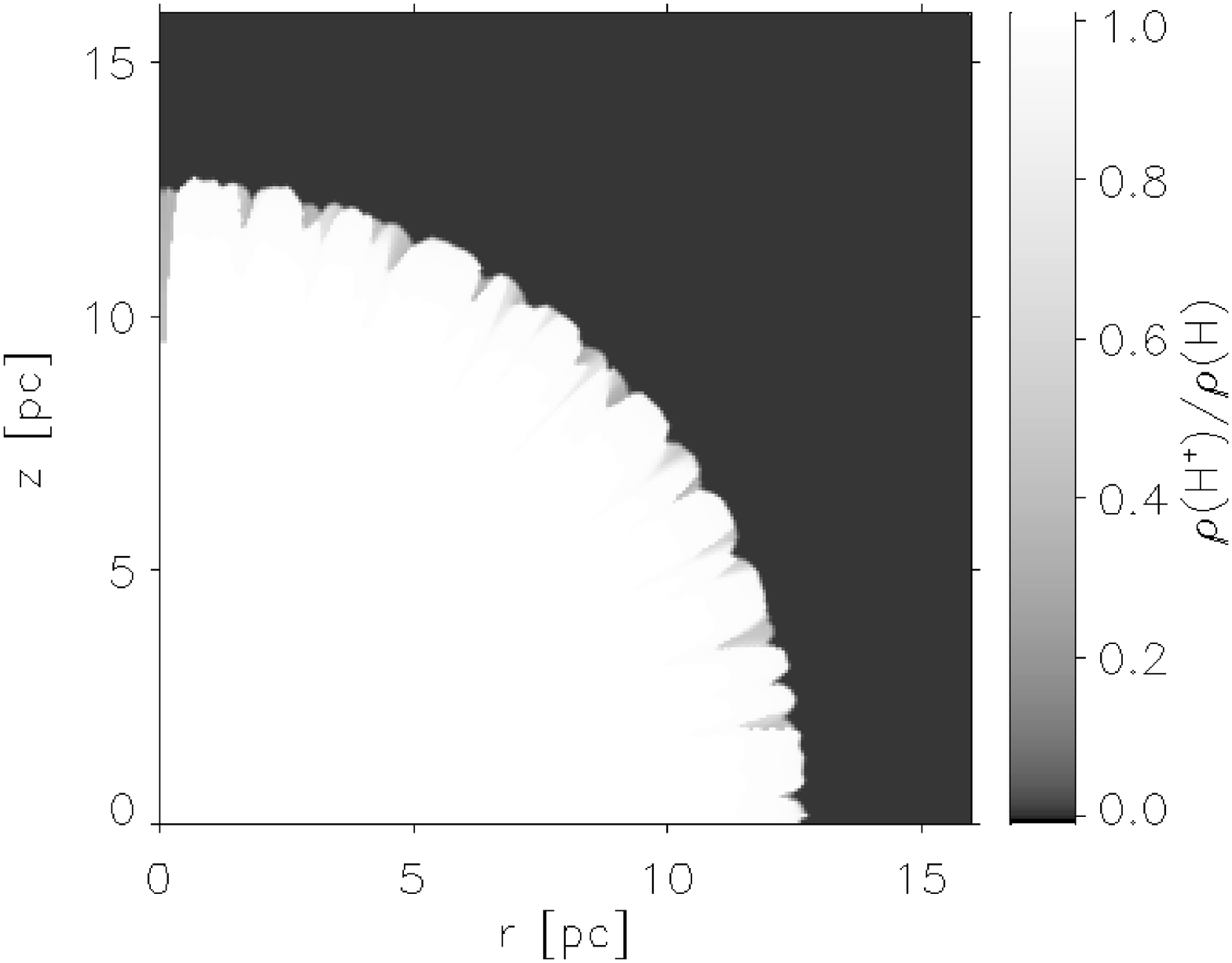}
  \caption{Same as Fig.~\ref{ion_uchii184.001.001.3.med.mono.eps} but at
           age $2\times 10^4\,\mathrm{yr}$. The velocity arrows in the
           free-flowing wind zone and in the hot bubble have been omitted
           to prevent confusion.
           \label{ion_uchii184.008.001.3.med.mono.eps}
          }
\end{figure}

\begin{figure}
  \epsscale{0.50}
  \plotone{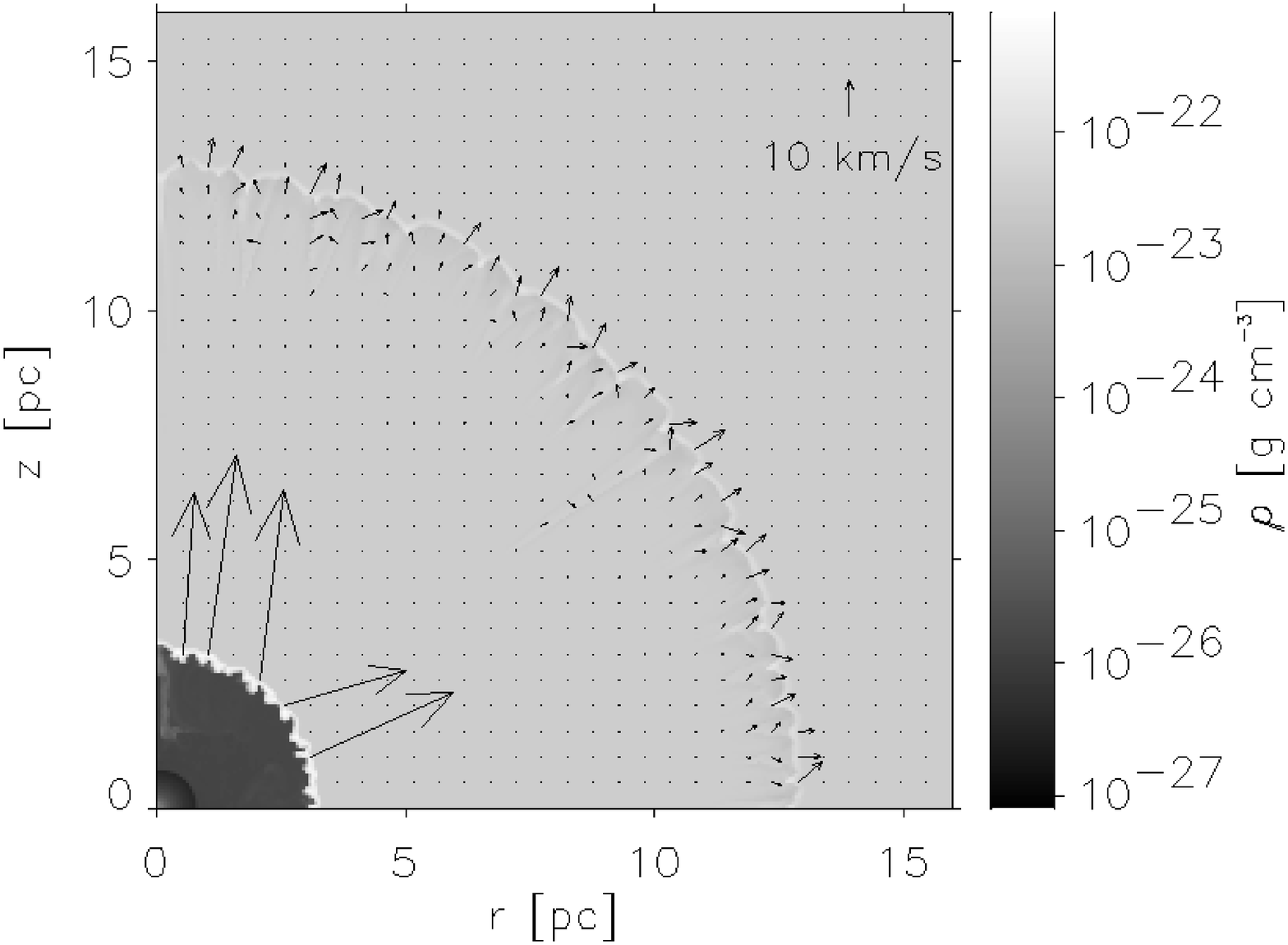}
  \plotone{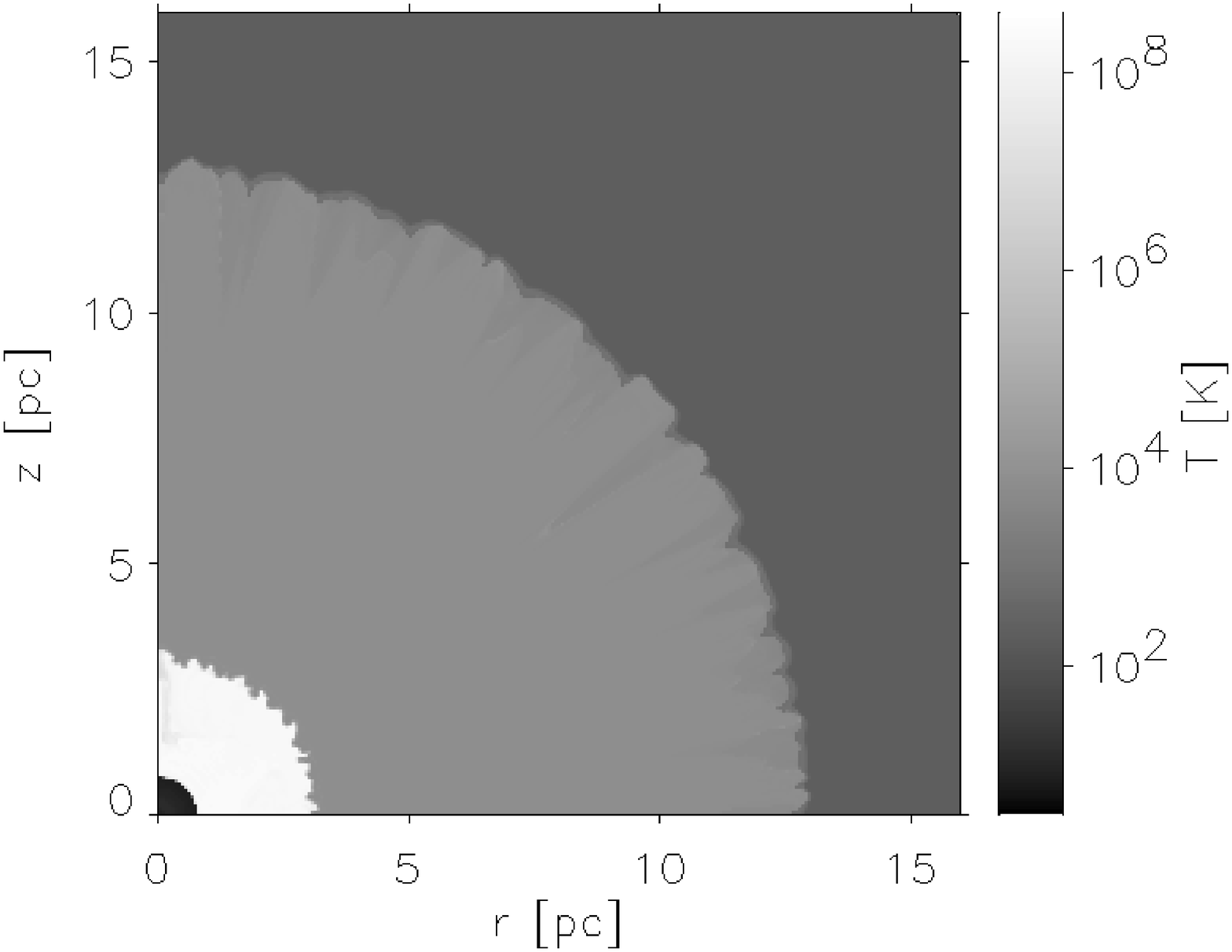}
  \plotone{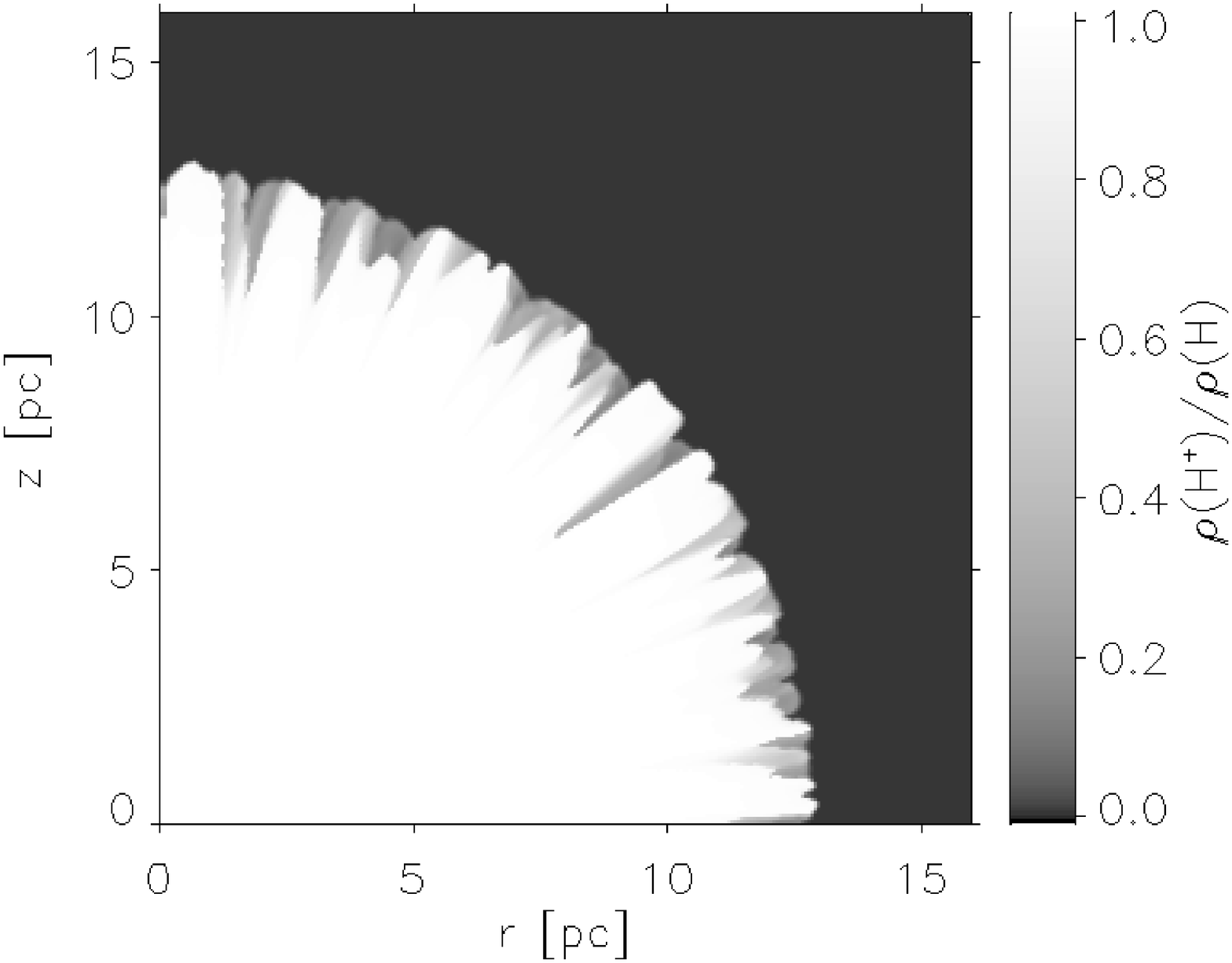}
  \caption{Same as Fig.~\ref{ion_uchii184.008.001.3.med.mono.eps}
           but at age $4 \times 10^4\,\mathrm{yr}$.
           \label{ion_uchii184.030.001.3.med.mono.eps}
          }
\end{figure}

\begin{figure}
  \epsscale{0.50}
  \plotone{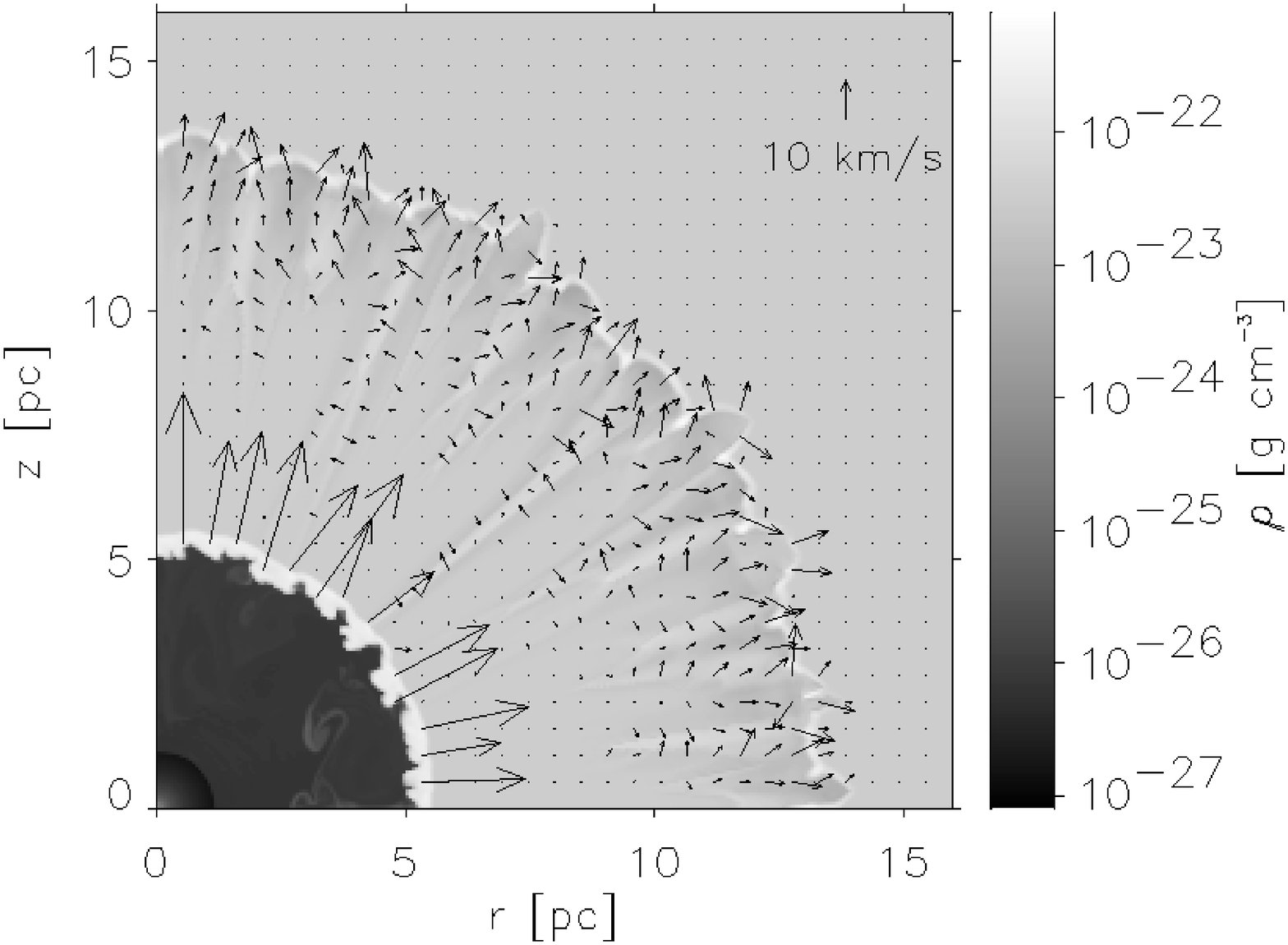}
  \plotone{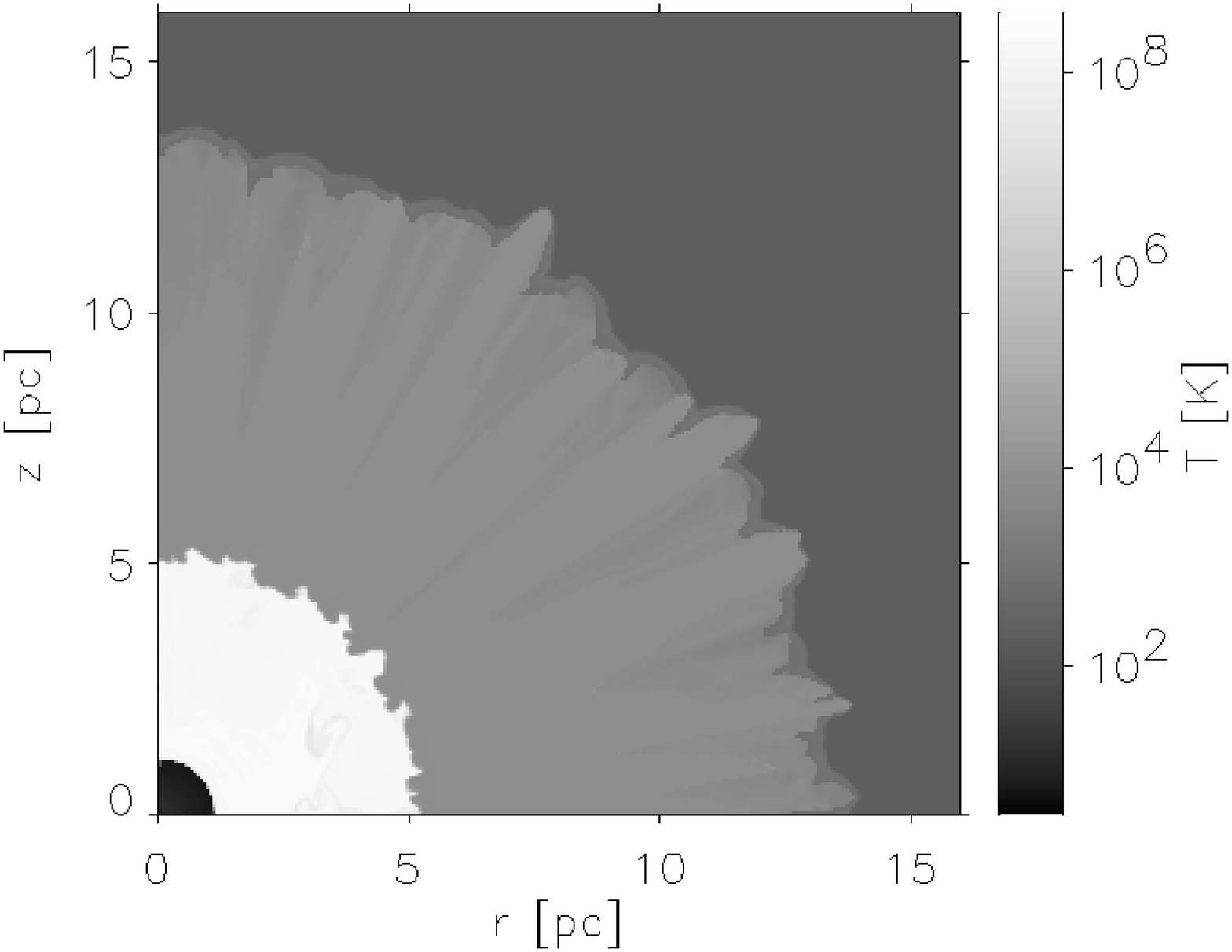}
  \plotone{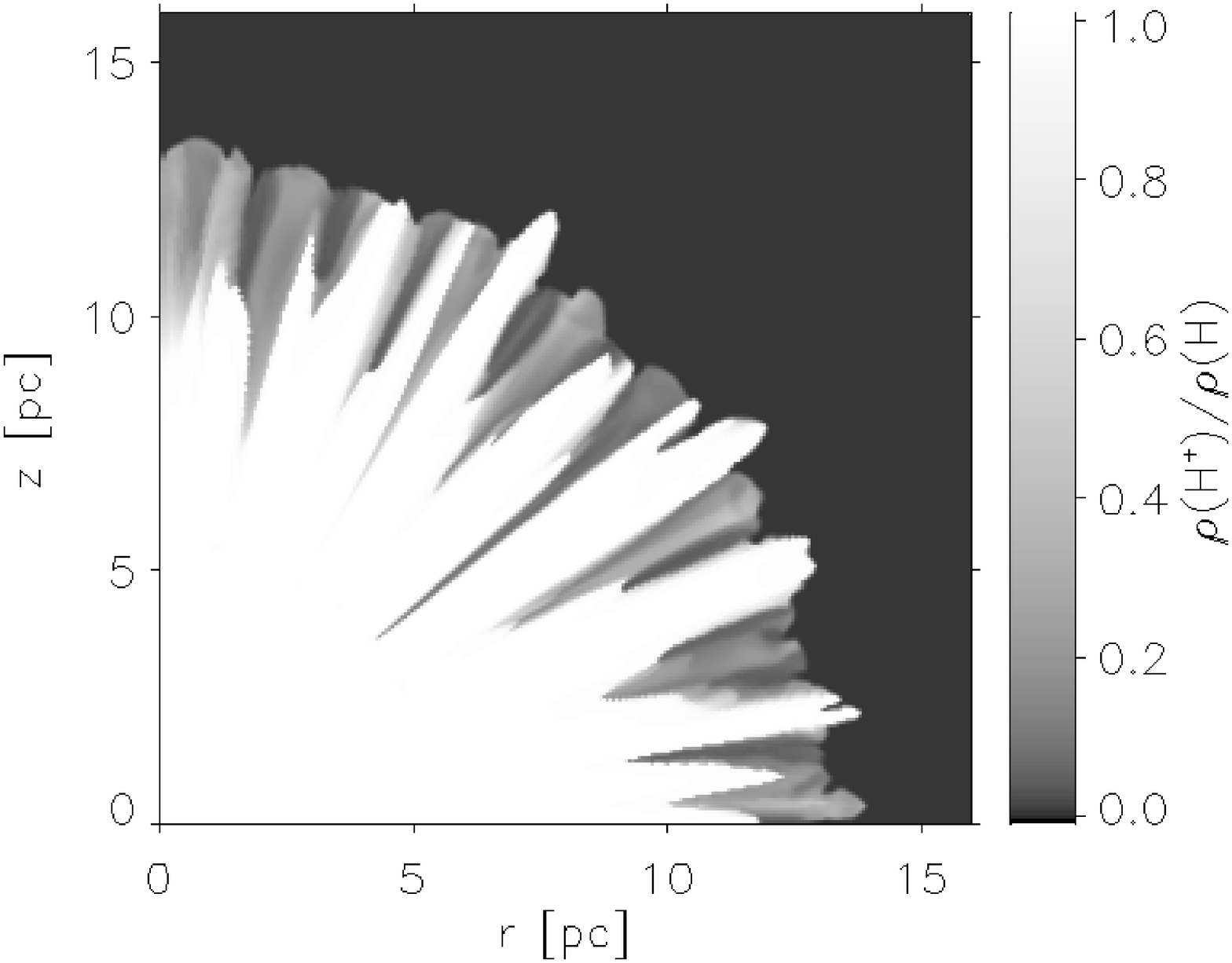}
  \caption{Same as Fig.~\ref{ion_uchii184.008.001.3.med.mono.eps}
           but at age 0.1\,Myr.
           \label{ion_uchii184.103.001.3.med.mono.eps}
          }
\end{figure}

\begin{figure}
  \epsscale{0.50}
  \plotone{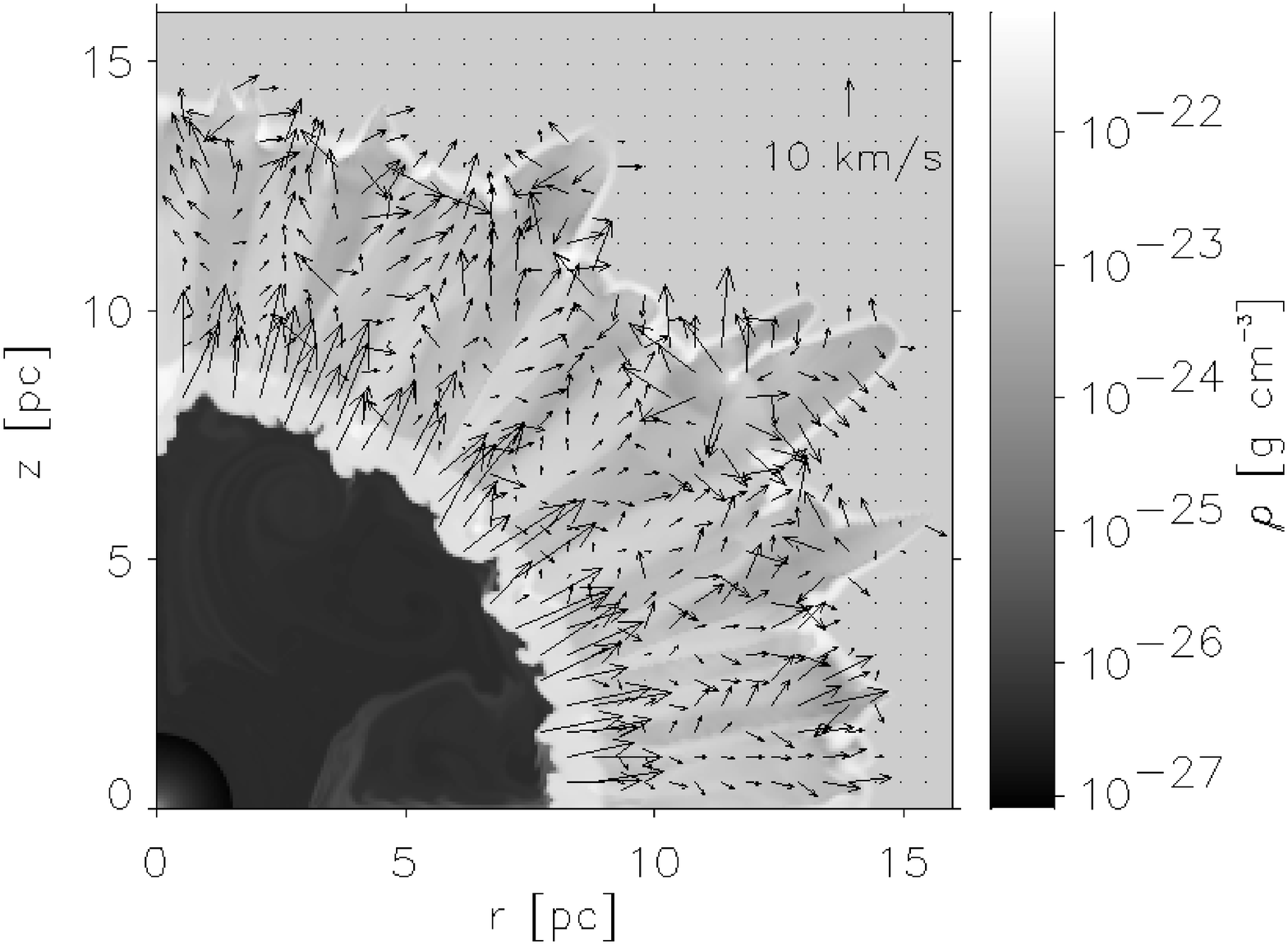}
  \plotone{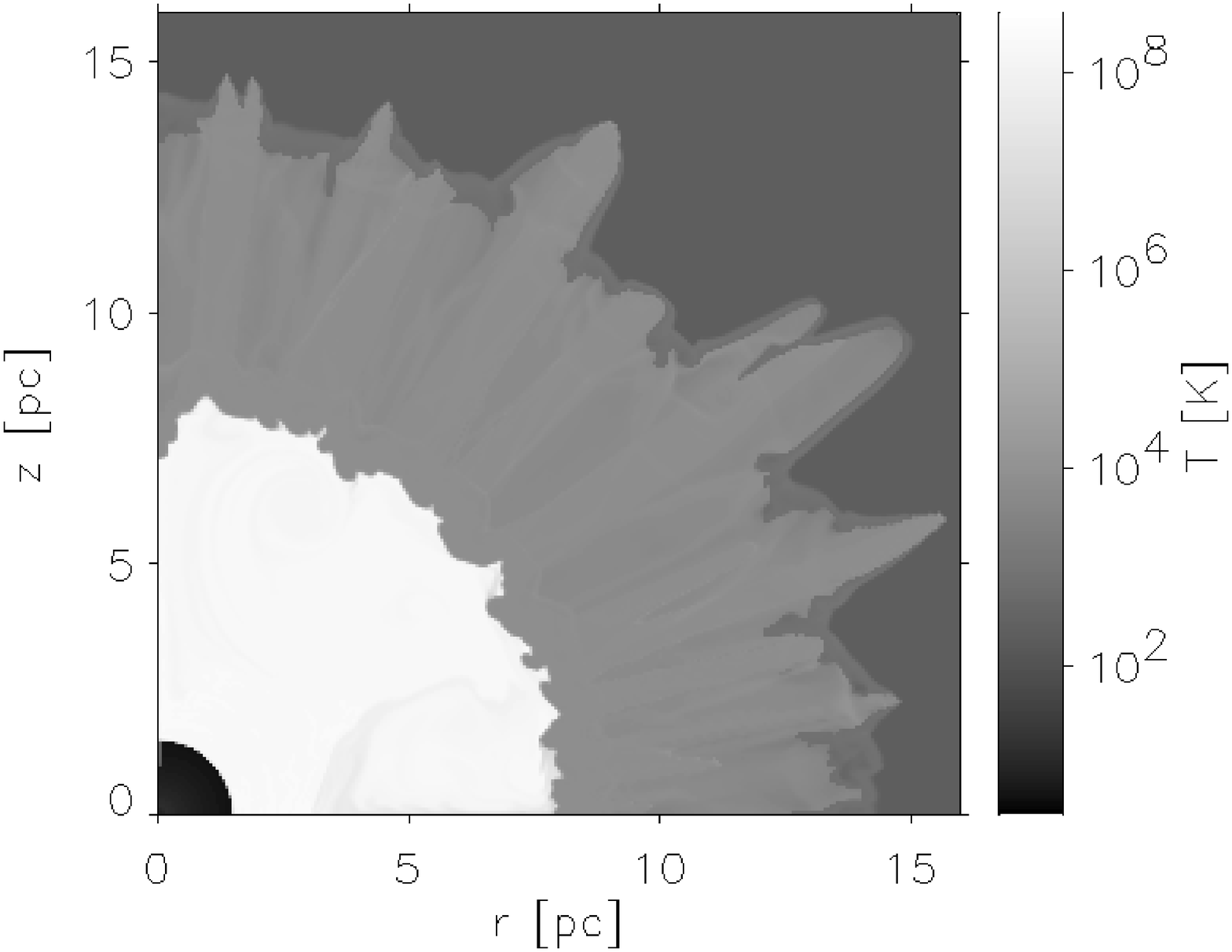}
  \plotone{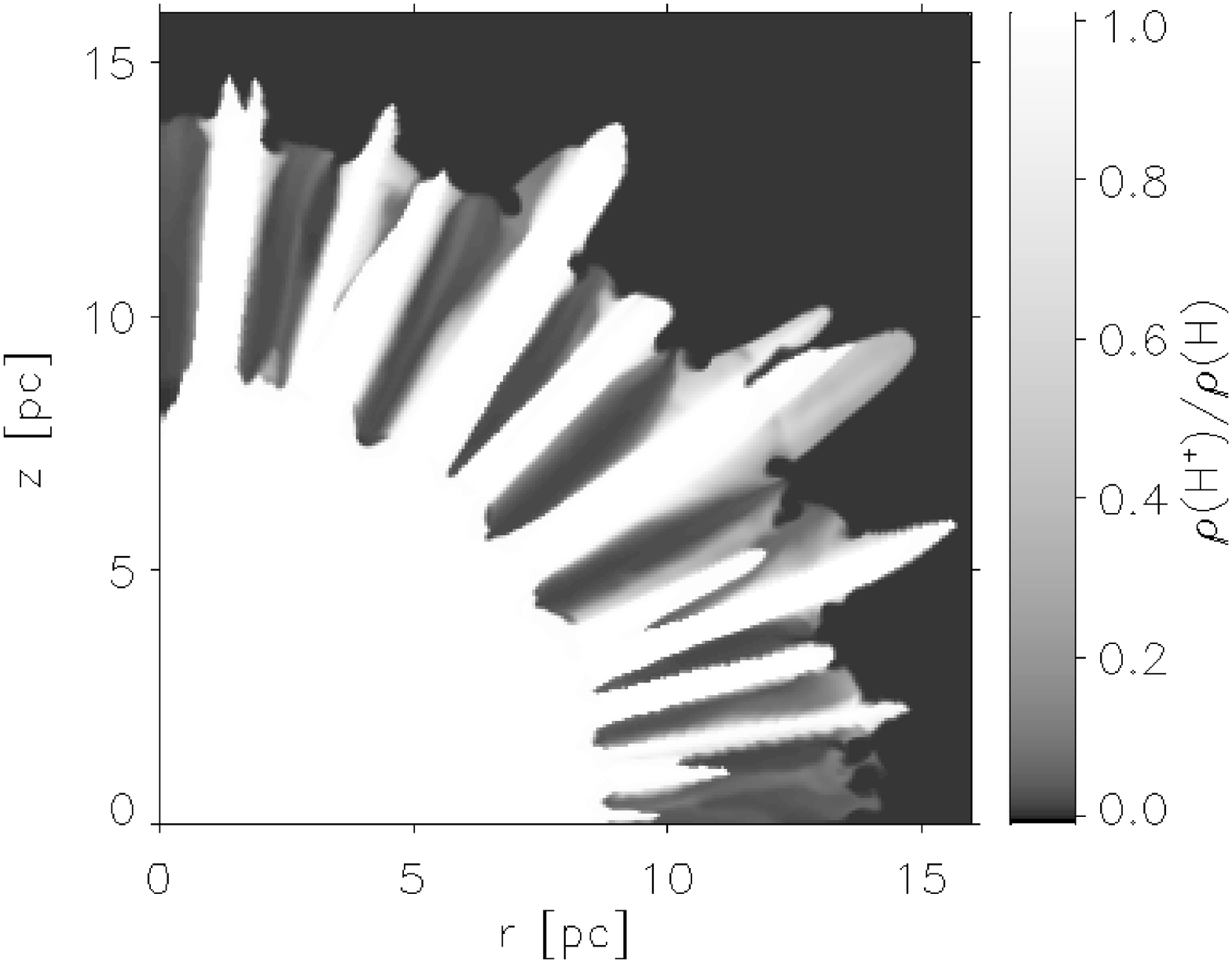}
  \caption{Same as Fig.~\ref{ion_uchii184.008.001.3.med.mono.eps}
           but at age 0.22\,Myr.
           \label{ion_uchii184.149.001.3.med.mono.eps}
          }
\end{figure}

\begin{figure}
  \epsscale{0.50}
  \plotone{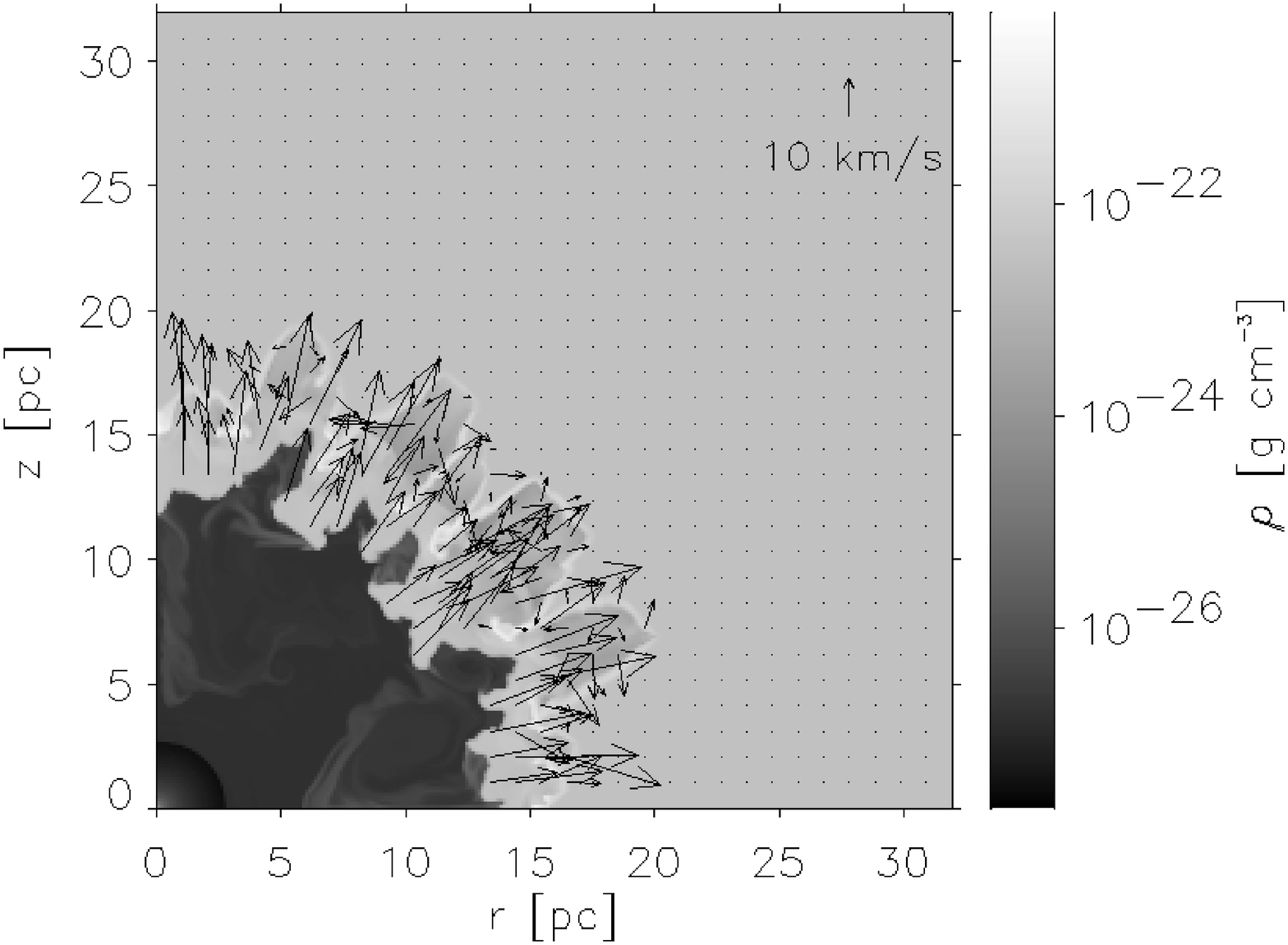}
  \plotone{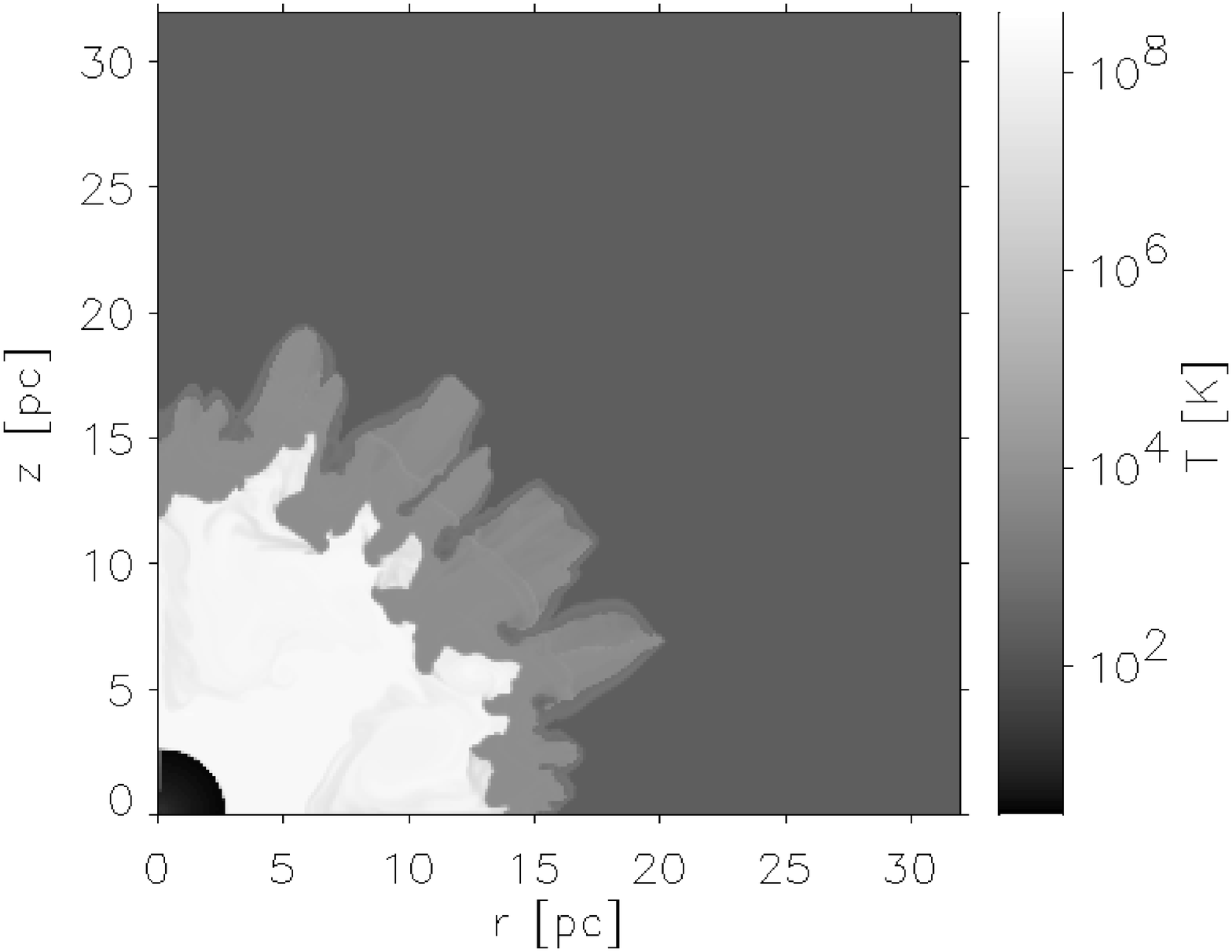}
  \plotone{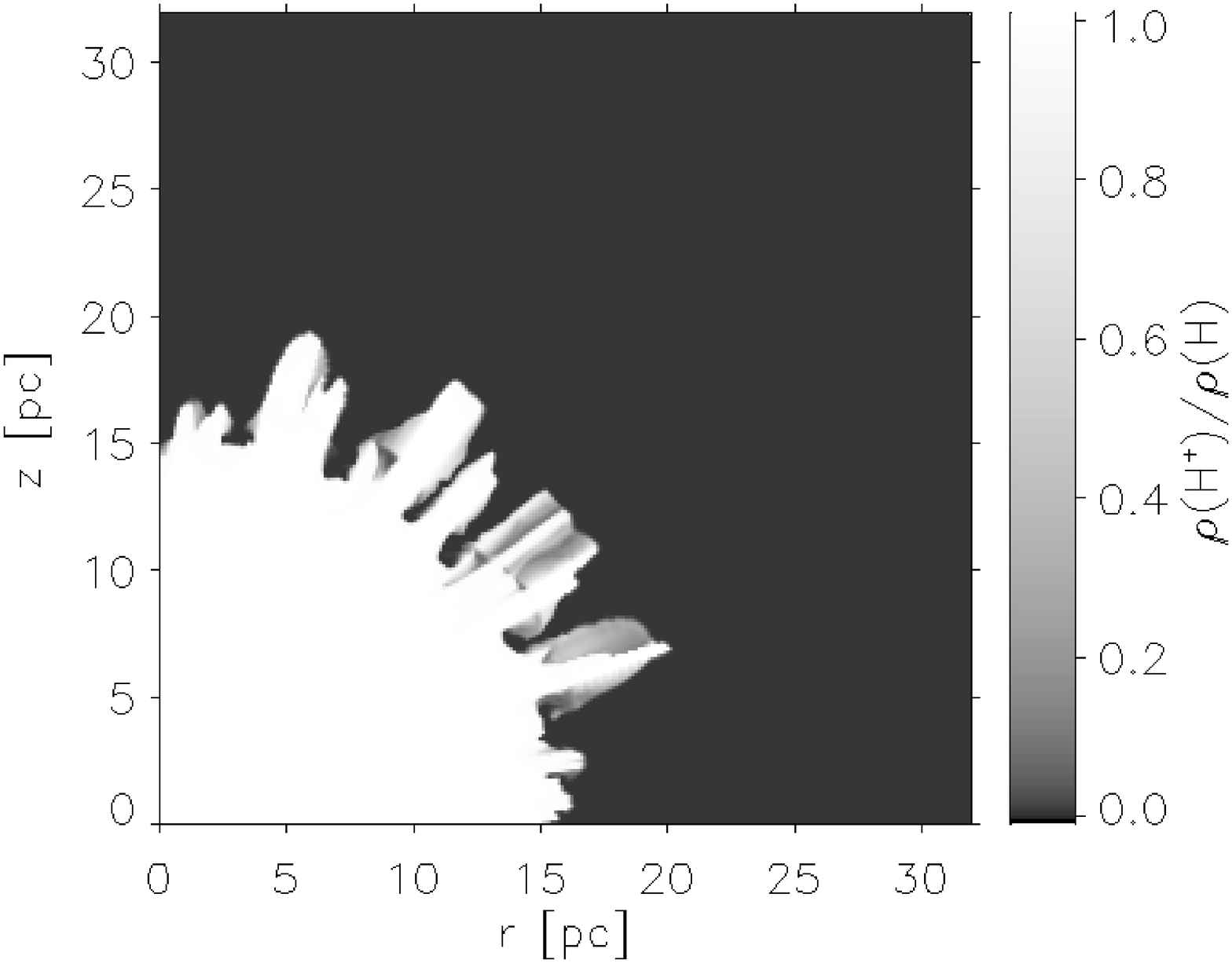}
  \caption{Same as Fig.~\ref{ion_uchii184.008.001.3.med.mono.eps}
           but at age 0.5\,Myr. Because of the bubble
           expansion a larger volume is shown.
           \label{ion_uchii184.244.001.2.med.mono.eps}
          }
\end{figure}

\begin{figure}
  \epsscale{0.50}
  \plotone{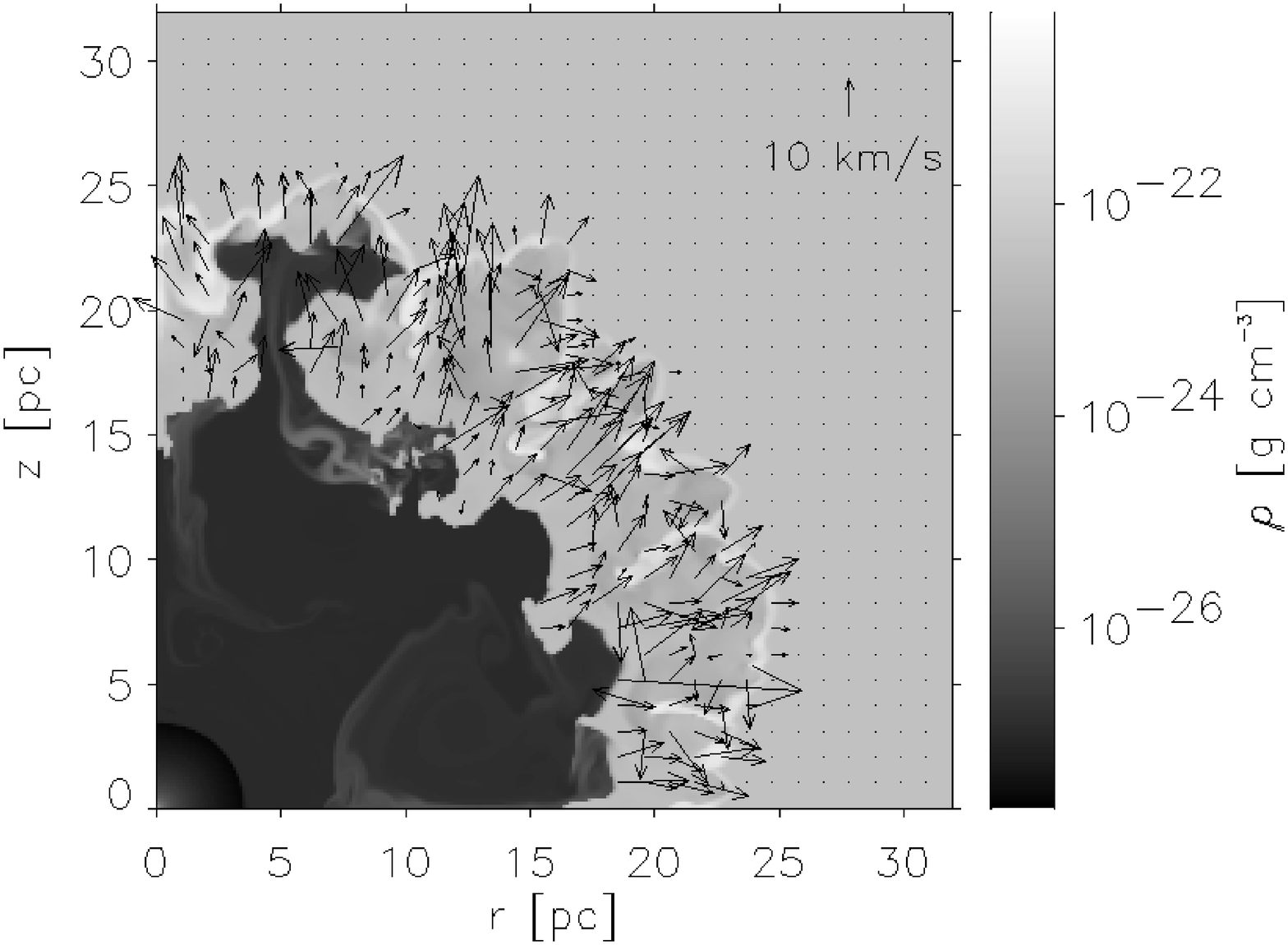}
  \plotone{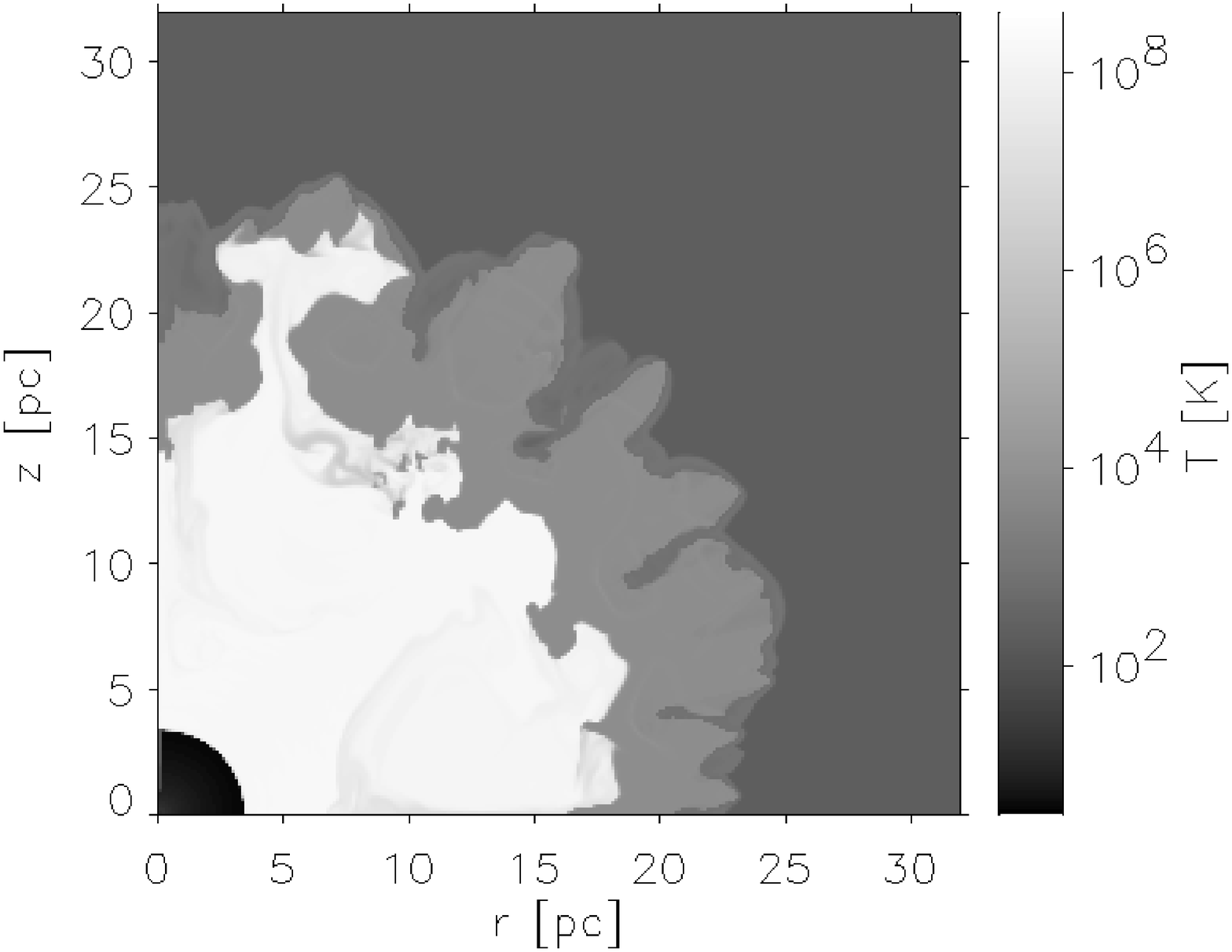}
  \plotone{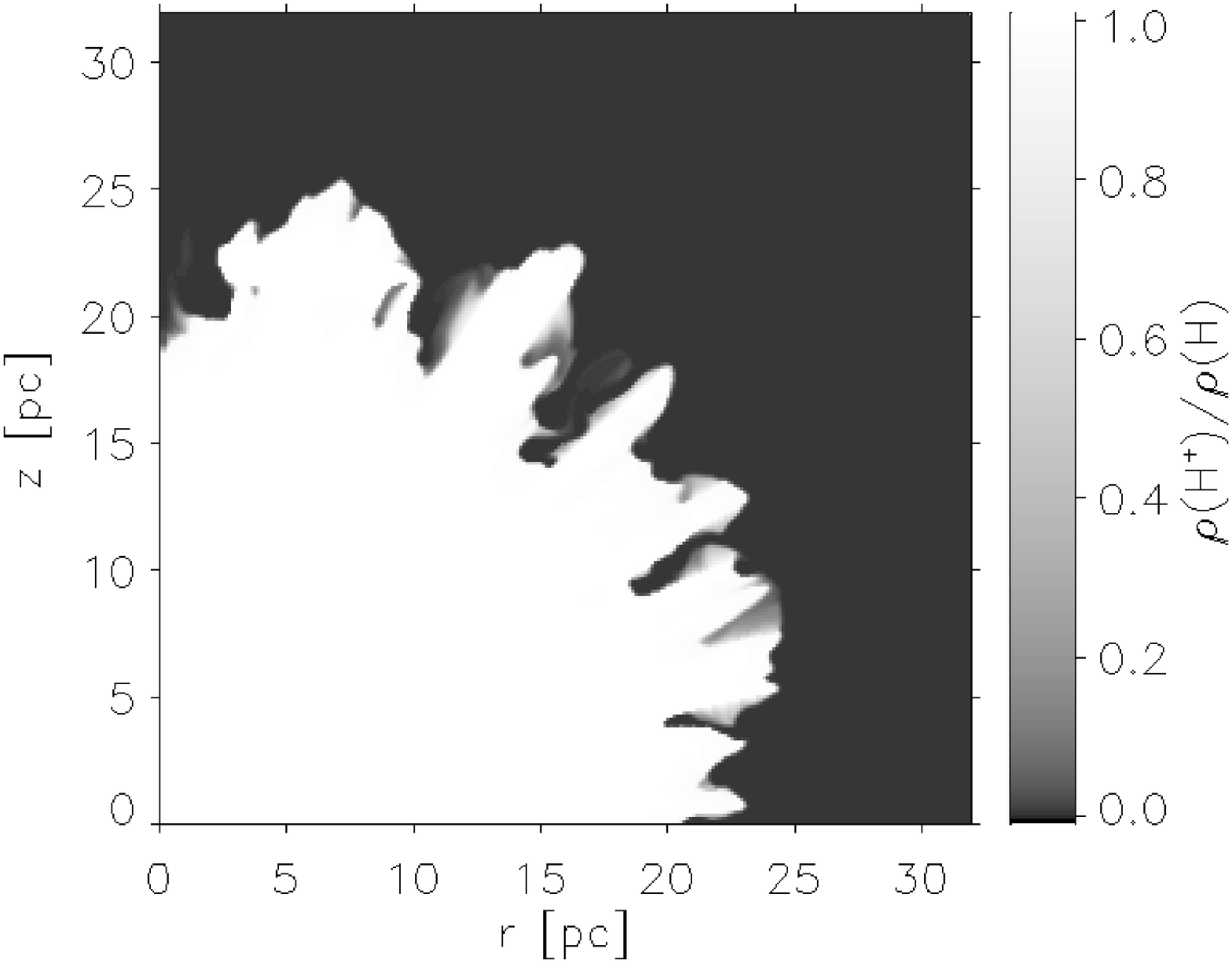}
  \caption{Same as Fig.~\ref{ion_uchii184.244.001.2.med.mono.eps}
           but at age 1\,Myr.
           \label{ion_uchii184.473.001.2.med.mono.eps}
          }
\end{figure}

\begin{figure}
  \epsscale{0.50}
  \plotone{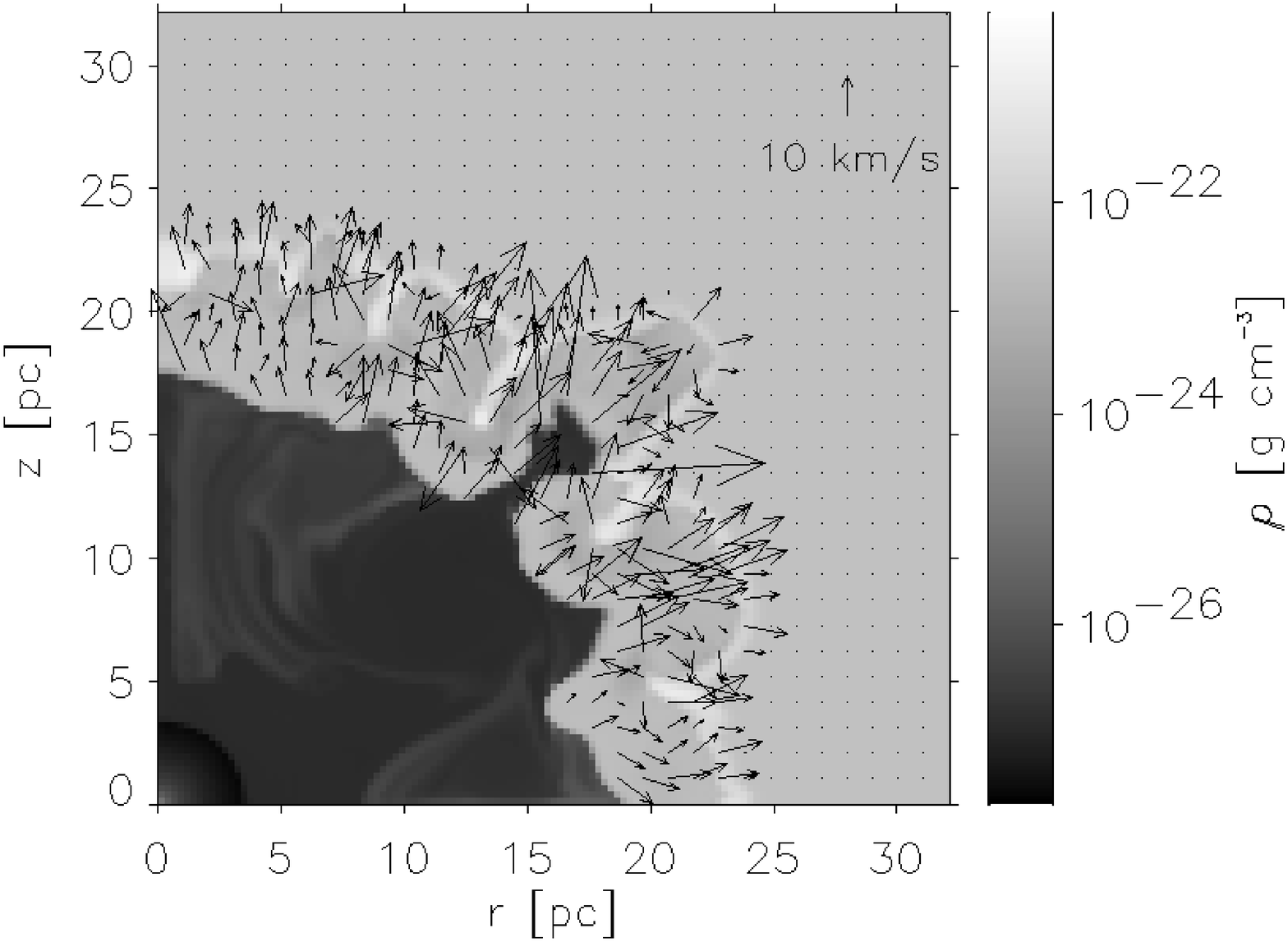}
  \plotone{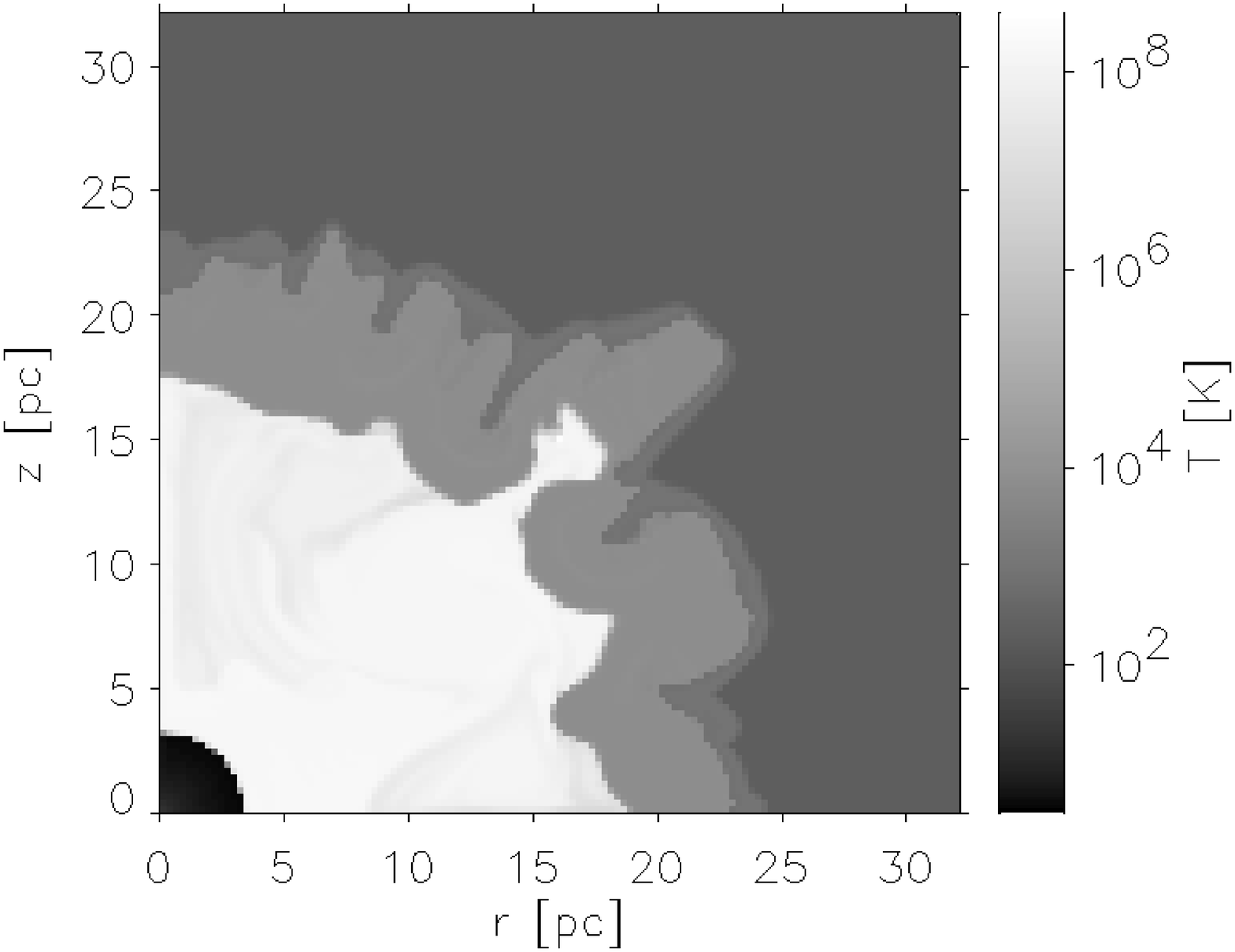}
  \plotone{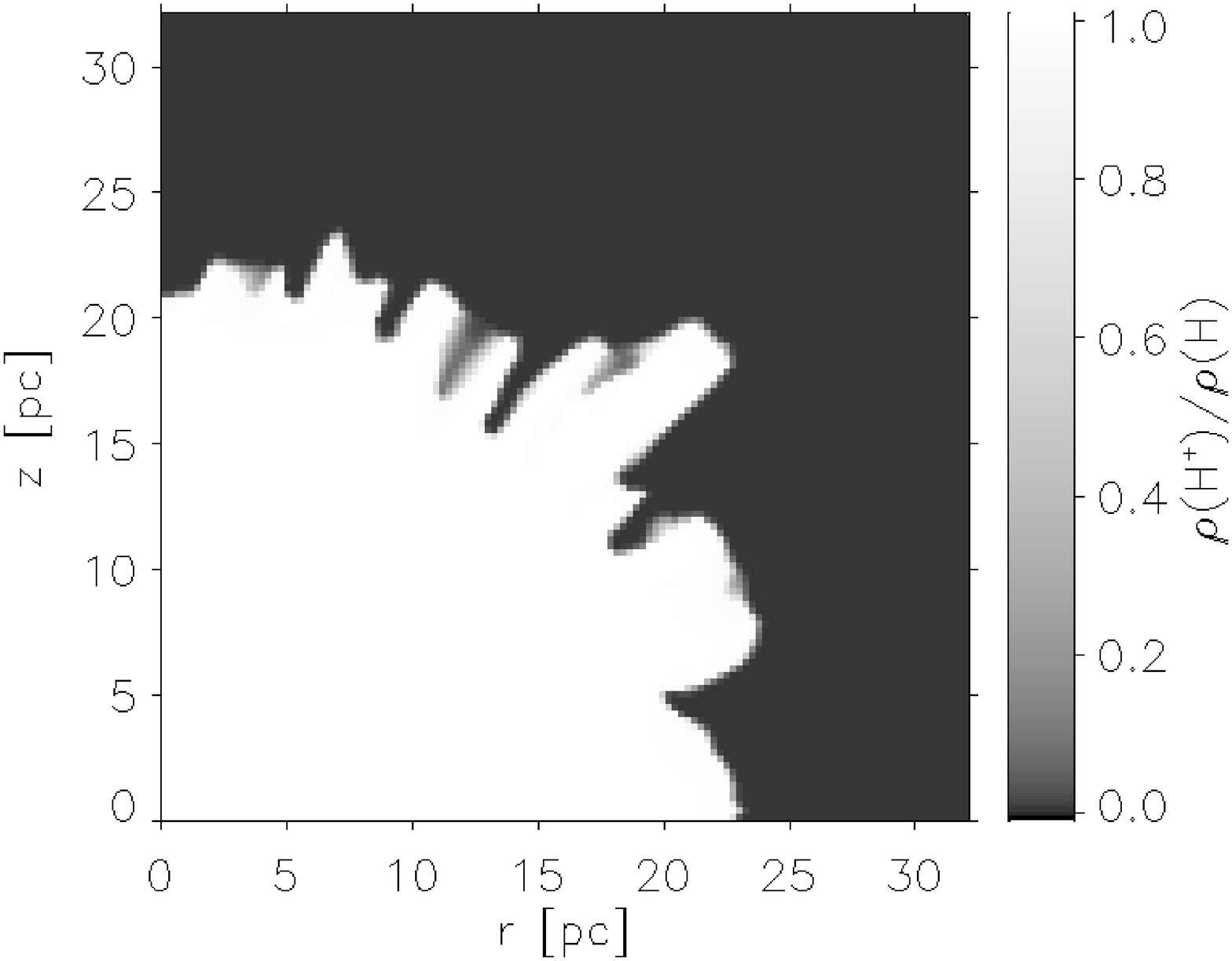}
  \caption{Same as Fig.~\ref{ion_uchii184.473.001.2.med.mono.eps}
           but for the medium resolution run.
           \label{ion_uchii227.283.001.2.med.mono.eps}
          }
\end{figure}

\begin{figure}
  \epsscale{1.10}
  \plottwo{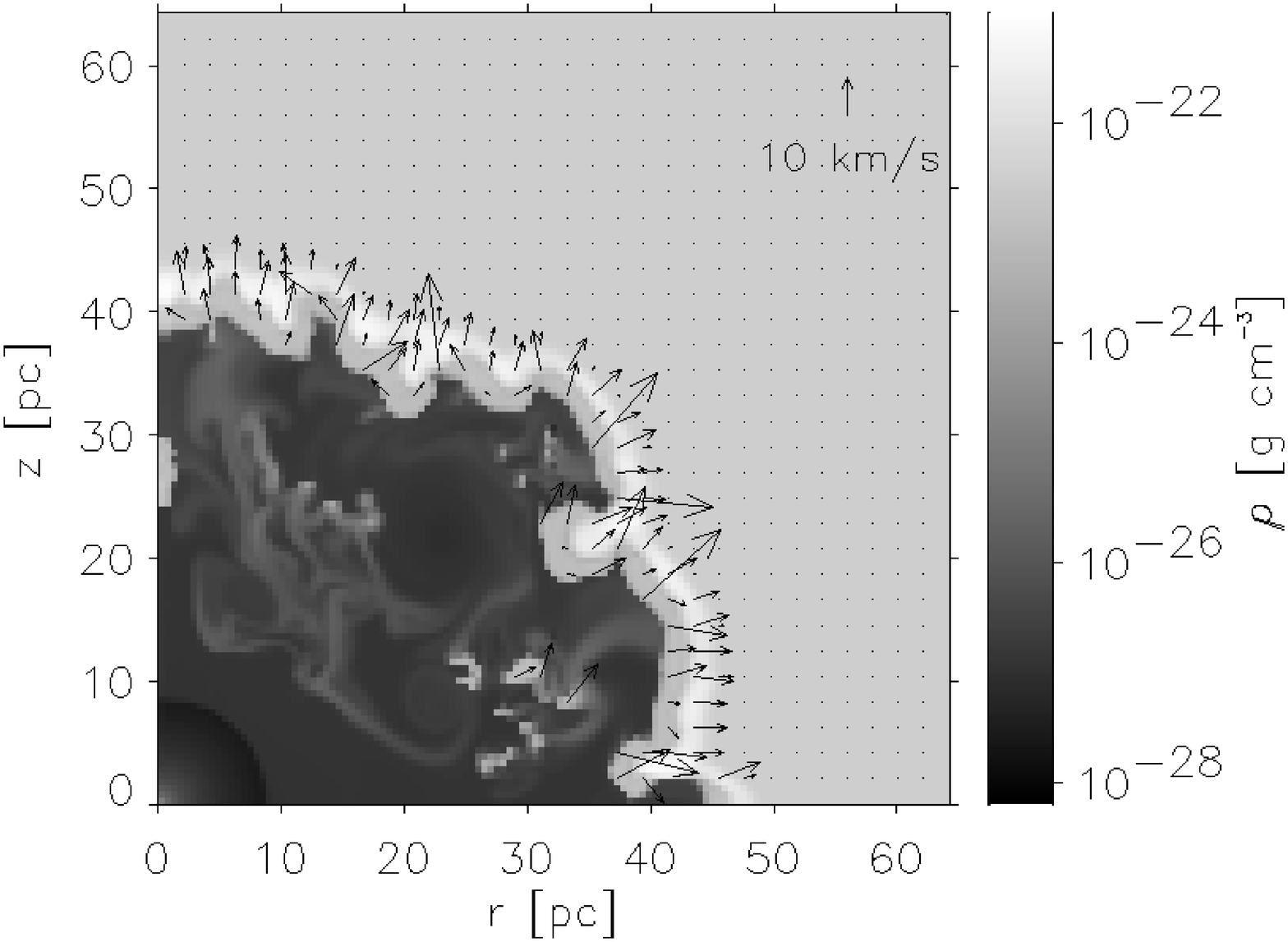}{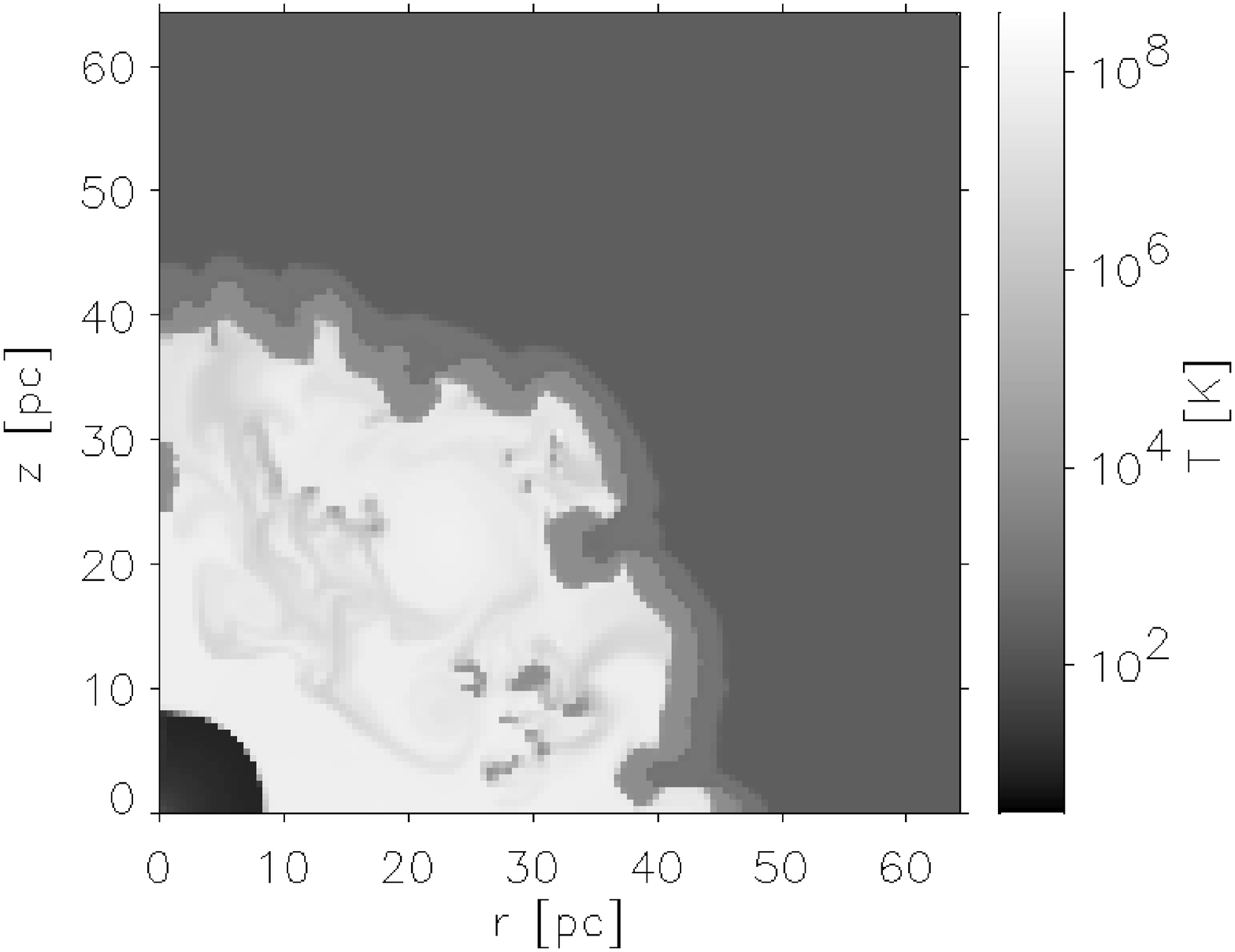}
  \caption{Same as Fig.~\ref{ion_uchii227.283.001.2.med.mono.eps}
           but at age 3.30\,Myr.
           Only mass density and velocity field (left panel) and
           temperature (right panel) are shown and the displayed area
           is enlarged once again.
           \label{tem_uchii227.758.002.1.med.mono.eps}
          }
\end{figure}

\begin{figure}
  \epsscale{1.10}
  \plottwo{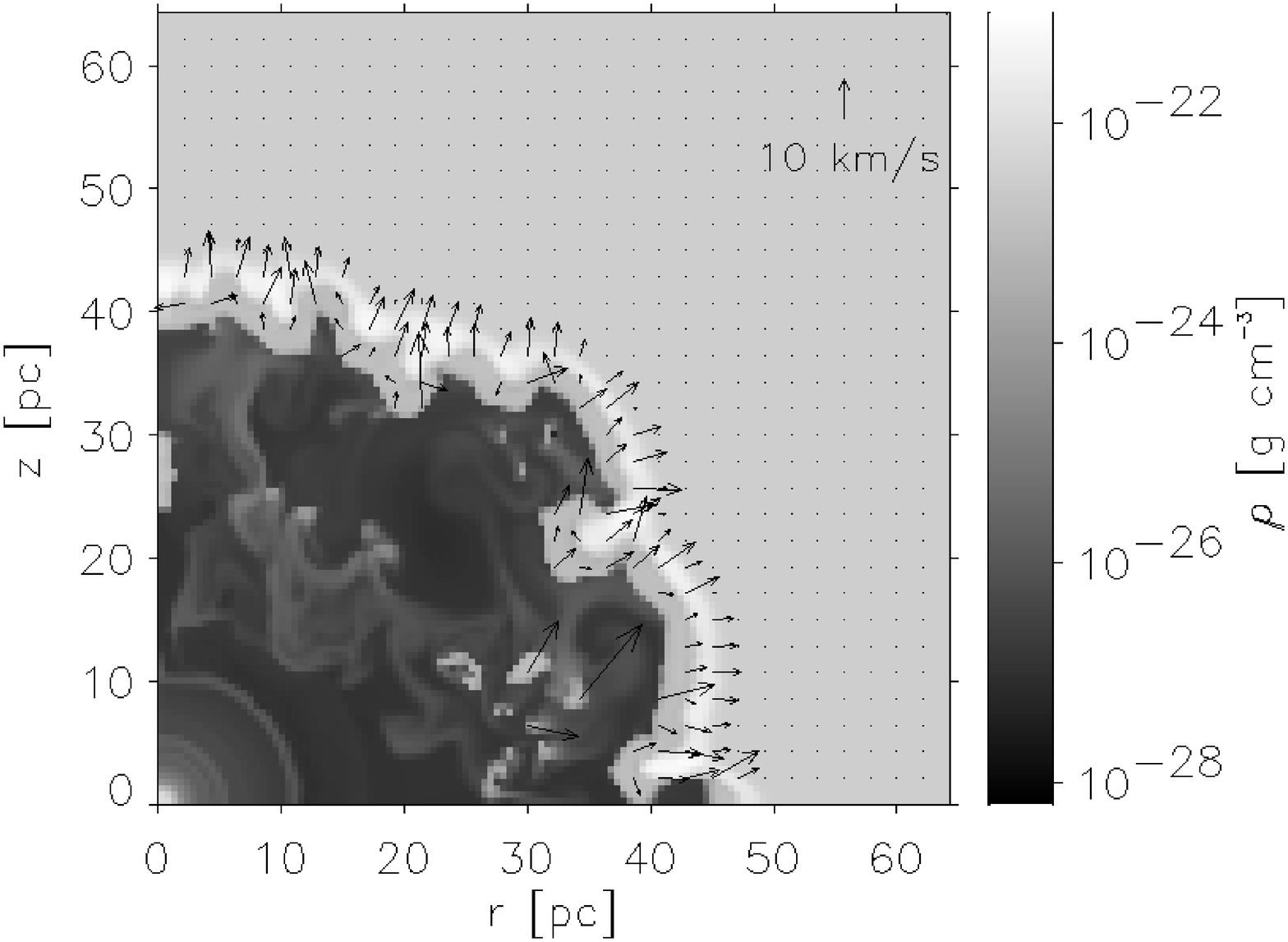}{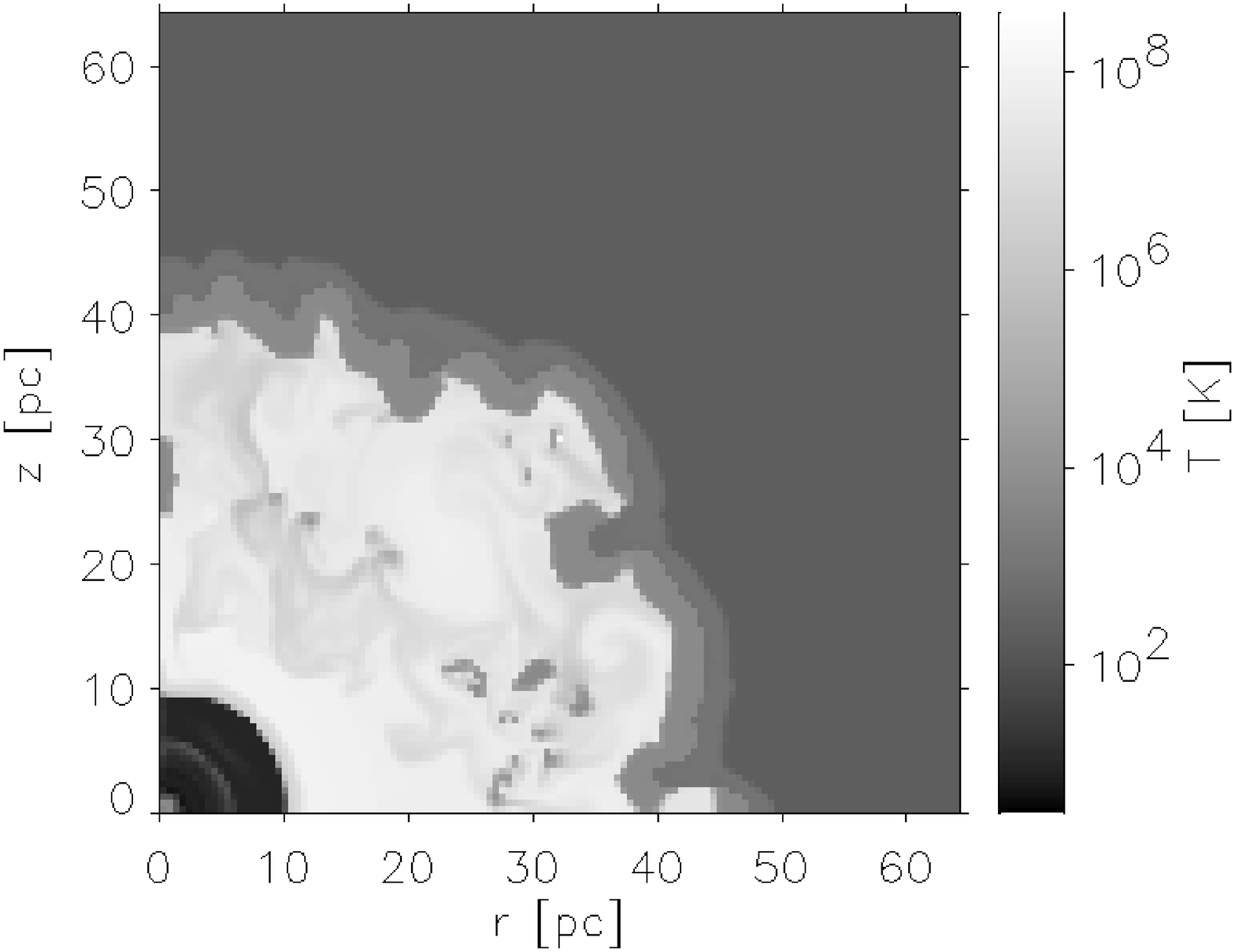}
  \caption{Same as Fig.~\ref{tem_uchii227.758.002.1.med.mono.eps}
           but at age 3.36\,Myr.
           \label{tem_uchii227.760.003.1.med.mono.eps}
          }
\end{figure}

\begin{figure}
  \epsscale{1.10}
  \plottwo{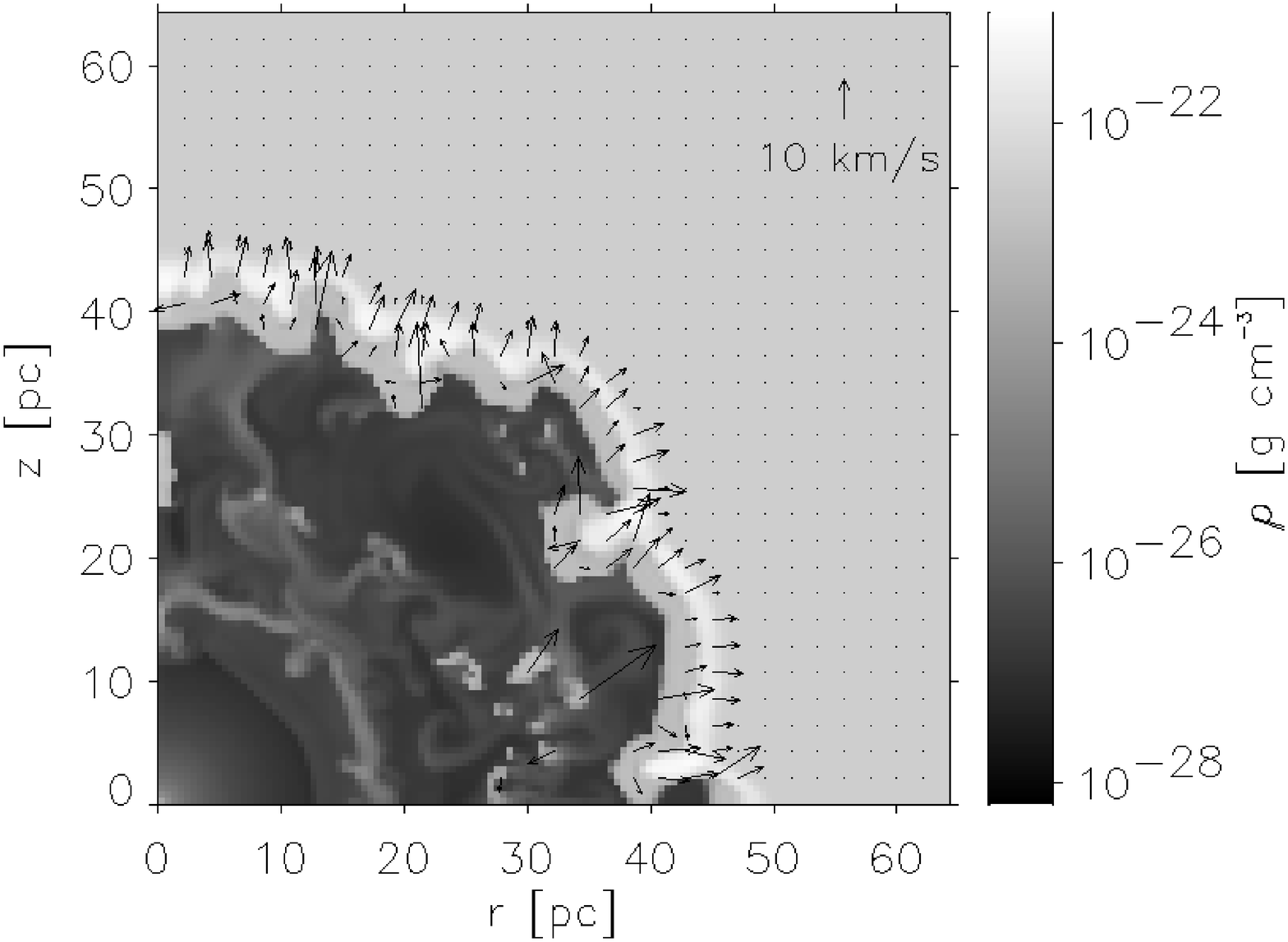}{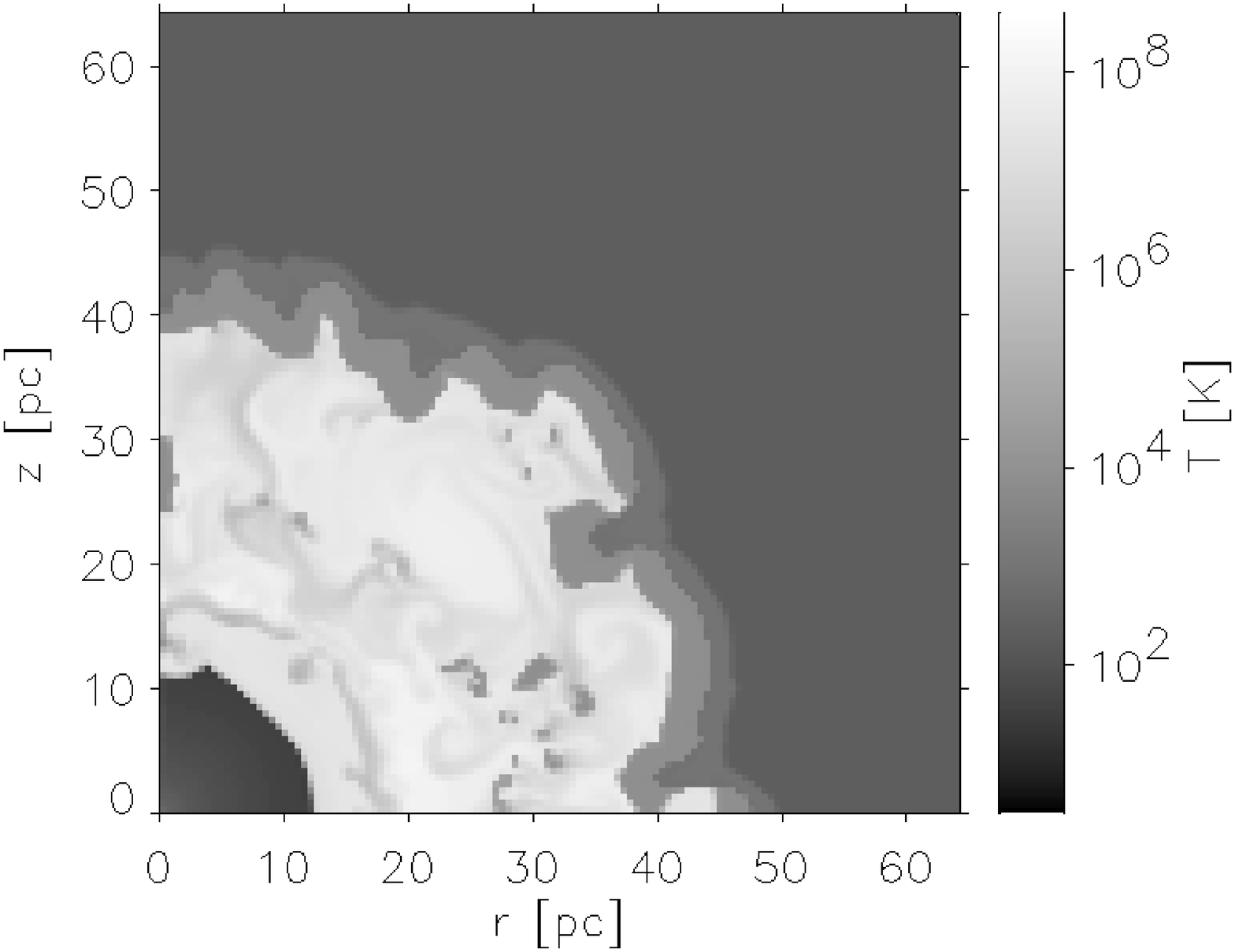}
  \caption{Same as Fig.~\ref{tem_uchii227.758.002.1.med.mono.eps}
           but at age 3.38\,Myr.
           \label{tem_uchii227.761.002.1.med.mono.eps}
          }
\end{figure}

\begin{figure}
  \epsscale{1.10}
  \plottwo{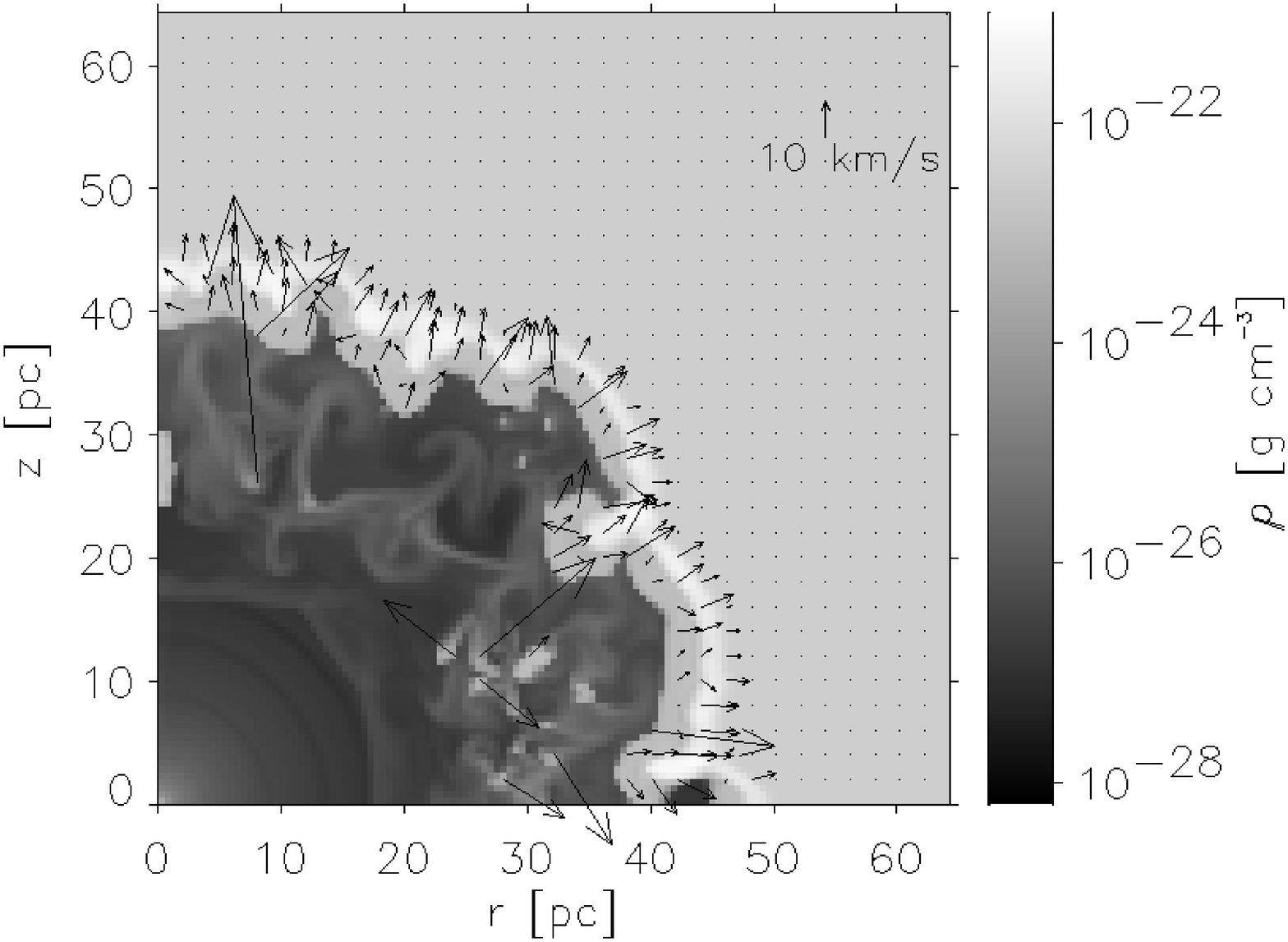}{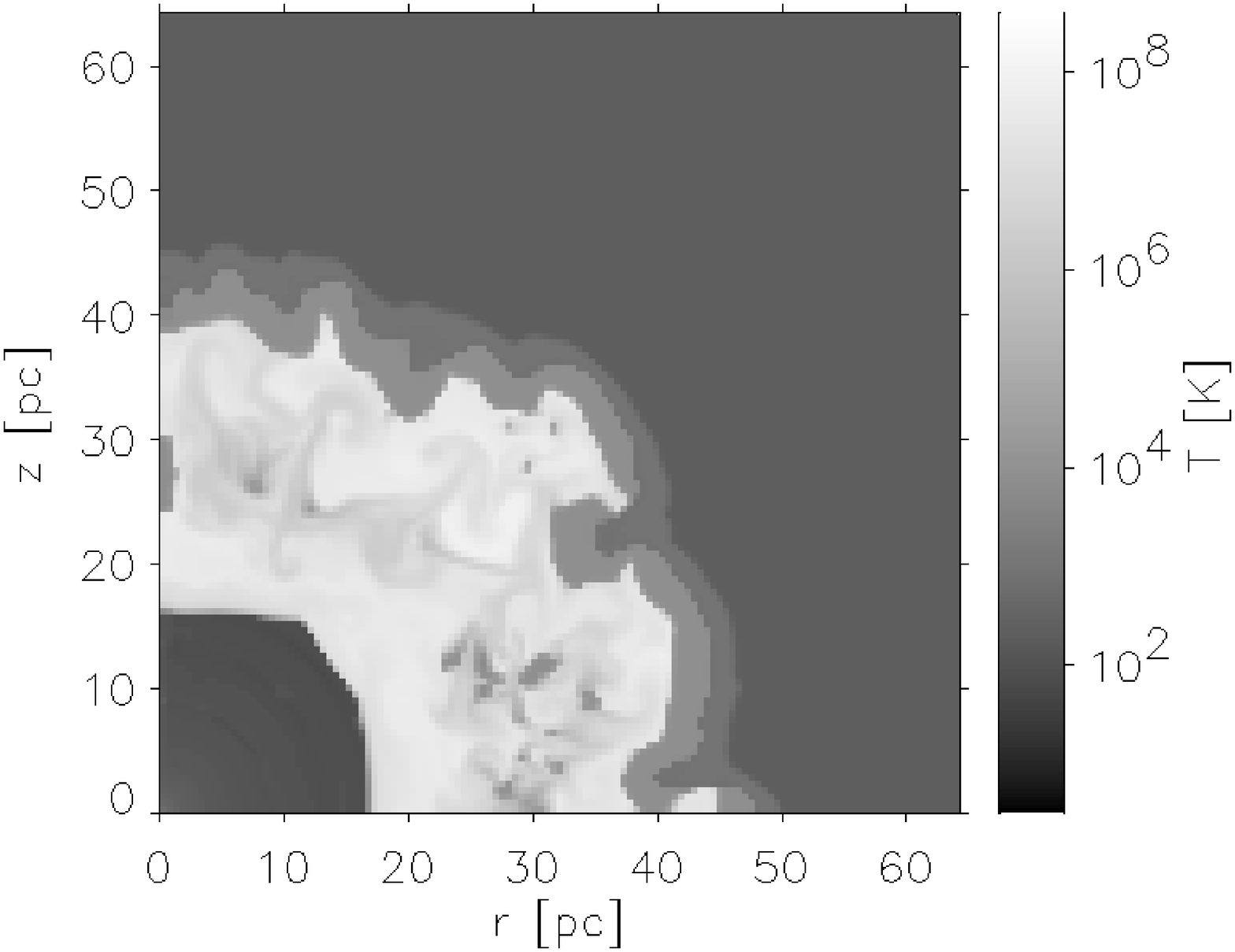}
  \caption{Same as Fig.~\ref{tem_uchii227.758.002.1.med.mono.eps}
           but at age 3.41\,Myr.
           \label{tem_uchii227.761.008.1.med.mono.eps}
          }
\end{figure}

\begin{figure}
  \epsscale{1.10}
  \plottwo{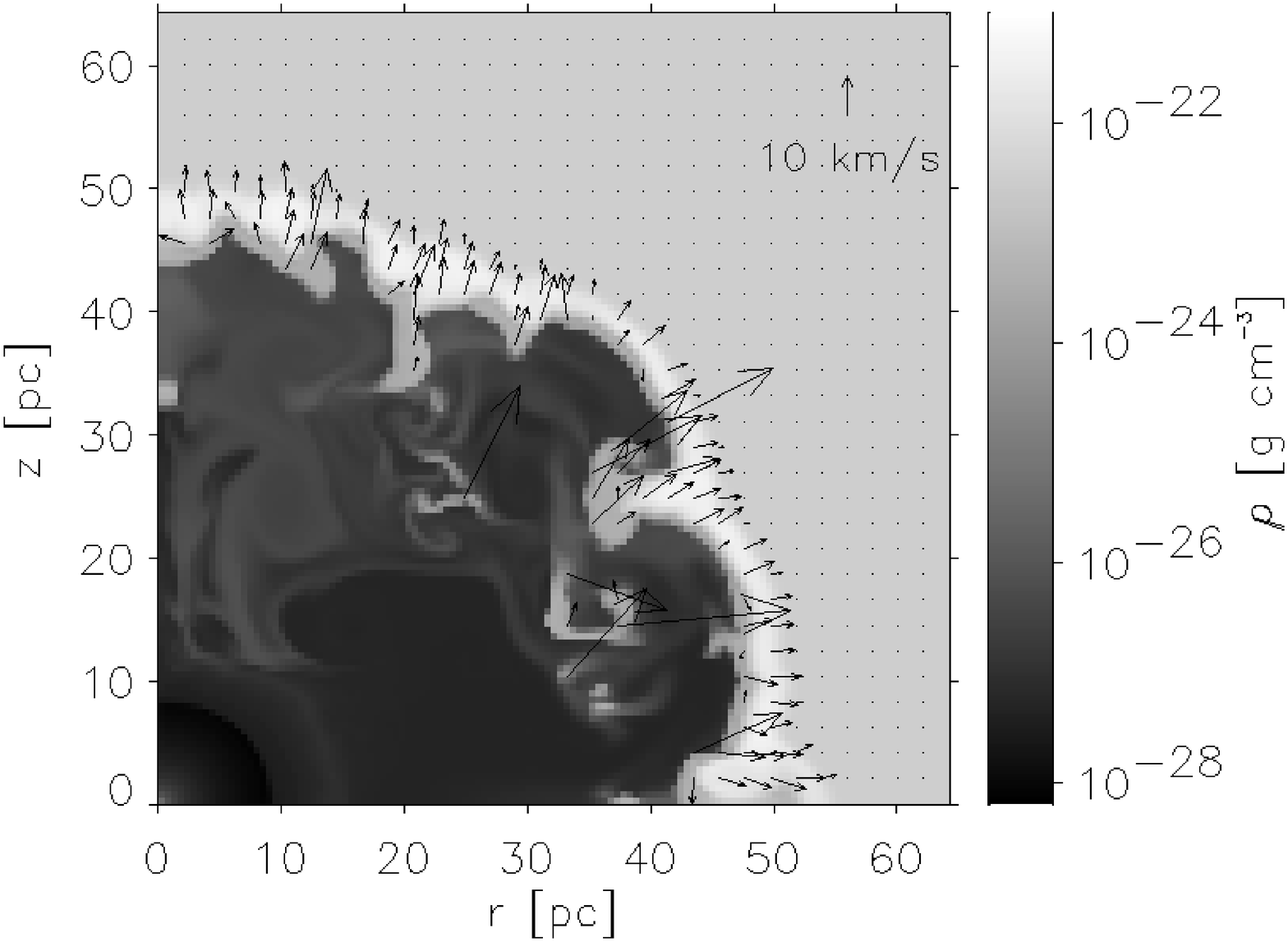}{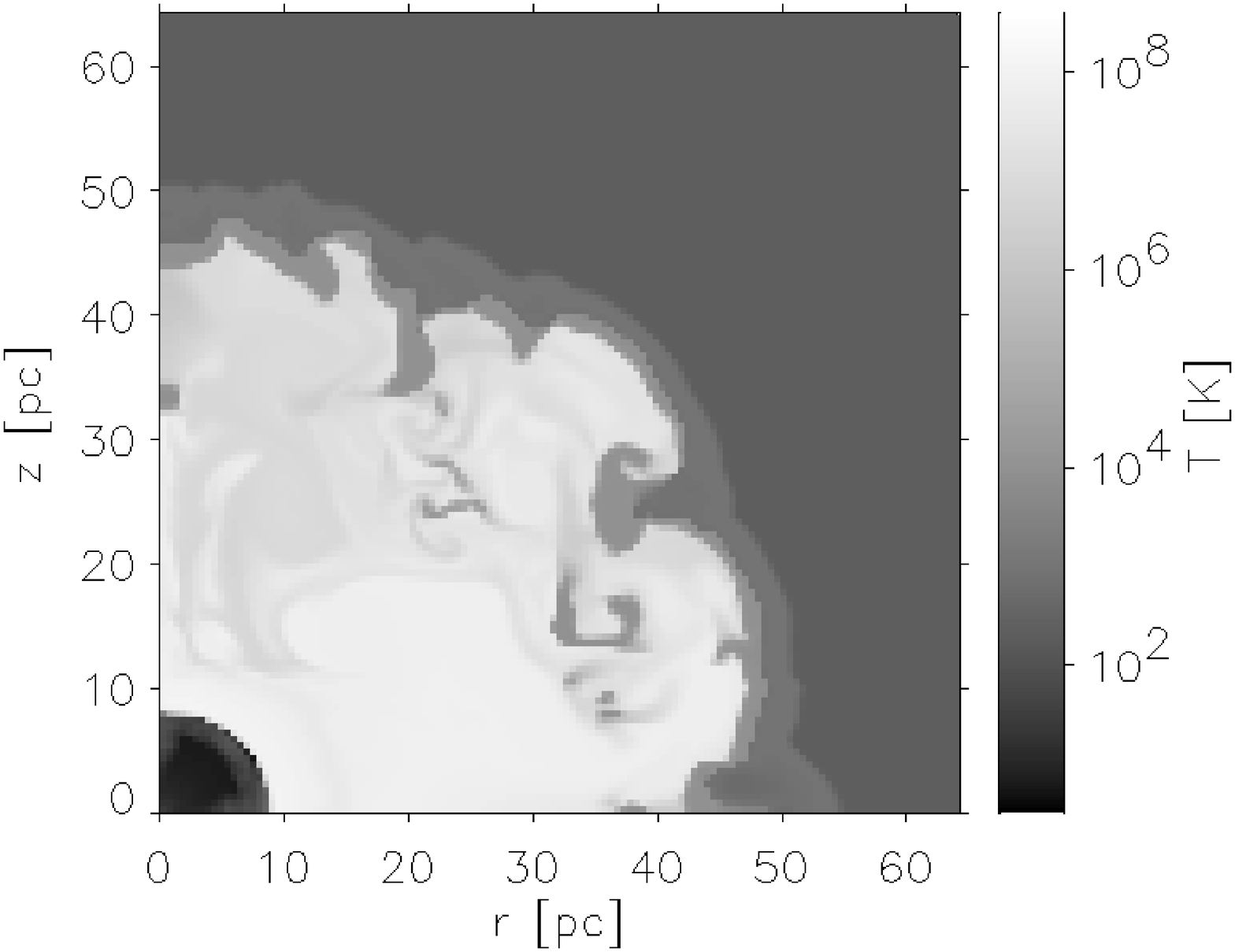}
  \caption{Same as Fig.~\ref{tem_uchii227.758.002.1.med.mono.eps}
           but at age 4.07\,Myr.
           \label{tem_uchii227.805.001.1.med.mono.eps}
          }
\end{figure}

\begin{figure}
  \epsscale{1.00}
  \plotone{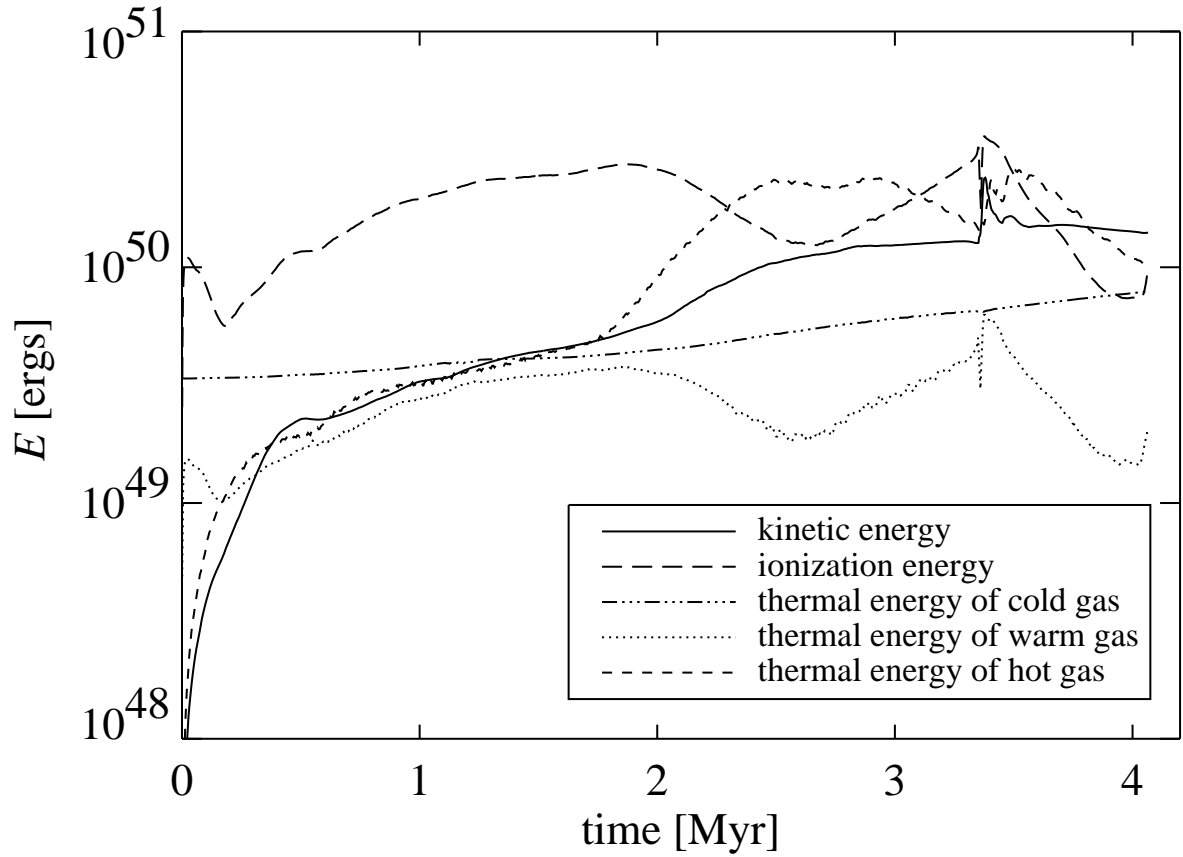}
  \caption{The temporal evolution of kinetic, thermal, and ionization
           energy in the 60\,$\Msun$ model. For details see text.
           \label{e_distri227_up.eps}
          }
\end{figure}

\begin{figure}
  \plotone{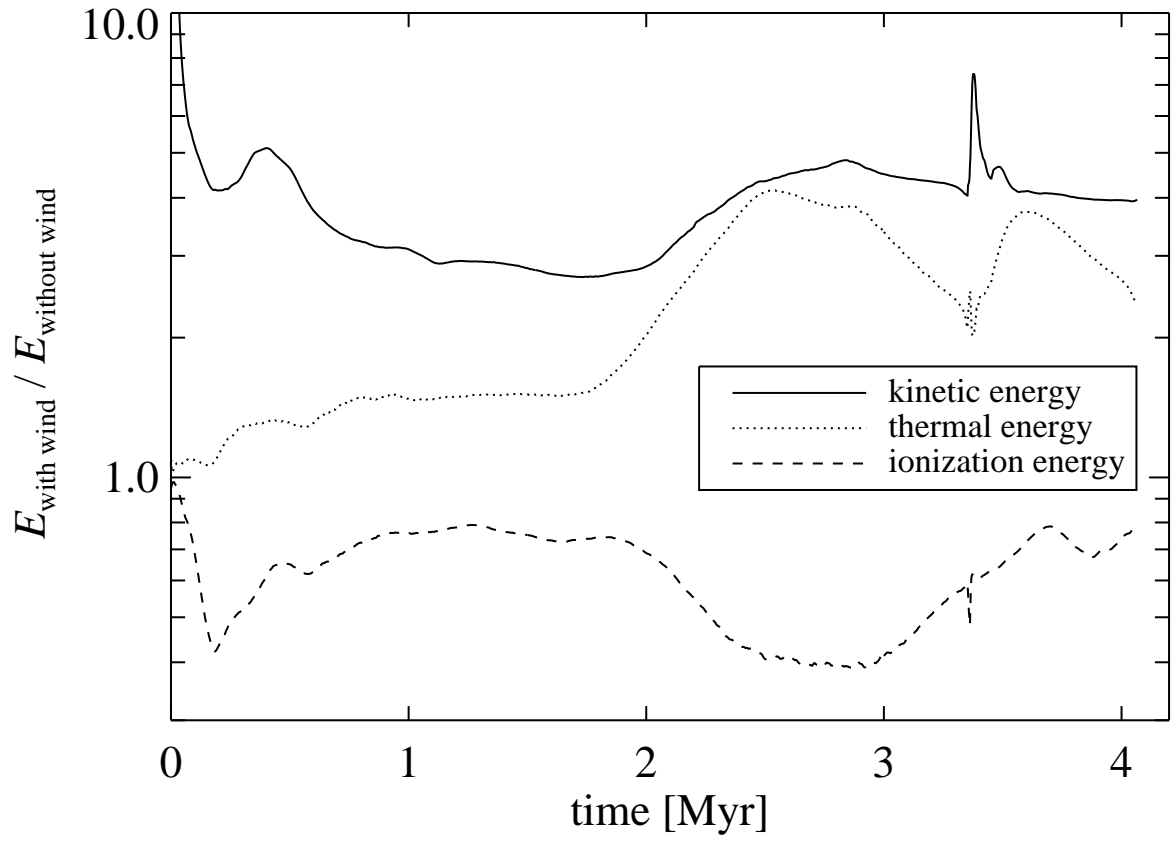}
  \caption{Ratio of energies in the 60\,$\Msun$ model with and without
           a stellar wind.
           \label{E_with_without_wind60_up.eps}
          }
\end{figure}

  As the star enters the first WN stage at $t \approx 1.7\,\mathrm{Myr}$, the
  photon luminosity in the Lyman continuum begins to decrease slightly while
  the mechanical wind luminosity starts to increase
  (see Fig.~\ref{60Msun_input4_up.eps}). The former directly reduces the rate
  of photons available for photoionization, whereas the latter enhances the
  pressure in the hot bubble, thus compressing the \HII region, which
  implies higher recombination rates. Thus, ionization energy and thermal
  energy of warm gas decrease by a factor of 2 and reach local minima at
  $t \approx 2.6\,\mathrm{Myr}$. They rise again when the photon luminosity in
  the Lyman continuum increases and the mechanical wind luminosity decreases
  until the star reaches the LBV stage.

  Though spectacular in appearance, the impact of the LBV phase on the energy
  balance is only limited due to a relatively brief period of time. We note
  here that the table of stellar parameters that we use is sufficiently
  time resolved even during the LBV phase, contrary to what could be suspected
  from Figure \ref{60Msun_input4_up.eps}. For a detailed plot of the stellar
  parameters during the LBV phase see \citeauthor{garcia96a}.
  The ionization energy in the circumstellar gas drops
  during the LBV stage, because the effective temperature of the star decreases
  significantly.  But immediately after the LBV eruption
  during the next WN stage the stellar UV flux drives the 
  circumstellar ionization energy to a global maximum at
  $3.4 \times 10^{50}\,\mathrm{ergs}$. The ionization energy follows
  the stellar UV flux. The final values of ionization energy and thermal
  energy of warm gas (i.e. at the end of our calculation)
  are $1.0 \times 10^{50}\,\mathrm{ergs}$ and
  $2.1 \times 10^{49}\,\mathrm{ergs}$, respectively.

  The kinetic energy of bulk motion in the circumstellar gas rises from zero
  at the beginning (because we started with a quiescent medium) as more
  and more circumstellar gas is accelerated by the expansion of the \HII region
  and the SWB. After some $0.3\,\mathrm{Myr}$ it reaches the same
  level as the thermal energy of warm gas and rises in the same manner. When
  the mechanical luminosity of the stellar wind increases in the first WR
  phase it lifts the kinetic energy above $10^{50}\,\mathrm{ergs}$, close
  to the ionization energy level.
  The ejection of the LBV nebula raises the kinetic energy for a short time
  almost by a factor of 2, but it drops back when the LBV material hits the
  outer shell and the kinetic energy is dissipated into thermal energy. 
  Its final value is $1.4 \times 10^{50}\,\mathrm{ergs}$.

  The thermal energy of hot gas accounts for the internal energy of the
  rarefied gas in the SWB that is heated to high temperatures by the
  reverse shock during the MS and WR phases of the star. Since there is
  no hot gas at the very beginning, the thermal energy of this gas phase rises
  from 0. The behavior of the curve is very similar to that of the kinetic
  energy but it increases faster during the first WN stage of the star. Between
  2.3 and 3.1\,Myr it is the dominant form of energy in our system and after
  some slight changes in and around the LBV phase it finally decays with the
  mechanical luminosity of the star to $9.7 \times 10^{49}\,\mathrm{ergs}$.

  The thermal energy of cold gas starts already with the internal energy of
  the quiescent ambient medium. The value increases smoothly from 
  $3.4 \times 10^{49}\,\mathrm{ergs}$ at the beginning to
  $7.9 \times 10^{49}\,\mathrm{ergs}$ at the end of the calculation, because
  more and more ambient gas becomes swept up by the outer shell. The weak
  shock heats it to several 100\,K whereas cooling of this neutral gas is very
  low.

  At the beginning the radiative energy input has the
  strongest influence on the energy distribution. The dynamical response of
  the circumstellar gas (the expansion of the \HII region and the acceleration
  of the material by the SWB) takes much longer. Thus, the
  ionization energy dominates the circumstellar energy distribution over the
  first $\approx 2\,\mathrm{Myr}$. After the kinetic energy and thermal
  energy of hot gas have caught up with the thermal energy of warm gas,
  these three
  forms of energy are comparable between 0.3 and 1.8\,Myr. Later, due to the
  accumulation of kinetic energy in the shell and thermal energy of hot gas in
  the bubble, these two become comparable with the ionization energy. Between
  2.2\,Myr and the end of the calculation these three forms of energy differ by
  not more than a factor of 3, though some changes occur. The LBV phase of the
  star induces very rapid changes in the relative energy ratios of at most a
  factor of 3 but they are transitory.

  It is interesting to compare the energy stored in the system for the two
  cases: 1) with wind and 2) the pure \HII region evolution without wind
  (Fig.~\ref{E_with_without_wind60_up.eps}). Obviously, the total kinetic
  energy in the calculation with wind is always higher than in
  the corresponding calculation without, because of the added kinetic energy
  of the SWB shell. At the
  end of the calculation, the kinetic energy in the model with stellar wind is
  4 times higher than in the model without.

  By contrast, the ionization energy reaches
  the same level of $\approx 1.1 \times 10^{50}\,\mathrm{ergs}$ very soon after
  the start of the calculation in both cases. As we have already seen in
  Figure \ref{e_distri227_up.eps}, the ionization energy in the model with wind
  drops by a factor of 2 during the first $0.2\,\mathrm{Myr}$ whereas
  in the model without wind it continues to grow.
  This supports our explanation that inhomogeneities in the \HII region
  (shell, spokes) result in shorter recombination times and thus lead to 
  fewer photoionized hydrogen atoms in the
  system. Although there is additional energy input by the stellar wind, the
  ionization energy is lower than in the calculation without wind by 0.1 to
  0.4\,dex for most of the evolution. This, however, does not imply that the
  $\Ha$ luminosity is different.  The $\Ha$ luminosity directly mirrors the
  stellar ionization input.

  Although the thermal energy of warm gas ($55\,\%$ below the respective value
  of the windless model at the end of the calculation) is similarly affected
  as the ionization energy, the total thermal energy is always higher in the
  model with wind, because we have additional energy deposited into the hot and
  into the cold gas phases.

\begin{figure}
  \plotone{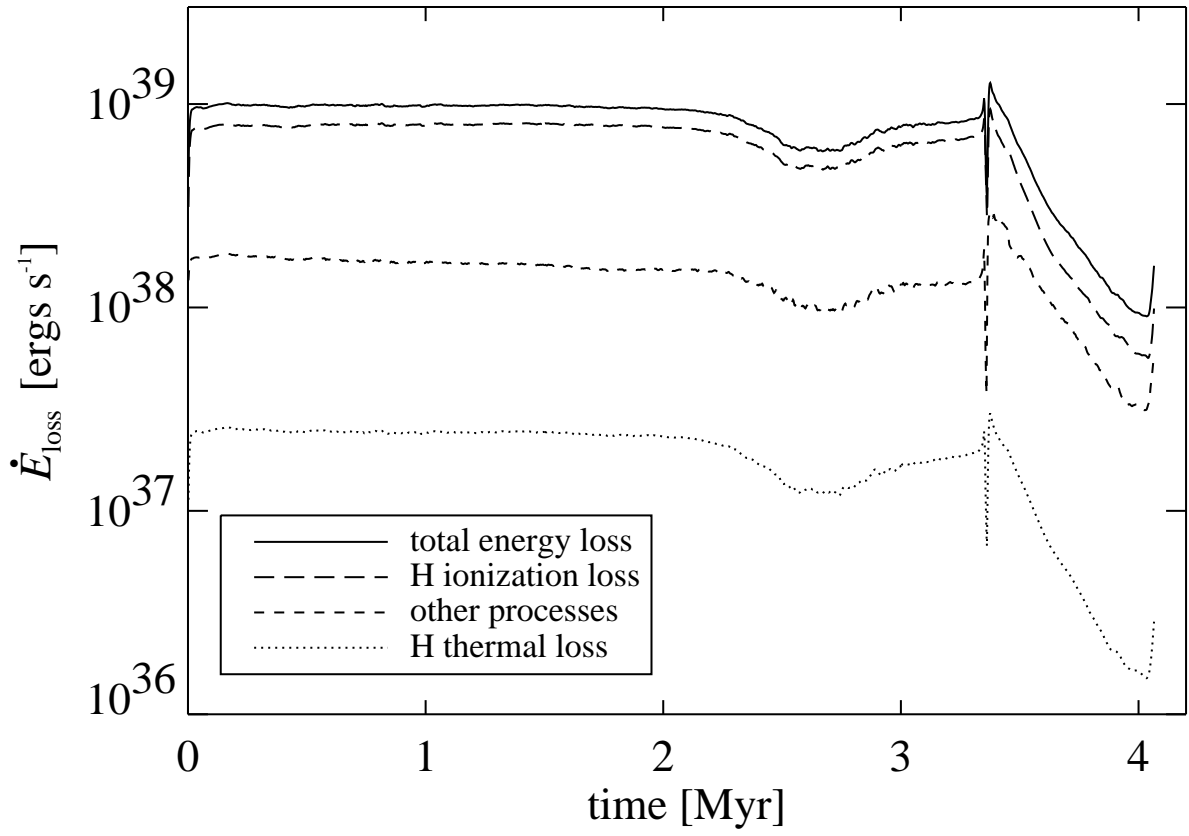}
  \caption{Energy loss due to cooling in the 60\,$\Msun$ model.
           \label{Lcool227_up.eps}
          }
\end{figure}

  Whereas the mechanical power of the stellar wind and the radiative power of
  the star are the energy sources in our system, the only energy sink
  term is radiative cooling of the gas. All other energetic processes lead to
  a redistribution of forms of energy within the system but not
  to a change of the total energy. Figure \ref{Lcool227_up.eps} shows the total
  energy loss of the system due to cooling of the gas and distinguishes 3
  different contributions: ``H ionization loss'' refers to
  the 13.6\,eV binding energy per hydrogen recombination into all states above
  the ground state, while ``H thermal loss''
  accounts for the thermal energy of the electrons lost by the same
  process. ``Other processes'' include all other cooling processes considered
  such as collisionally excited line emission and Bremsstrahlung.
  One can see from Figure \ref{Lcool227_up.eps} that the shape of all
  the curves follows the Lyman continuum photon luminosity of the star
  (see Fig.~\ref{60Msun_input4_up.eps}). The total energy-loss rate during the
  MS phase is close to $10^{39}\,\mathrm{ergs\,s^{-1}}$, approximately the same
  as the stellar Lyman continuum luminosity. Although the ionization
  energy in the system rises to $10^{50}\,\mathrm{ergs}$ during the first few
  10\,000\,yr, it is only a small fraction of the total input energy.

  ``H ionization loss'' is the dominant energy sink term during the 
  evolution. It accounts for roughly 80\,\% of the total emitted power during
  the MS and first WN stage of the star. After the LBV phase, the relative
  importance of ``other processes'' has risen because the LBV ejecta
  enhanced the density in the hot bubble where collisionally excited lines and
  Bremsstrahlung dominate the energy emission.

\begin{figure}
  \plotone{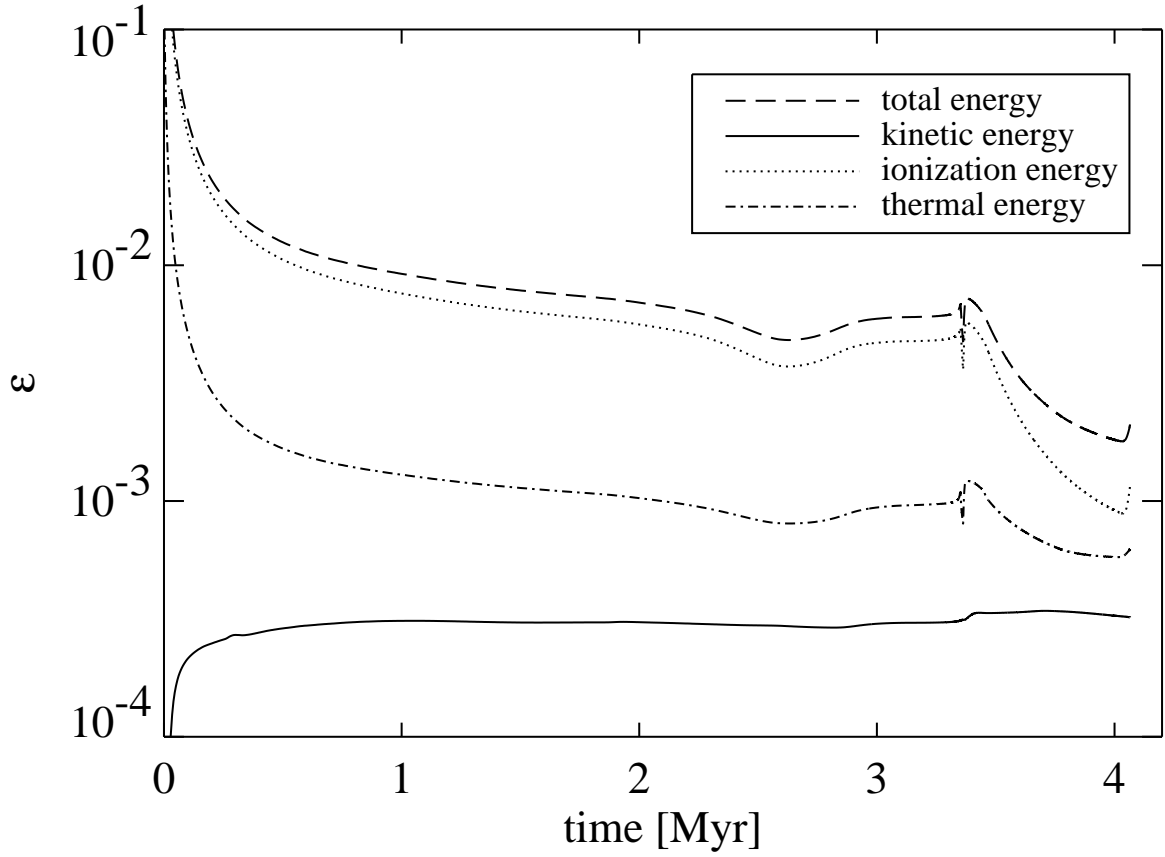}
  \caption{Energy transfer efficiency with respect to the total energy
           input in the 60\,$\Msun$ model without wind.
           \label{Ecompare242_up.eps}
          }
\end{figure}

  Figure \ref{Ecompare242_up.eps} shows the transfer
  efficiency into kinetic, ionization, and thermal energy and their sum
  for the model without wind. We define the transfer efficiency as the
  cumulative fraction of the input energy that has been converted into a
  particular form up to the time $t=\tau$. In this case we have only
  radiative input energy and all values are related to that.

  The transfer efficiency into ionization energy has the highest value
  during the whole calculation. The transfer efficiencies into ionization and
  thermal energy strongly drop at the beginning of the calculation. This is
  a consequence of the fact that the photoionization equilibrium has been
  quickly established and the ionization energy stays constant within an
  order of magnitude, whereas the transfer efficiency is
  related to the accumulated input energy that grows continuously in time.  
  The transfer efficiency into ionization energy reaches
  $\approx 5 \times 10^{-3}$ immediately before the LBV phase and drops due to
  the decreasing photon luminosity in the final WR phases by almost a factor
  of 5. The transfer efficiency into thermal energy reaches the level of
  $10^{-3}$ before the LBV phase, but decreases only by less than a factor
  of 2 during the remaining evolution. The discrepancy between the declines
  of both values can again be explained by the fact that cooling in formerly
  photoionized gas is greatly reduced due to missing electrons after
  recombination occurs.
  Thus, the recombined neutral gas has lost all its ionization energy but can
  retain a considerable portion of its thermal energy over a relatively long
  period.
  The transfer efficiency into kinetic energy has reached a nearly constant
  level of $\approx 3 \times 10^{-4}$ after 1\,Myr and remains at that level
  until the end of the calculation, i.e. after the first Myr the kinetic
  energy in the system grows almost proportionally to the total input energy.

\begin{figure}
  \plotone{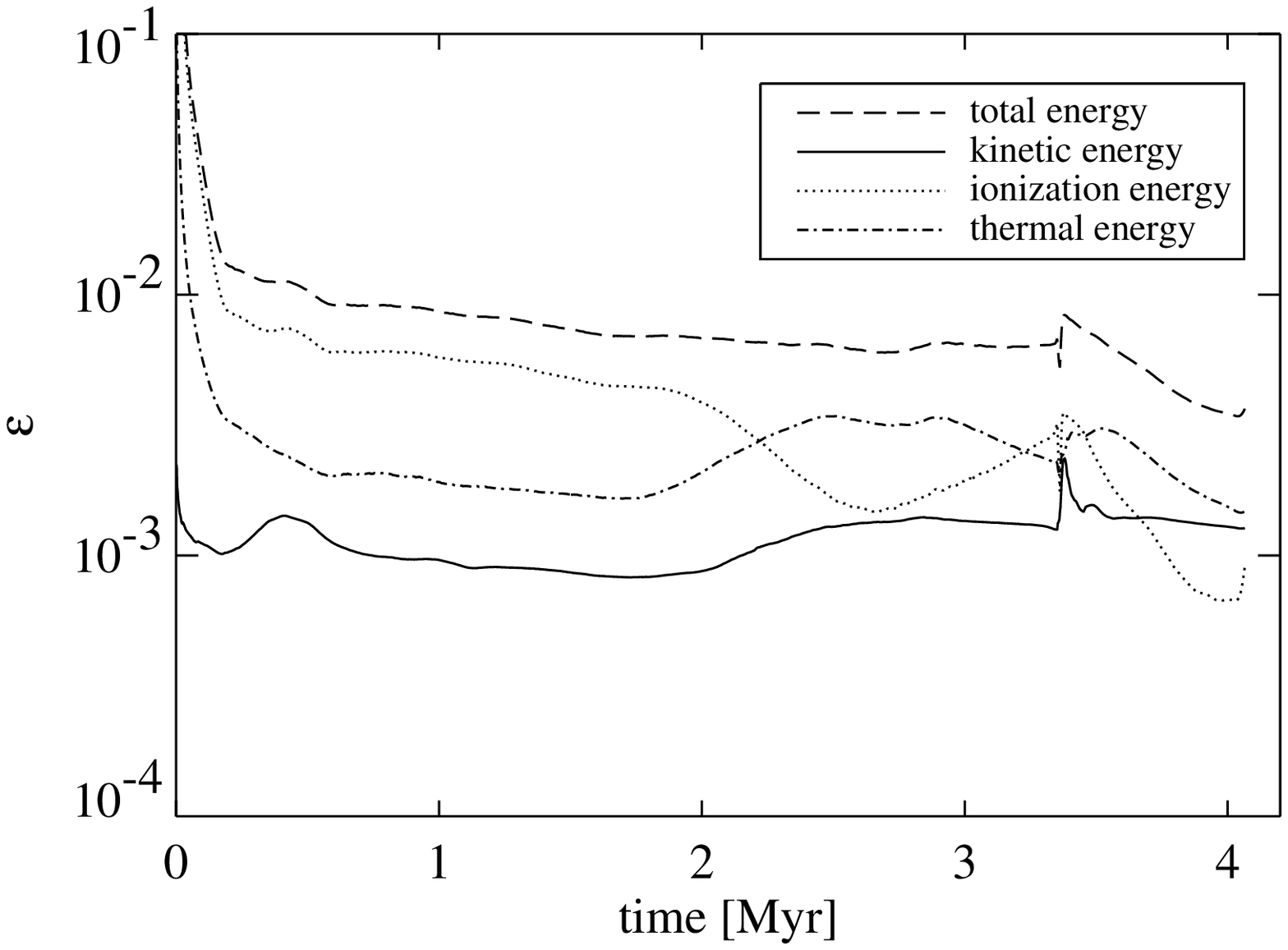}
  \caption{Energy transfer efficiency with respect to the total energy
           input in the 60\,$\Msun$ model.
           \label{Ecompare227_up.eps}
          }
\end{figure}

  Now we check the impact of the stellar wind on the energy transfer in the
  system. Figure \ref{Ecompare227_up.eps} shows the same transfer efficiencies
  as Figure \ref{Ecompare242_up.eps} but for the model with wind
  and radiation. Whereas for the model without wind the transfer efficiency
  into ionization energy dominated the total energy transfer efficiency,
  for the model with wind this is only true during the first
  $\approx 2.2\,\mathrm{Myr}$. In the first WN phase the transfer efficiency
  into thermal energy dominates for about 1\,Myr, because the photon luminosity
  in the Lyman continuum decreases and the mechanical wind luminosity
  increases, thus enhancing the production of hot gas. Curiously, in the LBV
  phase all three energy transfer efficiencies have the same value of about
  $2 \times 10^{-3}$. As in the case without wind, the transfer efficiency into
  ionization energy drops more strongly after the LBV phase than that into
  thermal energy. The drop of the latter one in this model
  is mostly due to energy loss of the hot gas 
  that dominates the thermal energy at late stages of the evolution. 

  The total energy transfer efficiency reaches approximately $4 \times 10^{-3}$
  at the end of the calculation, about twice as much as in the model
  without wind. This is interesting, because we know from Figure
  \ref{60Msun_input4_up.eps} that the mechanical wind luminosity integrated
  over the whole lifetime of the star is almost negligible compared to the
  total radiative energy input in the Lyman continuum. In other words,
  although the stellar wind itself does not actually inject a considerable
  amount of energy into the circumstellar gas (compared to the stellar
  radiation field), its presence almost doubles the total energy that is
  finally contained in the gas. This difference is due to the fact that the
  thermal energy deposited increased by a factor of $2.4$ and that the
  kinetic energy rose by a factor of 4 compared to the model without wind.
  The ionization energy actually decreased slightly by some $20-30\,\%$.

  To facilitate the discussion in the next section we summarize the values
  of the individual energy components at the end of the simulation in
  Table \ref{table_num_results}.
  Besides the kinetic energy of bulk motion ($E_{\mathrm{k}}$) and the
  ionization energy of hydrogen ($E_{\mathrm{i}}$) we list the
  thermal energy of the cold, warm, and hot gas components
  ($E_{\mathrm{t,cold}}, E_{\mathrm{t,warm}}$, and $E_{\mathrm{t,hot}}$)
  for the two models, one with and one without stellar wind.
  The values of the energy transfer efficiency into kinetic energy
  ($\varepsilon_{\mathrm{k}}$), ionization energy ($\varepsilon_{\mathrm{i}}$),
  and thermal energy ($\varepsilon_{\mathrm{t}}$) at the end of the simulation
  are given in Table \ref{table_num_eff}.
  The total energy of the Lyman continuum radiation emitted by the star is
  $E_{\mathrm{LyC}} = 1.07 \times 10^{53}\,\mathrm{ergs}$ and the mechanical
  energy injected into the system by the stellar wind, if considered,
  amounts to $E_{\mathrm{w}} = 3.31 \times 10^{51}\,\mathrm{ergs}$.

  \subsubsection{Comparison with analytical results}

  We now calculate the values for the kinetic-, ionization- and thermal energy
  in the system according to the analytical solutions given in
  section \ref{sec_theo_and_obs}. The analytical approach cannot handle
  time-dependent stellar parameters, therefore we simply choose mean
  values for effective temperature and luminosity in the Lyman continuum
  over the lifetime of our model star.
  With $\langle T_{\mathrm{eff}} \rangle = 5.03 \times 10^4\,\mathrm{K}$,
  $\langle L_{\mathrm{LyC}} \rangle =
  8.33 \times 10^{38}\,\mathrm{ergs\,s^{-1}}$, and
  using $\alpha_{\mathrm{B}} = 3.37\times 10^{-13}\,\mathrm{cm^3\,s^{-1}}$
  as hydrogen recombination coefficient and
  $c_{\mathrm{s,\scriptscriptstyle{II}}} =
  1.15\times 10^6\,\mathrm{cm\,s^{-1}}$
  for the isothermal sound speed in the \HII region (corresponding to
  $T_{\mathrm{\scriptscriptstyle{II}}} = 8000\,\mathrm{K}$), we obtain for
  $n_{\mathrm{0}} = 20\,\mathrm{cm^{-3}}$ after $\tau = 4.065\,\mathrm{Myr}$
  from equations (\ref{eq_E_k}), (\ref{eq_E_i}), and (\ref{eq_E_t}) for
  the model without wind:
  \begin{eqnarray}
    E_{\mathrm{k}} &=& 4.3 \times 10^{49}\,\mathrm{ergs}\ , \nonumber \\
    E_{\mathrm{i}} &=& 5.0 \times 10^{50}\,\mathrm{ergs}\ , \nonumber \\
    E_{\mathrm{t}} &=& 7.7 \times 10^{49}\,\mathrm{ergs}\ . \nonumber
  \end{eqnarray}

  The energy transfer efficiency in the analytical approach is then defined as
  \begin{equation}
    \varepsilon = \frac{E}{\tau\,\langle L_{\mathrm{LyC}} \rangle }\ ,
    \label{eq_eps_hii}
  \end{equation}
  where $E$ can be any of $E_{\mathrm{k}}$, $E_{\mathrm{i}}$, and
  $E_{\mathrm{t}}$, depending on the efficiency that is calculated.
  Thus, we obtain
  \begin{eqnarray}
    \varepsilon_{\mathrm{k}} &=& 4.0 \times 10^{-4}\ , \nonumber \\
    \varepsilon_{\mathrm{i}} &=& 4.7 \times 10^{-3}\ , \nonumber \\
    \varepsilon_{\mathrm{t}} &=& 7.2 \times 10^{-4}\ . \nonumber
  \end{eqnarray}

  Bearing in mind all the assumptions and approximations which have been made
  to obtain the analytical expression for the kinetic energy of bulk motion
  \citep[see][]{lasker67}, the deviation of less than 30\,\% in
  $\varepsilon_{\mathrm{k}}$ is not too bad. This difference is not due to the
  constant effective temperature and luminosity that we have chosen to obtain
  the analytical result. Recalculating $\varepsilon_{\mathrm{k}}$ for the
  first Myr, where the stellar parameters are about constant, makes the
  discrepancy between the analytical and numerical value of
  $\varepsilon_{\mathrm{k}}$ even larger. The difference is more likely
  due to temperature deviations from $8000\,\mathrm{K}$ in the \HII region of
  the model calculation.

  Comparing the analytical and numerical results for the transfer efficiency
  into ionization energy shows fairly good correspondence immediately before
  the LBV phase. At the end of the calculation the numerically determined
  transfer efficiency is about 0.6\,dex below the analytical value because
  the Lyman continuum luminosity of the star drops significantly during
  the final WR phases of the star and the ionization energy follows
  immediately. Of course, the analytical solution cannot reproduce this
  feature. The correlation between the drop of Lyman continuum luminosity
  and the thermal energy is weaker, because cooling ceases when the plasma
  becomes neutral, as already mentioned above. Thus, the correspondence
  between analytical and numerical transfer efficiency into thermal energy is
  much better and the deviation at the end of the calculation is less than
  15\,\%.

  It is more difficult to compare analytical with numerical results in the
  case of the combined \HII region / SWB model because we have only the
  analytical energy transfer efficiency solutions for the \HII region and SWB
  separately. We simply add up the energy contributions from the \HII region
  and the SWB bearing in mind that this is only a rough approximation, which
  actually neglects the interactions between both structures. Because we have
  not considered cooling in the hot bubble, the analytical energy transfer
  rates into kinetic and thermal energy are upper limits.

  We insert the mean value of the mechanical wind luminosity, 
  $\langle L_{\mathrm{w}} \rangle =
  2.58 \times 10^{37}\,\mathrm{ergs\,s^{-1}}$,
  into equations (\ref{eq_bubble_E_k}) and (\ref{eq_bubble_E_t}) and
  together with the results for the pure \HII region we obtain
  \begin{eqnarray}
    E_{\mathrm{k}} &=& 9.5 \times 10^{50}\,\mathrm{ergs}\ , \nonumber \\
    E_{\mathrm{i}} &=& 5.0 \times 10^{50}\,\mathrm{ergs}\ , \nonumber \\
    E_{\mathrm{t}} &=& 1.6 \times 10^{51}\,\mathrm{ergs}\ . \nonumber
  \end{eqnarray}
  To find the energy transfer efficiency, we relate these values to the sum
  of Lyman continuum radiation energy and the mechanical wind energy
  (which, of course, is almost negligible), 
  \begin{equation}
    \varepsilon = 
      \frac{E}{\tau\,\left( \langle L_{\mathrm{LyC}} \rangle +
      \langle L_{\mathrm{w}} \rangle \right)}\ ,
    \label{eq_eps_tot}
  \end{equation}
  where $E$ can again be any of $E_{\mathrm{k}}$, $E_{\mathrm{i}}$, and
  $E_{\mathrm{t}}$. We get
  \begin{eqnarray}
    \varepsilon_{\mathrm{k}} &=& 8.6 \times 10^{-3}\ , \nonumber \\
    \varepsilon_{\mathrm{i}} &=& 4.6 \times 10^{-3}\ , \nonumber \\
    \varepsilon_{\mathrm{t}} &=& 1.4 \times 10^{-2}\ . \nonumber
  \end{eqnarray}
  We have already discussed the fact that the ionization energy in the
  numerical model with wind is lower than in the model without
  wind. This slightly enhances the discrepancy between the analytical
  and the numerical results for the ionization energy in the case with
  stellar wind. On the other hand, there is no indication that the 
  neglect of the ionization energy in the hot bubble is a bad approximation.

  The increase of the thermal energy deposit from the windless \HII region
  model to the combined SWB/\HII region model (factor 2.4) is almost an
  order of magnitude below the analytical upper limit. Since the conversion
  of thermal energy by $P\mathrm{d}V$ work is the major source of kinetic
  energy, the increase of the kinetic energy deposit between the models
  without and with wind (factor of 4 at the end of the calculation)
  is also $82\,\%$ below the analytical upper limit.
  These findings are supported by the observation
  that the radius of the bubble at the end of the
  calculation $(\approx 50\,\mathrm{pc})$ is considerably smaller than the
  analytical result of $\approx 68\,\mathrm{pc}$ according to
  equation (\ref{eq_bubble_ph2_R2}). One reason for the lack of thermal energy
  in the numerical model is the variation of the mechanical luminosity of
  the star. The LBV wind enhances the density in the bubble which leads to
  stronger cooling, and the mechanical luminosity of the stellar wind
  generally
  decreases during the last $0.6\,\mathrm{Myr}$. But also during
  the other stages of evolution, cooling in the hot bubble is not
  completely negligible for the energetics of the system. 
  The thermal energy of the hot bubble calculated analytically at
  $t = 1\,\mathrm{Myr}$ is already a factor of 4.5 larger than the
  corresponding value in the numerical model. Although we have not implemented
  heat conduction, effects like turbulent mixing between hot and cold gas as
  well as numerical diffusion enhance the cooling rate.
  The resolution dependent numerical diffusion is not a key factor as we have
  shown in our resolution study. However, the cooling due to gas mixing is
  enhanced because the SWB expands at the beginning into the highly
  nonuniform \HII region, a fact that cannot be properly handled by
  the analytical description. In general, the analytical
  results for the upper limits of the transfer efficiency into thermal and thus
  kinetic energy of bulk motion are much higher than the values from the
  numerical simulations even without heat conduction.

  \subsection{Direct observational implications}
  \label{subsec_obs_imp}
  Finally, we briefly discuss some direct observational implications of
  our models. Since for a correct construction of intensity maps one would
  need 3D models, the results of our 2D calculations can only give a rough
  estimate of the observable intensities. Thus, we have calculated
  angle-averaged intensity profiles for $\Ha$ and soft X-ray emission based
  on the cylindrical symmetry in our models. 
 
  The $\Ha$ emissivity is calculated according to the table (case B) in
  \citet{osterbrock89}. We use a $T_{\mathrm{\scriptscriptstyle{II}}}^{-0.92}$
  fit for the temperature dependence and set the emissivity to zero where the
  degree of hydrogen ionization is below $10^{-4}$.
  We do not account for absorption.

\begin{figure}
  \plotone{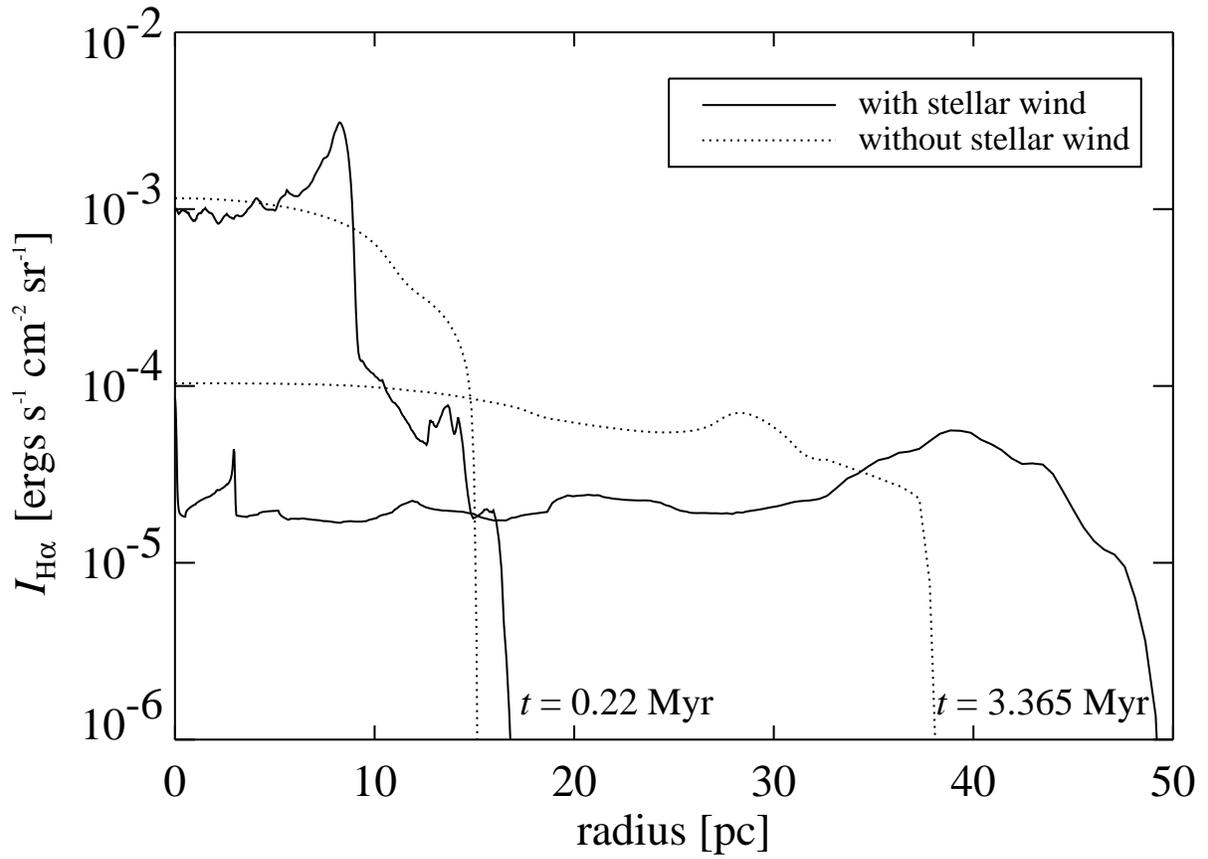}
  \caption{Angle-averaged $\Ha$ intensity for the 60\,$\Msun$ model
           with and without a stellar wind are compared at two
           evolutionary times.
           \label{IHa3_ng_plot_up.eps}
          }
\end{figure}

  In Figure \ref{IHa3_ng_plot_up.eps} we plot the angle-averaged
  $\Ha$ intensity profiles at two
  different times ($t = 0.22\,\mathrm{Myr}$, during which intense structure
  formation occurs in the \HII region, and $t = 3.365\,\mathrm{Myr}$,
  at a late stage after the LBV phase of the star). For comparison,
  we have also plotted the $\Ha$ intensity profiles of the corresponding
  models without stellar wind for the same times.

  At $t = 0.22\,\mathrm{Myr}$ the pure \HII region model without stellar wind
  shows the intensity profile of a spherical, emitting volume that has just
  started to expand. On the other hand, the intensity profile of the model
  with stellar wind shows a dominant peak between $r = 8$ and $9\,\mathrm{pc}$.
  This peak represents the global intensity maximum at that time and is
  produced by the emission of the dense shell swept up by the hot bubble of
  shocked stellar wind gas. A secondary, much smaller peak between $r = 13$ and
  $14\,\mathrm{pc}$ originates from the trapping of the ionization front in
  the dense outer shell fragments piled up by the expanding \HII region.
  
  Comparing these data with the intensity profiles at $t = 3.365\,\mathrm{Myr}$
  shows that the further expansion of the bubble and \HII region generally
  lowers the $\Ha$ surface brightness, as expected for comparable Lyman
  continuum fluxes. In particular, when the SWB completely
  overtakes the \HII region and enlarges the whole structure, the $\Ha$ surface
  brightness is lower than in the pure \HII region model at the same time.

  Due to the facts that the photoionized region is more or less limited to the
  illuminated inner part of the shell and that the $\Ha$ emissivity in the hot
  cavity is very low, the intensity profile of the combined
  SWB / \HII region appears slightly limb brightened.
  The $\Ha$ emission from the WR bubble is barely visible above the
  $\Ha$ background from the MS bubble. It might be that in our simulation
  the shell of the WR bubble is too hot (e.g. due to the neglect of heat
  conduction) and thus emits more strongly in X-rays (see below) than in
  the optical.

\begin{figure}
  \plotone{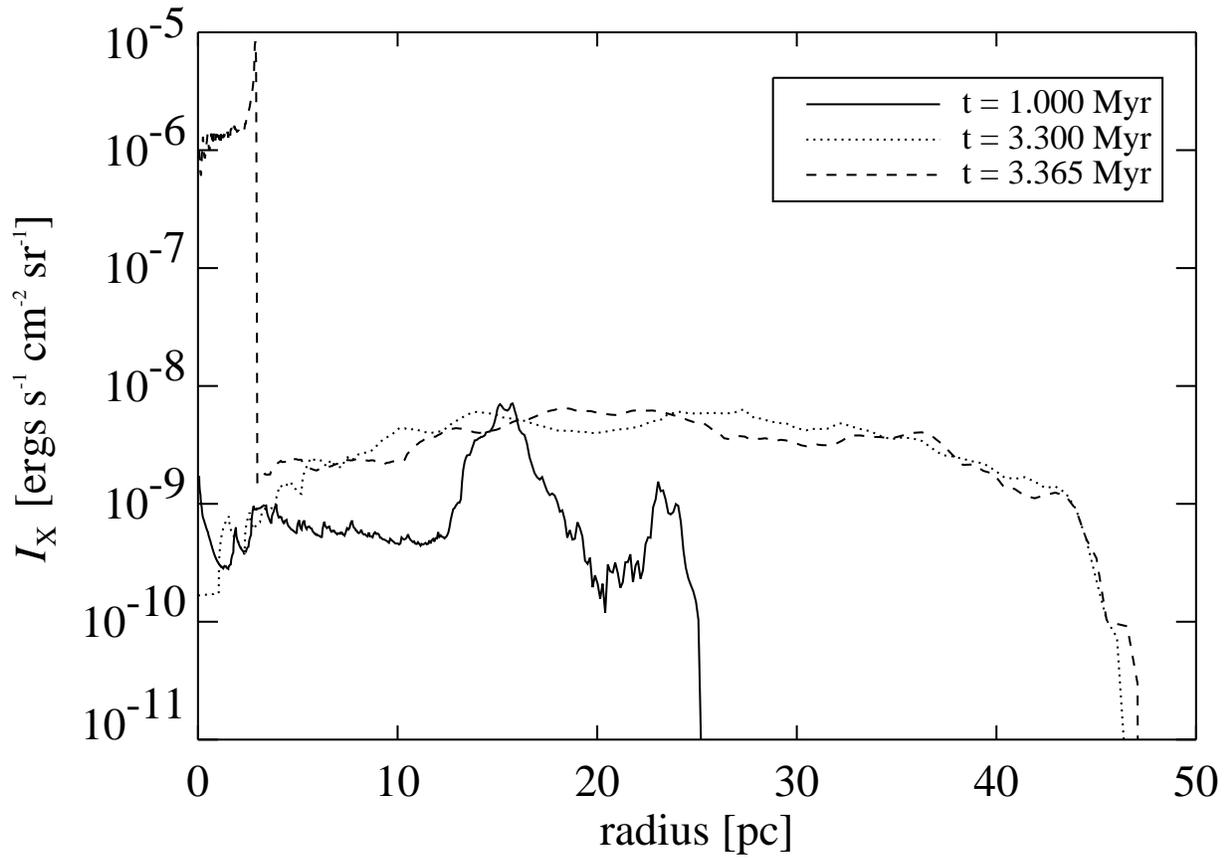}
  \caption{Angle-averaged soft (0.5 \ldots~3.0\,keV) X-ray intensity 
           for the 60\,$\Msun$ model with stellar wind at selected
           evolutionary times.
           \label{IX3_ng_500_3000ev_plot_up.eps}
          }
\end{figure}

  For the X-ray intensity profiles in the energy range from 0.5\,keV to
  3.0\,keV we use the emissivity calculated with the \citet{raymond77}
  program for cosmic chemical composition
  \citep{allen73}. The emissivity is set to zero for temperatures below
  $10^5\,\mathrm{K}$; absorption is not considered.
  In Figure \ref{IX3_ng_500_3000ev_plot_up.eps} we display the angle-averaged
  soft X-ray intensity profiles at $t = 1.0\,\mathrm{Myr}, 3.30\,\mathrm{Myr}$
  (before the LBV-phase), and $3.365\,\mathrm{Myr}$ (after the LBV-phase).
  The two snapshots before the LBV stage show intensities between
  $10^{-10}$ and $10^{-8}\,\mathrm{ergs\,s^{-1}\,cm^{-2}\,sr^{-1}}$ for
  lines of sight through the hot bubble. After the LBV stage, the soft X-ray
  intensity close to the center of the bubble is strongly enhanced
  by more than three orders of magnitude to values above
  $10^{-6}\,\mathrm{ergs\,s^{-1}\,cm^{-2}\,sr^{-1}}$ when the shocked WR wind
  hits the LBV ejecta. The soft X-ray emission comes mainly from the inside
  of the shell of the swept-up LBV wind, before it breaks apart.

  The WR nebula RCW 58 around HD 96548 (= WR 40) is a possible candidate
  for an observed counterpart of the WR nebula that forms in our model
  calculation at $t = 3.365\,\mathrm{Myr}.$ The WR star is currently of
  type WN8 and there are indications that the star passed the LBV stage
  (\citeauthor{garcia96a}; \citealt{humphreys91}). The mean radius of the
  optical nebula is 3.5\,pc \citep{chu82} or 2.5\,pc \citep*{arthur96}
  depending on the estimation of the distance to HD 96548.
  Thus, the evolutionary state of the nebula should be approximately
  comparable to the nebula in our model.

  Though the strong increase of the surface brightness in our model
  after the LBV phase is basically in agreement with the fact that
  up to now only WR bubbles (and no MS bubbles) have been observed in X-rays,
  the comparison with RCW 58 shows that our model also suffers from the
  same problem that all analytical and numerical models thus far have:
  The X-ray luminosity is much higher than observed. In our model,
  the X-ray luminosity of the WR bubble in the energy range from
  0.5\,keV to 3.0\,keV is $7 \times 10^{33}\,\mathrm{ergs\,s^{-1}}$ whereas
  \citet{moffat82} report that no X-ray emission from
  RCW 58 was detected with the EINSTEIN satellite, meaning the X-ray
  luminosity of RCW 58 in this energy range
  must be below $10^{32}\,\mathrm{ergs\,s^{-1}}$
  \citep{arthur96}.

  \subsection{The impact of the ambient medium, the stellar parameters,
              and unconsidered physics}
  \label{subsec_impact_changes}

  Due to the immense computational efforts we can only show results for
  one set of ambient medium parameters ($n_{\mathrm{0}} = 20\,\mathrm{cm^{-3}}$
  and $T_{\mathrm{0}} = 200\,\mathrm{K}$).  We shall utilize the scaling
  laws for \HII regions in ionization/recombination and heating/cooling
  equilibrium with negligible gravity \citep[see table 2 of][]{yorke82}
  and briefly discuss the
  changes of the results that we expect if other values for the ambient medium
  are used.  Separating the evolution of the \HII region and SWB, we note
  that if the density in the ambient medium is modified
  by a factor $\tilde\delta$ while assuming the same equilibrium temperature
  and same ionizing luminosity, all linear scales $\tilde\lambda$
  and time scales $\tilde\tau$ transform according to
  $\tilde\lambda \propto \tilde\tau \propto \tilde\delta^{-2/3}$, velocities
  remain unchanged, and mass outflow rates and mechanical luminosity (as well
  as other rates of energy changes) transform as $\tilde\delta^{-1/3}$.
  Values for the various energies given at a particular time
  scale as $\tilde\delta^{-1}$. Thus, an increase in ambient density results
  in smaller \HII regions, lower energy injection rates and proportionally
  lower values for the ionization, thermal, and kinetic energy.  These same
  scaling laws can be derived in a less rigorous manner from our equations
  \ref{eq_r_i} to \ref{eq_E_t}.

  Modifying the temperature of the ambient medium does not strongly affect
  the evolution as long as the gas is not ionized and the pressure in the
  \HII region is much higher than the pressure of the ambient medium.

  With respect to the SWB, equation \ref{eq_bubble_ph2_R2} shows that the
  radius of the SWB is proportional to $\rho_{0}^{-1/5}$, i.e. at comparable
  times it is smaller (larger) by $\tilde\delta^{1/5}$ in the case of an
  ambient density higher (lower) by a factor $\tilde\delta$.
  In contrast to the \HII region, the growth of the thermal and kinetic
  energy of the SWB does not depend on the ambient density, as long as the
  assumptions under that equations \ref{eq_bubble_E_k} and \ref{eq_bubble_E_t}
  were derived remain fulfilled (e.g. thermal pressure in the hot bubble is
  much higher than the thermal pressure in ambient medium).

  It is difficult to estimate the impact of a variation of the ambient
  conditions on the highly non-linear \HII region / SWB interaction effects.
  Basically, an increase of the density in the ambient medium leads to both
  a smaller SWB and a thicker shell of swept up material.
  Thus, the ratio of shell thickness to SWB diameter grows which increases
  the unstable length scales and
  tends to inhibit thin shell instabilities with all their associated effects.
  On the other hand, clumpiness in the circumstellar medium may exist anyway,
  abolishing the need to trigger the interaction effects by shell
  instabilities. It is clear that numerical simulations under a variety of
  different ambient conditions are highly desirable in order to study these
  effects in more detail. They will become available with increasing
  computer power.

  It should also be noted here that the stellar parameters which we use as
  time-dependent boundary conditions in our simulations are still somewhat
  uncertain. E.g. \citet*{martins02} showed that the determination of the
  effective temperature of O dwarfs based on non-LTE line blanketed atmosphere
  models including stellar winds results in differences to previous
  calibrations up to a few 1000\,K. The mass-loss rate and its temporal
  variation in the short and violent evolutionary stages like the LBV phase
  are probably even less well known. We have chosen the set of stellar
  parameters from \citeauthor{garcia96a} because it is current state-of-the-art,
  sufficiently time-resolved, and we wanted to directly compare our 2D
  simulations to their 1D calculations. 

  Nevertheless, deviations of the stellar parameters from this set will
  certainly influence the results of the model calculations. Returning to
  the scaling laws discussed by \citet{yorke82} and scaling the
  hydrogen-ionizing photon luminosity by $\tilde l$ but keeping density and
  temperature constant, we find that lengths and times scale according
  to $\tilde\lambda \propto \tilde\tau \propto \tilde l^{1/3}$, velocities
  remain unchanged, rates of energy changes scale like $\tilde l^{2/3}$,
  and values for energy injected into the ISM at a given time like $\tilde l$.
  In a similar manner, a higher stellar wind luminosity will increase the size
  of the SWB and the energy transferred to the shell at a given time.
  Note that although the
  analytical description of the SWB evolution depends on the stellar wind
  luminosity, i.e. only on the product of mass-loss rate and wind
  terminal velocity squared, this does not account for cooling
  in the hot bubble which will be enhanced when the mass-loss rate rises
  without a proportional increase in wind velocity.

  The impact of changes in the temporal evolution of the stellar parameters
  (e.g. a different number of outbursts of varying length) on the results of
  our calculations are difficult to estimate. Highly non-linear effects appear
  when e.g. a slow shell is overtaken by a fast wind and instabilities arise.
  Interaction processes between the SWB and the \HII region may also be
  amplified or damped by variations of the stellar parameters. Numerical
  simulations for different stellar evolutionary histories are necessary to
  study these effects in detail.

  Some physical processes are not yet considered in our models and we want to
  briefly discuss their possible influence here. One of these processes is
  heat conduction which may play a role for the energy transfer in the system.
  Though classical theory as well as our numerical models without
  heat conduction predict that the cooling time for the interior of the
  MS bubble is longer than the lifetime of the star, there is indirect
  evidence that, at least in NGC6888, the MS bubble is cold at the time when
  the WR wind blows out into the MS bubble \citep{maclow00}. 
  A possible explanation is mass loading of the MS bubble: Heat conduction
  evaporates cold material from the shell or from immersed clouds which
  enhances the density in the bubble. The higher density reduces
  the cooling time of the bubble. Such evaporation from a cloud embedded in
  hot interstellar gas has recently been observed \citep{chu00}.    
  We will discuss the case of NGC6888 in a forthcoming paper that describes
  numerical simulations of the circumstellar medium around a 35\,$\Msun$ star,
  a model which is more appropriate for NGC6888.

  On the other hand, magnetic fields provide an efficient mechanism to reduce
  the efficiency of heat conduction. In a plasma the electrons are constrained
  to move along
  magnetic field lines. If the magnetic field lines are aligned with the shell
  or if they are tangled, the effective path lengths for fast electrons
  moving from the hot to the cold gas are much longer and heat conduction by
  electrons is less efficient.
  Saturation effects of heat conduction due to electric fields 
  arising from charge separation further complicates the situation.
  All these effects are worthy to be studied in detail, but the simultaneous
  inclusion into multidimensional hydrodynamical simulations remains a
  computational challenge for the future.

  We only allowed for isotropic winds in our models. However, line-driving
  theory for rotating stars suggests that mass outflow from the poles should
  be both more vigorous and faster. As already mentioned in section
  \ref{sec_numerics_so_far}, \citet{brighenti97} and \citet{frank98}
  showed that anisotropic winds can indeed influence the formation and
  appearance of circumstellar shells, producing lobes or bipolar outflows.
  Similar effects may be expected for our models when anisotropic winds are
  employed, but details remain to be explored.

  Other physical effects not yet considered in our models include
  variations of the metallicity in the circumstellar gas and the
  photodissociation of molecular hydrogen. In its later stages, the star
  is expected to eject gas that is enriched with metals. This enhancement
  of the metallicity in the circumstellar medium leads to stronger cooling
  and thus influences the circumstellar evolution.
  The dissociation of molecular hydrogen by stellar photons
  with energies below the Lyman threshold adds additional (exterior) layers
  to the basic structure shown in Figure \ref{bubble_structure_new_up.eps}. 
  The \HII region evolves into the photodissociated region which itself
  expands into the molecular gas. Interaction processes between
  the photodissociated region and the \HII region similar to
  those between the \HII region and the SWB may arise.

  Finally, large scale density gradients such as those encountered at the
  edge of the clouds will give rise to a variety of additional dynamical
  effects, e.g. champagne flows of \HII regions and break-out of hot gas as
  well as the interactions between these two phenomena.

\section{Summary and conclusions}
\label{sec_conclusions}

  Our numerical models for the evolution of circumstellar gas around a 
  60\,$\Msun$ star show that the interaction of the stellar wind bubble
  with the stellar radiation field can strongly influence the morphology of
  the circumstellar medium. The rearrangement of circumstellar gas by the
  stellar wind influences the way it reacts to the ionizing radiation.
  Density clumps formed in the shell of gas swept up by the stellar wind
  bubble cast shadows into the \HII region. The resulting pressure gradients
  force material into the shadowed regions enhancing their density and forming
  neutral ``spokes'', which subsequently raise the mass of the clumps in the
  shell. The \HII region is extended in directions free of clumps.
  Thus, the formation of these elongated \HII region ``fingers'' in our model
  occurs not only due to the ionization front instability described by
  \citet{garcia96c}, but is triggered and amplified by the redistribution of
  mass by the action of the stellar wind shell.

  These results also shed light on the open question whether the complex
  structures that can be found in \HII regions are primordial, i.e. relics
  from the time before the gas became ionized, or formed by dynamical
  processes in the course of the \HII region evolution. While there are strong
  observational indications that the former plays an important role, our
  results give support to the idea that the latter cannot be completely
  excluded: Intense structure formation in \HII regions with strong stellar
  winds can occur even if the neutral ambient medium was initially
  homogeneous.

  Nevertheless, these structures are only a temporary phenomenon because the
  extended \HII region is eventually swept up by the stellar wind shell.
  If we compare our results at $t = 3.30\,\mathrm{Myr}$ with the
  1D solution of \citeauthor{garcia96a}, we see that, from a
  morphological point of view, there is basically little difference in
  the overall structure except for the appearance of an \HII region at the
  inner part of the remnant shell and a few filaments in the hot bubble. 
  Thus, the approach of \citeauthor{garcia96a} to use the 1D solution for the
  estimation of the initial conditions for closer studies of the LBV and
  subsequent WR stage appears to be valid.

  The structure formation induced by the interaction of the stellar radiation
  field with the SWB also has implications for the energy
  balance of the circumstellar gas, mostly via the effect that denser gas has
  a shorter recombination time and a higher cooling efficiency.
  This can be seen in the decrease of ionization and thermal
  energy of warm photoionized gas in the circumstellar medium during the
  time when the structure formation occurs in the first few
  $10^5\,\mathrm{yr}$. Furthermore, the ionization energy
  lags behind the corresponding value in the model without wind by
  $0.1 - 0.4\,\mathrm{dex}$ and the thermal energy of warm gas by
  $0.2 - 0.6\,\mathrm{dex}$ for almost the entire lifetime of the star.
  This also implies that the \HII regions around stars with winds can have
  a higher emission measure than undisturbed ones. 

  The ionization energy and the thermal energy of warm gas in the
  circumstellar medium are lower compared to the models without wind. Though
  the total mechanical wind energy is negligible compared to the accumulated
  energy of the Lyman continuum photons, the simulation with a
  stellar wind results in a kinetic energy in the circumstellar medium which
  is 4 times higher than in the model without wind. The total thermal energy
  is almost twice as high
  (or, subtracting the initial thermal energy of the background gas, the
  thermal energy that is added to the system is enhanced by a factor of 2.4).
  The energy transfer efficiency of the stellar Lyman continuum radiation
  over long time scales 
  is so low because this radiative energy is mostly used to maintain the
  photoionization of hydrogen; it is lost from the system when the hydrogen
  recombines into levels above the ground state.  By contrast, most of the
  energy of the stellar wind is converted into thermal
  energy of hot gas that accelerates the shell. The kinetic energy can be
  accumulated in the system for a long time unless it is dissipated, and the
  thermal energy of hot gas can also be saved if the density in the bubble is
  sufficiently low. For a plasma of solar chemical composition with
  $T = 10^7\,\mathrm{K}$ and $n = 10^{-2}\,\mathrm{cm^{-3}}$, the cooling time
  in collisional ionization equilibrium is several $10^7\,\mathrm{yr}$.
  Although the LBV phase of the star induces very rapid changes in the
  energy balance of the circumstellar medium, its impact over longer
  time scales is limited due to the brevity of the LBV eruption.

  The above conclusions regarding the inefficiency of converting the energy
  flux of photoionizing UV photons into kinetic energy are modified in the
  presence of large-scale density gradients (i.e. at the edge of a
  molecular cloud).  Generally speaking, the resulting ``champagne flow'' can
  lead to an efficiency of about 1\,\% for converting the stellar UV flux into
  kinetic energy of expansion \citep[see][]{yorke86}.  The exact value
  for efficiency depends on the details of the problem, however; it is
  conceivable in some cases that the champagne phase is negligibly short
  compared to the lifetime of the star.
  We note that simulations which assess the role of stellar winds in the
  presence of champagne flows are currently rare. \citet{comeron97} examined
  the evolution of wind-driven \HII regions across a density discontinuity
  for various parameters but without the consideration of energy transfer
  efficiencies.

  For the simulations of HII region evolution without a stellar wind the
  analytical prediction for the kinetic energy in the system at the end of the
  star's lifetime differs from our numerical result by less than 30\,\%.
  The analytical solution for the energy transfer efficiency in the case of
  the combined \HII region / SWB is just an upper limit and overestimates the
  transfer efficiencies into kinetic energy and thermal energy by factors of
  6.7 and 9.7, respectively. 

  This detailed examination is suitable to improve studies of the energization
  of the ISM in the solar neighborhood \citep[e.g.][]{abbott82}, which use
  energy transfer efficiencies as theoretical input. It is also important
  when considering the heating of the galactic disk as a global phenomenon
  self-regulated by star formation. To address this the effects of
  overlapping \HII regions and SWBs in OB clusters and associations still
  need to be assessed, the subject of a future investigation.

\acknowledgments

  We thank Sabine Richling, Don Cox, Mordecai-Mark Mac Low, Jos\'e Franco,
  Ralf-J\"urgen Dettmar, and Matthias Wrigge for helpful comments and
  interesting discussions. This work was supported by the Deutsche
  Forschungsgemeinschaft (DFG) under grant number He~1487/17, by the
  Graduiertenkolleg GRK 118 in Bochum, and by the
  National Aeronautics and Space Administration (NASA) under grant
  NRA-99-01-ATP-065.  Part of the research described in this paper was
  conducted at the Jet Propulsion Laboratory (JPL), California Institute 
  of Technology. TF also gratefully
  acknowledges financial support by the German Academic Exchange Service
  (Doktorandenstipendium HSP\,\,II\,/ AUFE) and the hospitality of the
  Department of Physics at the University of Wisconsin--Madison during a
  research visit. The computations were performed at the Rechenzentrum der
  Universit\"at Kiel, the Konrad-Zuse-Zentrum f\"ur Informationstechnik in
  Berlin, and the John von Neumann-Institut f\"ur Computing in J\"ulich.
  We also wish to thank an anonymous referee for valuable comments.




\clearpage


\clearpage
\clearpage
\clearpage

\bf

\rm

\clearpage
%
%






\begin{table}
\begin{center}
\footnotesize
\caption{Hydrodynamical models of SWBs / \HII regions around single
         early-type stars. \label{table_num_models}}
\begin{tabular}{ccccc}
\tableline\tableline
Ref. & Dimensions & Geometry & Linear size & Resolution\tablenotemark{a} \\
     &            &          & (pc)        & $(10^{-3})$                 \\
\tableline
1 & 2 & cylindrical & $\mathrm{7.8\ldots(3.2\times10^{-3})}$ & 5 \\
2 & 2 & spherical & 0.1 & 5 \\
3 & 1 / 2 & spherical & $\mathrm{70\ldots45~/~19\ldots3}$ &
 $2.5~/~2.5\ldots1.25$ \\
4 & 2 & spherical & $\mathrm{0.1\ldots2}$ & 5 \\
5 & 2 & spherical & 6.6 / 1 & 1.3 \\
6 & 2 & cylindrical & 0.2 & 3.9 \\
7 & 2 & cylindrical & 6.5 & 2.5 \\
8 & 2 & cylindrical & 64 & $8.0\ldots0.125$ \\
\tableline\tableline
Ref. & Duration & Ionization & Heating & Cooling \\
     & (Myr)    &            &         &         \\
\tableline
1 & $\mathrm{0.02\ldots(1.5\times 10^{-5})}$ & thermal & mechanical &
 CIE\tablenotemark{b} \\
2 & $10^{-3}$ & no & mechanical & isothermal EOS\tablenotemark{c} \\
3 & $\mathrm{4\ldots5~/~(28\ldots7.2)\times 10^{-3}}$ & no & mechanical &
 CIE + cutoff \\
4 & $0.02\ldots0.33$ & PIE\tablenotemark{d} & mech. + PIE & CIE + cutoff \\
5 & $0.023$ & no & mechanical & CIE \\
6 & $8\times 10^{-4}$ & no & mechanical & CIE \\
7 & 0.015 & no & mechanical & CIE + cutoff \\
8 & 4 & time dependent\tablenotemark{e} & mech. + radiative &
 explicit or CIE\tablenotemark{f} \\
\tableline\tableline
Ref. & Wind asymmetry & $L_{\mathrm{w}}(t)$ & $L_{\mathrm{LyC}}(t)$ &
 $n_{\mathrm{0}}(r)$ \\
\tableline
1 & no & fixed / two-level & 0 & const. / composite \\
2 & no & two-level & 0 & const. \\
3 & no & variable & variable & const. / from 1D \\
4 & no & 0 & fixed & const. / power law \\
5 & yes & two-level with trans. & 0 & pre WR \\
6 & yes & three-level & 0 & const. \\
7 & no & const. & 0 & const. \\
8 & no & variable & variable & const. \\
\tableline
\end{tabular}


\tablenotetext{a}{in units of linear size}
\tablenotetext{b}{cooling function based on the assumption of
                  collisional ionization equilibrium}
\tablenotetext{c}{equation of state}
\tablenotetext{d}{photo ionization equilibrium calculated assuming
                  that the gas is fully ionized inside the \HII region}
\tablenotetext{e}{thermal and radiative ionization of hydrogen}
\tablenotetext{f}{explicit calculation of important cooling processes for
                  lower temperatures and CIE for high temperatures}


\tablerefs{
(1) \citet{rozyczka85a}, \citet{rozyczka85b, rozyczka85c};
(2) \citet{stone95};
(3) \citeauthor{garcia96a}, \citet{garcia96b};
(4) \citet{garcia96c};
(5) \citet{brighenti97};
(6) \citet{frank98};
(7) \citet{strickland98};
(8) This work.}

\end{center}
\end{table}

\bf
\clearpage

\begin{table}
\begin{center}
\caption{The energy components at the end of the simulation.
         \label{table_num_results}}
\begin{tabular}{cccccc}
\tableline\tableline
Model                    & $E_{\mathrm{k}}$         & $E_{\mathrm{i}}$        &
$E_{\mathrm{t,cold}}$    & $E_{\mathrm{t,warm}}$    & $E_{\mathrm{t,hot}}$   \\
                         & $(10^{49}\mathrm{ergs})$ & $(10^{49}\mathrm{ergs})$&
$(10^{49}\mathrm{ergs})$ & $(10^{49}\mathrm{ergs})$ & $(10^{49}\mathrm{ergs})$
\\
\tableline
windless  & $3.5$  & $13$  &
$5.5$     & $4.7$  & $0$   \\
with wind & $14$   & $10$  &
$7.9$     & $2.1$  & $9.7$ \\
\tableline
\end{tabular}
\tablecomments{The thermal energy of the cold component, $E_{\mathrm{t,cold}}$,
   contains the internal energy of the initially unperturbed ambient medium
   ($3.4 \times 10^{49}\mathrm{ergs}$) that has to be subtracted whenever
   the input of thermal energy into the system is considered.}
\end{center}
\end{table}

\rm

\bf
\clearpage

\begin{table}
\begin{center}
\caption{The energy transfer efficiencies at the end of the simulation.
         \label{table_num_eff}}
\begin{tabular}{cccc}
\tableline\tableline
Model                      & $\varepsilon_{\mathrm{k}}$ &
$\varepsilon_{\mathrm{i}}$ & $\varepsilon_{\mathrm{t}}$ \\
                           & $(10^{-4})$                &
$(10^{-4})$                & $(10^{-4})$                \\
\tableline
windless  & $3.3$ &
$12$      & $6.4$ \\
with wind & $13$  &
$9.1$     & $15$  \\
\tableline
\end{tabular}
\end{center}
\end{table}

\rm






\end{document}